\documentclass[11pt,a4paper]{article}
\pdfoutput=1

\usepackage{amssymb,amsmath,amsfonts, mathtools, mathrsfs}
\usepackage[utf8]{inputenc} 
\usepackage[dvipsnames]{xcolor}
\usepackage{comment}
\usepackage[justification=justified]{subcaption}
\newif\ifnatbibsort\natbibsorttrue

\relax

\ifnatbibsort\RequirePackage[numbers,sort&compress]{natbib}\else\RequirePackage[numbers,compress]{natbib}\fi
\RequirePackage[colorlinks=true
,urlcolor=blue
,anchorcolor=blue
,citecolor=blue
,filecolor=blue
,linkcolor=blue
,menucolor=blue
,pagecolor=blue
,linktocpage=true
,pdfproducer=medialab
,pdfa=true
]{hyperref}

\usepackage[labelfont=bf,justification=centering,
   format=plain]{caption}
\usepackage[]{graphicx,geometry}
\usepackage{tensor}
\usepackage{enumerate}
\usepackage{multirow}
\usepackage{dsfont}

\usepackage{verbatim}
\usepackage{amsfonts,euscript,amssymb,stmaryrd,braket}
\usepackage{tikz}
\usepackage{import}
\usetikzlibrary{arrows,decorations.markings,patterns}
\usepackage{slashed}

\def\clock{{\count0=\time
		\divide\count0 60
		\ifnum\count0<10 0\fi\the\count0
		\multiply\count0 -60 \advance\count0 \time
		:\ifnum\count0<10 0\fi \the\count0
}}
\newcommand{\timestamp}{{\small\vbox{\hbox{\tt\jobname.tex}
			\hbox{\the\day/\the\month/\the\year, \clock}}}}

\newcommand{\bea}{\begin{eqnarray}}
\newcommand{\eea}{\end{eqnarray}}

\DeclareMathOperator{\Tr}{Tr}

\newcommand{\be}{\begin{equation}}
\newcommand{\ee}{\end{equation}}

\makeatletter
\let\old@startsection=\@startsection
\let\oldl@section=\l@section
\renewcommand{\@startsection}[6]{\old@startsection{#1}{#2}{#3}{#4}{#5}{#6\mathversion{bold}}}
\renewcommand{\l@section}[2]{\oldl@section{\mathversion{bold}#1}{#2}}
\makeatother

\numberwithin{equation}{section}

\newsavebox{\largestimage}

\usepackage{color}

\def\rd {{\rm d}}
\def\e {{\rm e}}

\begin{document}
	\renewcommand{\thefootnote}{\arabic{footnote}}

	\overfullrule=0pt
	\parskip=2pt
	\parindent=12pt
	\headheight=0in \headsep=0in \topmargin=0in \oddsidemargin=0in

	\vspace{ -3cm} \thispagestyle{empty} \vspace{-1cm}
	\begin{flushright} 
		\footnotesize
		\textcolor{red}{\phantom{print-report}}
	\end{flushright}

\begin{center}
	\vspace{.0cm}

	{\Large\bf \mathversion{bold}
	A contour for the entanglement negativity
	}
\\
	\vspace{.25cm}
	\noindent
	{\Large\bf \mathversion{bold}
	of bosonic Gaussian states}

	\vskip  0.8cm
	{
		Gioele Zambotti
		and 
        Erik Tonni
	}
	\vskip  1.cm
	
	\small
	{\em
        SISSA, via Bonomea 265, 34136, Trieste, Italy 
        \vskip 0.09cm
		INFN, Sezione di Trieste, via Valerio 2, 34127 Trieste, Italy
	}
	\normalsize

\end{center}

\vspace{0.3cm}
\begin{abstract} 
We construct a contour function 
for the logarithmic negativity 
and the logarithm of the moments of the partial transpose of the
reduced density matrix 
for multimode bosonic Gaussian states 
of a free lattice model. 
In one spatial dimension, numerical results are obtained
for harmonic chains 
either in the ground state or at finite temperature,
by considering, respectively, either a subsystem made by 
two adjacent or disjoint blocks on the line
or a bipartition of the circle. 
The contour function of the logarithmic negativity 
 diverges only at the entangling points,
while the contour function for the logarithm of the moments of the partial transpose is divergent 
also at the boundary of the bipartite subsystem,
as functions of the position. 
In a two-dimensional conformal field theory, 
analytic expressions that describe these divergencies are discussed. 
In one spatial dimension, we explore 
the partial derivative of the logarithmic negativity of two adjacent intervals
with respect to the logarithm of the harmonic ratio of their lengths
while their ratio and the other parameters are kept fixed.
%
Considering the ground state 
of the harmonic chain on the line and in the massive regime, 
we report numerical results 
showing that this quantity displays a monotonically decreasing behaviour.
\end{abstract}

\newpage
\tableofcontents

\section{Introduction}
\label{sec:intro}

The bipartite entanglement in quantum many-body systems and quantum field theories has been widely investigated during the past few decades. 
For pure states, the standard measure of bipartite entanglement is the entanglement entropy 
\cite{Bombelli1986,Srednicki1993,Callan:1994py,
Holzhey:1994we,Calabrese:2004eu}
(see also the reviews \cite{Casini:2009sr,Eisler:2009vye,Calabrese:2009qy,Eisert2010,Weedbrook2012}).
Given a quantum system in a pure state $| \psi \rangle$ (like e.g. its ground state), 
let us consider the case 
where the Hilbert space $\mathcal{H}$ can be factorized as 
$\mathcal{H} = \mathcal{H}_A \otimes \mathcal{H}_B$, 
where $A$ and $B$ are the regions providing a spatial bipartition of a slice of the spacetime at fixed time. 
The reduced density matrix of the subsystem $A$ is 
defined as $\rho_A \equiv \textrm{Tr}_{\mathcal{H}_B} 
(| \psi \rangle \langle \psi |)$, 
normalized by the condition 
$\textrm{Tr}_{\mathcal{H}_A} \rho_A = 1$,
and it characterizes a mixed state. 
The entanglement entropy and the R\'enyi entropies of $A$
are defined respectively as 
\begin{equation}
\label{EntEnt-def-intro}
    S_{A} \equiv  - \Tr\!\big( \rho_{A} \log \rho_{A}\big)
\;\;\;\;\;\qquad\;\;\;
    S_A^{(n)} \equiv  \frac{1}{1-n} \, \log \Tr  \rho_{A}^n
\end{equation}
where the integer $n \geqslant 2$ is the R\'enyi index
and we have omitted the subindex for the trace, assuming that it is taken over the proper vector space
($\mathcal{H}_A$ in this case). 
The entanglement entropy can be obtained from the 
 R\'enyi entropies through the replica limit
$S_{A} = \lim_{n \to 1} S_{A}^{(n)}$. 
In two spacetime dimensions,
a fundamental result involving  
the entanglement entropy $S_A$ 
when the whole system is in its ground state
is the entropic $C$-theorem 
for the renormalization group (RG) flows,
found in \cite{Casini:2004bw}. 
Considering an RG flow in two spacetime dimensions, 
when the spatial direction is an infinite line 
bipartite by an interval $A$ of length $\ell$ and its complement, this theorem states that the UV finite quantity 
$C_S \equiv \ell \, \partial_\ell S_A$ 
monotonically decreases along the RG flow. 
Consistency checks for this important result
have been performed in free fermionic and bosonic lattice models 
\cite{Casini:2005rm, Casini:2005zv}.

In a quantum many-body system on a lattice, 
the spatial structure of the bipartite entanglement 
of a pure state 
\cite{Botero:2004vpl, ChenVidal2014,Fr_rot_2015, Coser:2017dtb}
can be probed by introducing the functions  
$\mathsf{S}_A : A \to \mathbb{R}$
and
$\mathsf{S}_A^{(n)}: A \to \mathbb{R}$,
assigning a real number to every site in $A$, 
that provide the entanglement entropies (\ref{EntEnt-def-intro})
through the normalization condition 
\begin{equation}
\label{entropy-density-property}
    S_A = \sum_{i \,\in \, A} \mathsf{S}_A(i)
\;\;\;\;\;\qquad\;\;\;\;\;
    S_A^{(n)} = \sum_{i \,\in \, A} \mathsf{S}_A^{(n)}(i)
\end{equation}
and satisfy also the positivity condition 
\begin{equation}
\label{entropy-contour-positivity}
\mathsf{S}_A(i) \geqslant 0
\;\;\;\;\;\qquad\;\;\;\;\;
    \mathsf{S}_A^{(n)}(i) \geqslant 0
    \;\;\;\;\;\qquad\;\;\;\;\;
    \forall \, i \in A \,.
\end{equation}
The functions $\mathsf{S}_A(i)$ and $\mathsf{S}_A^{(n)}(i)$,
which are dubbed the contour functions for 
the entanglement entropy and 
the R\'enyi entropies respectively 
\cite{ChenVidal2014},
are naturally interpreted as 
densities for $S_A$ and $S_A^{(n)}$ respectively. 
We remark that many functions satisfying 
(\ref{entropy-density-property}) and (\ref{entropy-contour-positivity}) can be constructed. 
Further properties have been introduced in \cite{ChenVidal2014}
to constraint the construction of $\mathsf{S}_A(i)$ and $\mathsf{S}_A^{(n)}(i)$,
but they are still not enough to identify the contour functions for the entanglement entropies in a unique way. 
The definition of contour functions for the entanglement entropies in quantum field theories,
where the discrete sums in (\ref{entropy-density-property}) become integrals, requires some care because the entanglement entropies are divergent quantities as the ultraviolet (UV) cutoff vanishes. 
This is relevant for quantum many-body systems at criticality, 
whose  continuum limit is described 
by conformal field theories (CFT)
\cite{DiFrancesco:1997nk,Ginsparg:1988ui}.
The symmetry of the CFT
allows to obtain various insightful analytic results
also for the entanglement entropies 
where $A$ is the union of two or more disjoint intervals
\cite{Calabrese:2004eu, Caraglio:2008pk,Furukawa:2008uk, Casini:2009vk, Calabrese:2009ez,Calabrese:2010he, Coser:2013qda,Grava:2021yjp}.
In quantum field theories, the contour functions for the entanglement entropies have been explored in 
\cite{Nozaki:2013wia,Nozaki:2013vta,Wong:2013gua,Bhattacharya:2014vja,Cardy:2016fqc,Tonni:2017jom,Tonni18:bariloche-talk,Wen:2018whg,Kudler-Flam:2019oru,Wen:2020ech,Han:2019scu,Mintchev:2022fcp, Caggioli:2024uza,Caminiti:2025hjq},
by employing also the AdS/CFT correspondence 
\cite{Ryu:2006bv,Ryu:2006ef,Freedman:2016zud,Agon:2018lwq}
and the algebraic approach \cite{Haag:1996hvx}.

In this manuscript we mainly focus on multimode Gaussian states of continuous-variable systems
\cite{Weedbrook2012, Serafini:2017rrn, Holevo2011} which
provide a very insightful and tractable arena to explore free bosonic quantum many-body systems.
In continuous-variable systems, the dimension of the Hilbert space is infinite 
and the spectra of the observables are labeled by continuous parameters. 
The prototypical continuous-variable system is given by 
$N$ bosonic modes associated to the Hilbert space 
$\mathcal{H} = \otimes_{k=1}^N \mathcal{H}_k$.
An important example is 
the harmonic lattice in $d$ spatial dimensions 
(each direction is discretized in the same way)
made by $N$ sites interacting through a nearest neighbors spring-like interaction, 
which is described by the following Hamiltonian \cite{Weedbrook2012}
\begin{equation}
\label{hamiltonian}
    H = \sum_{i=1}^N \left( \frac{1}{2m_0} \, \hat{p}_i^2 
    + \frac{m_0 \,\omega_0^2}{2} \,\hat{q}_i^2\right) 
    + 
    \frac{\kappa}{2} \sum _{\langle i,j\rangle}\left( \hat{q}_{j} - \hat{q}_i \right)^2 
\end{equation}
with certain boundary conditions, 
where the canonically conjugate operators $\hat{q}_i$ and $\hat{p}_i$
satisfy the well known canonical commutation relations 
$\big[\hat{q}_i ,\hat{p}_j\big] = \textrm{i} \delta_{i,j}$
and the sum over $\langle i,j\rangle$ involve the neighbours sites 
at relative distance of a single lattice unit. 
The Hamiltonian \eqref{hamiltonian} can be interpreted as the discretization of the free scalar boson with mass $m = \omega_0$  on a lattice with spacing 
$\mathsf{a} \equiv \sqrt{m_0/\kappa}$.
The invariance under canonical transformations implies that the only 
relevant parameter is the dimensionless parameter 
$\omega \equiv \sqrt{m_0\,\omega_0^2/\kappa} = m\,\mathsf{a}$. 
When $d=1$, the Hamiltonian \eqref{hamiltonian} 
describes the harmonic chain 
and its continuum limit $\mathsf{a} \to 0$ gives the Hamiltonian of 
the massive scalar field on the infinite line,
whose massless limit  is a conformal field theory with central charge $c=1$.
The bipartite entanglement in harmonic lattices 
has been widely explored in the literature
\cite{Plenio_2005,Cramer_2006,Eisert2010,Weedbrook2012,
Bombelli1986,Srednicki1993,Peschel1999,
Audenaert_2002,Botero:2004vpl}.
Our numerical analyses focus on 
(\ref{hamiltonian}) in one spatial dimension,
i.e. on harmonic chains, 
and only some Gaussian states 
(the ground state and the thermal state).

In the setup given by the 
multimode bosonic Gaussian states of continuous-variable systems,
a proposal for the contour functions of the entanglement entropies
fulfilling
the normalization condition (\ref{entropy-density-property}), 
the positivity condition (\ref{entropy-contour-positivity}) 
and a weaker version of the properties introduced in \cite{ChenVidal2014}
has been introduced in \cite{Coser:2017dtb}
(see \cite{Fr_rot_2015} for a proposal where 
(\ref{entropy-density-property}) is satisfied, 
while (\ref{entropy-contour-positivity}) is not guaranteed).
For free fermionic systems on the lattice, 
the construction of the contour functions for the entanglement entropies
proposed in \cite{ChenVidal2014} has been mainly explored.

Quantifying the bipartite entanglement when the whole quantum system is in a mixed state (e.g. the thermal state) is a problem of crucial importance in quantum information theory. 
In contrast to the case of the pure states, for mixed states 
a unique quantity measuring the bipartite entanglement does not exist. 
Different entanglement measures have been proposed \cite{Bennett:1995ra,Bennett:1995tk,Bennett:1996gf,Wootters:1997id,Vedral:1997qn,Terhal:2000zqp,Plenio:2007zz},
but many of them are difficult to evaluate because their definition involves an extremization procedure.
A measure whose evaluation does not require
an extremization procedure is the logarithmic negativity 
\cite{Peres:1996dw,Vidal:2002zz,Plenio:2005cwa,Eisert:2006kue}.

Consider a mixed state for a quantum system defined 
on the spatial region $A=A_1 \cup A_2$, 
bipartite into the subsystem in $A_1$ 
and the one in its complement $A_2 = A\setminus A_1$, 
and denote by $\rho_A$ the density matrix characterizing 
this mixed state. 
The partial transpose e.g. with respect to $A_2$ 
is the operator whose generic element in the basis 
$\big\{\, \big|e_k^{(1)}e_l^{(2)} \big\rangle \, \big\}$ 
of $\mathcal{H}_A = \mathcal{H}_1 \otimes \mathcal{H}_2$ 
is defined as follows
\begin{equation}
\label{rhoA-T2-def-intro}
\big\langle e_{i}^{(1)}e_{j}^{(2)} \big| \, 
    \rho_{A_1 \cup A_2}^{\textrm{\tiny $\Gamma_2$}}
\, \big|e_k^{(1)}e_l^{(2)} \big\rangle
    \equiv 
\big\langle e_{i}^{(1)}e_{l}^{(2)} \big| \, 
    \rho_{A_1 \cup A_2}
\, \big|e_k^{(1)}e_j^{(2)} \big\rangle \,.
\end{equation}
The operator $\rho_{A_1 \cup A_2}^{\textrm{\tiny $\Gamma_2$}}$ 
is not positive definite, in general, 
but still satisfies the normalization condition
$\textrm{Tr}_{\mathcal{H}_A} 
\rho_{A_1 \cup A_2}^{\textrm{\tiny $\Gamma_2$}} = 1$. 
The logarithmic negativity is defined 
by taking the logarithm of the trace norm of 
$\rho_{A_1 \cup A_2}^{\textrm{\tiny $\Gamma_2$}}$, 
namely 
\begin{equation}
\label{log-neg-def-intro}
    \mathcal{E} 
    \equiv \,
    \log \big|\!\big|
    \rho_{A_1 \cup A_2}^{\textrm{\tiny $\Gamma_2$}}
    \big|\!\big|
    \,= \,
    \log \Tr 
    \big|
    \rho_{A_1 \cup A_2}^{\textrm{\tiny $\Gamma_2$}}
    \big| 
\end{equation}
and depends only on the negative eigenvalues of 
$\rho_{A_1 \cup A_2}^{\textrm{\tiny $\Gamma_2$}}$.
A replica trick to evaluate (\ref{log-neg-def-intro})
has been introduced in 
\cite{Calabrese:2012ew, Calabrese:2012nk},
by first taking the logarithm of the moments 
of the partial transpose, namely
\begin{equation}
\label{neg-moments-def-intro}
    \mathcal{E}^{(n)} 
    \equiv \,
    \log\!  
    \left[\Tr \!\big( \rho_{A_1 \cup A_2}^{\textrm{\tiny $\Gamma_2$}}\big)^{n} \right]
\end{equation}
and then considering the sequence given by 
the even integers $n=n_e$. 
The analytic continuation $n_{\textrm{\tiny e}} \to 1$
in this sequence labeled by the even integers $n_e \geqslant 2$
gives the logarithmic negativity (\ref{log-neg-def-intro}), i.e.
\be
\label{neg-replica-limit-intro}
\mathcal{E} = \lim_{n_{\textrm{\tiny e}} \to 1} 
\mathcal{E}^{(n_{\textrm{\tiny e}})} \,.
\ee

Analytic results for (\ref{log-neg-def-intro}) and 
(\ref{neg-moments-def-intro})
in a CFT in two spacetime dimensions (CFT$_2$) have been found in 
\cite{Calabrese:2012ew, Calabrese:2012nk, Calabrese:2014yza}.
In particular, for a CFT$_2$ on the line and in its ground state,
when the bipartite subsystem $A$ is the union of two finite and adjacent intervals $A_1 = (a,p)$, and $A_2 = (p,b)$,
the logarithmic negativity $\mathcal{E}$ diverges logarithmically in a universal way; 
indeed, its explicit expression reads 
\begin{equation}
\label{logneg-CFT-adj-intro}
    \mathcal{E} = 
    \frac{c}{4} \, 
    \log \! \left( 
    \frac{(b-p)(p-a)}{(b-a)\, \epsilon}
    \right) 
    + \textrm{const}
    \end{equation}
where $c$ is the central charge of the CFT$_2$.

Considering Gaussian states of quadratic models on the lattice, the action of the partial transpose depends on whether the underlying system is bosonic or fermionic. 
Indeed, while for bosonic Gaussian states the partial transpose preserves  
the Gaussian nature of the state
\cite{Audenaert_2002,Simon:1999lfr}, 
the partial transpose of a fermionic Gaussian state is not Gaussian anymore \cite{Eisler:2015tgq}.
For this reason, for free fermionic systems
the partial time-reversal of $\rho_A$
has been introduced in \cite{Shapourian:2016cqu}.
The partial time-reversal of $\rho_A$ is still Gaussian, 
and therefore it is more accessible from the computational viewpoint
\cite{Shapourian:2018ozl,Ruggiero:2016yjt,Shapourian:2018lsz,
Shapourian:2019xfi,Murciano:2022vhe,Rottoli:2022plr, Arias:2026bqh}.
In this paper we consider only 
multimode bosonic Gaussian states of quadratic lattice models, 
whose Gaussian nature is preserved by the partial transposition.
Our numerical analyses involves the special cases in one spatial dimension
given by some Gaussian states for the harmonic chains.

In this paper we mainly investigate 
contour functions for the logarithmic negativity 
and for the logarithm of the moments of the partial transpose 
(see (\ref{log-neg-def-intro}) and (\ref{neg-moments-def-intro}) respectively). 
An explicit construction of the contour function 
for the logarithmic negativity for Gaussian states in harmonic lattices
has been proposed in \cite{DeNobiliThesis},
while another construction  
for the logarithmic negativity associated 
to the partial time-reversal of the reduced density
matrix for free Gaussian fermionic states 
has been proposed in \cite{Kudler-Flam:2019nhr}.
These constructions have been inspired from the ones 
proposed in \cite{Fr_rot_2015} and \cite{ChenVidal2014} respectively,
for the contour function of the entanglement entropy.

Given a quantum many-body system on a lattice 
defined on the spatial domain $A$,
either in a pure or mixed state, 
which is bipartite in $A_1$ and $A_2$
(see the setup introduced in the text 
above (\ref{rhoA-T2-def-intro})), 
we study functions  
$\mathsf{E}_A : A \to \mathbb{R}$
for the logarithmic negativity
and
$\mathsf{E}_A^{(n)}: A \to \mathbb{R}$
for the logarithm of the moments of the partial transpose
that satisfy  the normalization condition 
\begin{equation}
\label{neg-density-property-intro}
    \mathcal{E} = \sum_{i \,\in \, A} 
    \mathsf{E}_A(i)
\;\;\;\;\qquad\;\;\;
    \mathcal{E}^{(n)} = \sum_{i \,\in \, A} 
    \mathsf{E}_A^{(n)}(i)
    \;\;\;\;\qquad\;\;\;
    A = A_1  \cup A_2
\end{equation}
and the positivity condition
\begin{equation}
\label{neg-density-positivity-property-intro}
    \mathsf{E}_A(i) \geqslant 0
    \;\;\;\;\;\qquad\;\;\;\;\;
    \forall \, i \in A \,.
\end{equation}
The positivity condition is not imposed on 
the contour functions $\mathsf{E}_A^{(n)}(i)$
because $\mathcal{E}^{(n)}$ do not have a definite sign.

The main result of this paper is an explicit construction 
of the contour function $\mathsf{E}_A(i)$ and $\mathsf{E}_A^{(n)}(i)$ for the 
multimode bosonic Gaussian states of quadratic lattice models 
in a generic spacetime dimensionality 
that satisfy (\ref{neg-density-property-intro}) and (\ref{neg-density-positivity-property-intro}).
We remark that 
the positivity condition (\ref{neg-density-positivity-property-intro}) 
does not hold  in general for the proposal made in \cite{DeNobiliThesis}; 
as highlighted by the explicit violation of this condition 
shown in Fig.\,4.9 and Fig.\,4.10 of \cite{DeNobiliThesis}.
Our results are obtained by adapting to the partial transpose of the reduced density matrix the procedure followed in 
\cite{Coser:2017dtb} for the reduced density matrix 
of a Gaussian state, and both the conditions 
(\ref{neg-density-property-intro}) 
and (\ref{neg-density-positivity-property-intro}) 
are always satisfied, by construction. 
In the continuum limit, for the ground state of a CFT$_2$ 
we discuss some universal features of the contour functions 
for $\mathcal{E}$ and $\mathcal{E}^{(n)}$.
Numerical results for $\mathsf{E}_A(i)$ and $\mathsf{E}_A^{(n)}(i)$
are discussed in the case of harmonic chains, either on the line or on the circle, for the Gaussian mixed states given by 
either some reduced density matrices 
obtained from the  ground state 
or the thermal state. Inspired by the CFT$_2$ result (\ref{logneg-CFT-adj-intro}),
we also introduce a UV finite quantity 
that might be useful to study RG flows
(see Sec.\,\ref{subsec-GS-line-adjacent-logneg}) and
show that the corresponding numerical results obtained from the harmonic chain in the massive regime display the expected features. 

The outline of this paper is as follows. 
In Sec.\,\ref{sec:contour-entropies} 
we review the construction of the contour functions of the entanglement entropies for multimode bosonic Gaussian states of quadratic lattice models introduced in \cite{Coser:2017dtb}.
In Sec.\,\ref{sec:contour-neg}, 
this construction is adapted to the partial transpose of the reduced density matrix and explicit expressions for contour functions 
$\mathsf{E}_A(i)$ and $\mathsf{E}_A^{(n)}(i)$ 
are provided.
As for the continuum limit, 
considering the ground state of a CFT$_2$,
in Sec.\,\ref{sec:cft} we discuss 
some features of the contour functions 
for the logarithmic negativity 
and the logarithm of the moments of the partial transpose
for the reduced density matrices of two adjacent and disjoint intervals. 
Numerical results for $\mathsf{E}_A(i)$ and $\mathsf{E}_A^{(n)}(i)$ 
for the ground state of the harmonic chain in the line 
when $A$ is made by either two adjacent or two disjoint blocks 
are discussed in 
Sec.\,\ref{sec-GS-line-adjacent} and 
Sec.\,\ref{sec-GS-line-disjoint} respectively. 
%
The case of the harmonic chain on a circle at finite temperature is considered in Sec.\,\ref{sec-thermal-state},
where $\mathsf{E}_A(i)$ for 
two adjacent blocks partitioning the circle is briefly explored. 
Finally, in Sec.\,\ref{sec:conclusions} we draw some conclusions. 
In Appendices\;\ref{app-contour-properties}, 
\ref{app-relation} and \ref{app-W-matrix}, 
some technical details and further results 
supporting the discussion in the main text are reported.

\section{A contour function for the entanglement entropies}
\label{sec:contour-entropies}

In this section we review the definition of the contour functions 
for the entanglement entropies, their properties \cite{ChenVidal2014}
and a specific construction proposed in \cite{Coser:2017dtb}
for the multimode bosonic Gaussian states of free lattice models.

In continuous-variable systems, one considers
$\hat{\boldsymbol{r}}\equiv 
\big(\hat{q}_1, \dots \hat{q}_N , \hat{p}_1, \dots \hat{p}_N \big)^{\textrm{t}}$,
dubbed the vector of the quadrature operators,
whose elements satisfy the canonical commutation relations,
given by 
$\big[ \hat{r}_i , \hat{r}_j \big] = \textrm{i} \, J_{i,j}$,
where $J$ is the standard symplectic matrix 
\be
\label{J-mat-def}
J \equiv 
 \bigg( \,
\begin{array}{cc}
\boldsymbol{0}  \; \,& \boldsymbol{1}
\\
-\boldsymbol{1}   \;\, & \boldsymbol{0}
\end{array} 
\,\bigg)
\ee
where $\boldsymbol{1}$ and $\boldsymbol{0}$ 
are the $N \times N$ identity matrix and the $N \times N$ matrix made by zeros
respectively. 
Notice that $J^2 = - \boldsymbol{1}$ and $J^{\textrm{t}}=-J$.
The real symplectic group $\textrm{Sp}(N)$ is made by the $(2N)\times (2N)$ 
real matrices $M$ satisfying $M J M^{\textrm{t}} = J$.
For any $M \in \textrm{Sp}(N)$, 
we have that $\textrm{det}(M)=1$ and $M^{\textrm{t}} \in \textrm{Sp}(N)$ \cite{de2006symplectic}.
These matrices play a crucial role because the linear transformations $\hat{\boldsymbol{r}} \mapsto \hat{\boldsymbol{r}}' = M \hat{\boldsymbol{r}}$ with $M \in \textrm{Sp}(N)$ preserve the canonical commutation relations. 

%
A Gaussian state is fully characterized by the correlators 
$\langle \hat{r}_i \rangle$ and $\langle \hat{r}_i\, \hat{r}_j \rangle$,
i.e. the first and second moments respectively.
A proper unitary transformation provides a shift in the first moments
and preserves the Gaussian nature of the state; hence we restrict to Gaussian states having vanishing first moments, which are fully described by the $(2N)\times (2N)$
covariance matrix $\gamma \equiv \textrm{Re} \big( \langle \hat{\boldsymbol{r}} \,\hat{\boldsymbol{r}}^{\textrm{t}} \rangle\big)$ \cite{de2006symplectic}.
Since the covariance matrix $\gamma$ is real, symmetric and positive definite,
the Williamson theorem allows us to introduce its symplectic eigenvalues,
which are larger or equal to $1/2$ because of the uncertainty principle
\cite{de2006symplectic,Simon1994}.
For any pure state, we have that 
$(\textrm{i} J \gamma)^2 = \tfrac{1}{4} \boldsymbol{1}$
and all the symplectic eigenvalues of its covariance matrix are equal to $1/2$ \cite{Lindblad2000}.
A canonical change of coordinates is implemented by the linear map $\hat{\boldsymbol{r}} \mapsto \hat{\boldsymbol{r}}' = M \hat{\boldsymbol{r}}$ 
with $M \in \textrm{Sp}(N)$ 
and it induces on the covariance matrix 
the transformation $\gamma \mapsto \gamma' = M \gamma M^{\textrm{t}}$.

Considering a bipartition of the system provided by the domain $A$ 
and its complement $B$, 
which contain $L_A$ and $L_B = N- L_A$ sites respectively,
the reduced density matrix $\rho_A$  is fully described by the $(2L_A)\times (2L_A)$ reduced covariance matrix $\gamma_A$, obtained by extracting from $\gamma$ the elements corresponding to the sites in $A$, 
namely $(\gamma_A)_{i,j} \equiv \gamma_{i,j}$ for $i,j \in A$.
Since $\gamma_A$ is real, symmetric and positive definite, 
its Williamson decomposition 
\cite{Williamson1936,Dragt1992,Simon1999} 
reads
\begin{equation}
\label{williamson-dec-gamma-A}
    \gamma_A = W^{\textrm{t}} 
    \big(  D \oplus D \big) W
\end{equation}
where $W \in \textrm{Sp}(L_A)$ and $D = \textrm{diag}(\sigma_1 , 
\dots , \sigma_{L_A})$ is the $L_A \times L_A$ diagonal matrix 
having the symplectic eigenvalues of $\gamma_A$ on the diagonal. 
In our numerical analyses, the symplecitic eigenvalues are ordered in the increasing way, i.e. 
$\sigma_1 \leqslant \sigma_2 \leqslant \dots \leqslant \sigma_{L_A}$.

The entanglement entropy $S_A$ and the R\'enyi entropies $S_A^{(n)}$ 
with $n \geqslant 2$ in (\ref{EntEnt-def-intro})
are obtained from the symplectic spectrum of $\gamma_A$ 
as follows 
\cite{Bombelli1986,Srednicki1993,Peschel1999,Audenaert_2002,Plenio_2005,Cramer_2006,Botero:2004vpl,Amico2008,Eisert2010} 
\be
\label{ee-from-sigma-k}
S_{A} = \sum_{k=1}^{L_A} F_1\left(\sigma_k\right)
\;\;\;\;\qquad\;\;\;
S_{A}^{(n)} = \sum_{k=1}^{L_A} F_n\left(\sigma_k\right)
\ee
where
\bea
\label{F1-entropy-def}
F_1(x) 
&\equiv& 
(x+1/2) \,\log(x+1/2) - (x-1/2)\, \log(x-1/2)
\\
\label{Fn-entropy-def}
\rule{0pt}{.8cm}
F_n(x) 
&\equiv& 
\frac{1}{n-1}\, \log \! \Big[  \big( x+1/2 \big)^n - \big( x -1/2 \big)^n \Big]
\hspace{2.5cm}
n\neq 1 \,.
\eea
Notice  that the entanglement entropies in (\ref{ee-from-sigma-k}) can be also written respectively as 
\begin{equation}
    \label{ee-from-O}
S_{A} = \frac{1}{2} \, 
\textrm{Tr} \big[ F_1(\Upsilon_A) \big]
\;\;\qquad\;\;
S^{(n)}_{A} = \frac{1}{2} \, 
\textrm{Tr} \big[ F_n(\Upsilon_A) \big]
\;\;\;\;\qquad\;\;\;\;
\Upsilon_A \equiv 
O^{\textrm{t}} \big(  D \oplus D \big) O
\end{equation}
where $O$ is a generic $(2L_A)\times (2L_A)$ orthogonal matrix.  

Beside the basic features of $\mathsf{S}_A(i)$ 
given by (\ref{entropy-density-property}) 
and (\ref{entropy-contour-positivity}),
further reasonable constraints have been proposed 
in \cite{ChenVidal2014}.
However, a list of properties that identifies the contour function 
$\mathsf{S}_A(i)$ in a unique way is not available in the literature. 
In order to describe the other properties of  $\mathsf{S}_A(i)$,
we need to introduce the following quantity
\begin{equation}
\label{ent-contour-subregion}
    \mathsf{S}_A(\tilde{A}) \equiv 
    \sum_{i\in\tilde{A}} \, \mathsf{S}_A(i)
    \;\;\;\;\qquad\;\;\;\;
    \tilde{A} \subseteq A
\end{equation}
which satisfies $\mathsf{S}_A(A) = S_A$
and has been studied numerically e.g. in \cite{DiGiulio:2019lpb}.
The conditions introduced in \cite{ChenVidal2014} are:
\newline
(a) {\it Symmetries.}
A transformation of the reduced density matrix  
$\rho_{A} \mapsto \rho'_{A} = U^{\dagger} \rho_{A} U $,
characterized by the unitary matrix $U$, 
is a symmetry when $\rho'_{A}= \rho_{A}$.
The contour function 
$\mathsf{S}_A(i)$ should be consistent with any 
spatial symmetry of $\rho_{A}$ associated to the geometry of $A$; 
hence, if there exists a unitary transformation 
sending the site $i$ onto the site $j$, 
for $i,j \in A$, 
that leaves $A$ invariant, 
then $\mathsf{S}_{A}(i)=\mathsf{S}_{A}(j)$ must hold. 
\\
(b) {\it Local unitary transformations.}
The amount of entanglement localized in a subregion $\tilde{A} \subseteq A$ cannot change when a local unitary transformation acting 
non trivially only in $\tilde{A}$ 
is applied to $\rho_{A}$.
\\
(c) {\it Upper bound.}
Assuming the Hilbert space decompositions given by 
$\mathcal{H} = \mathcal{H}_{A} \otimes \mathcal{H}_{B} $
and $\mathcal{H}_{j} = \mathcal{H}_{\Omega_j}\otimes\mathcal{H}_{\widetilde{\Omega}_j}$,
with $j \in \big\{A, B \big\}$,
if the factorization $\rho = \rho_{\Omega_A\Omega_B}
\otimes \rho_{\widetilde{\Omega}_A \widetilde{\Omega}_B}$ holds 
for the state of the whole system,
where $\rho_{\Omega_A\Omega_B}$ and $\rho_{\widetilde{\Omega}_A \widetilde{\Omega}_B}$ denote respectively the density matrix resulting from the trace over $\mathcal{H}_{\widetilde{\Omega}_A} \otimes \mathcal{H}_{\widetilde{\Omega}_B}$ in $\mathcal{H}$ and over
$\mathcal{H}_{\Omega_A} \otimes \mathcal{H}_{\Omega_B}$ in $\mathcal{H}$,
then $S_A=S_{\Omega_A}+S_{\widetilde{\Omega}_A}$.
In general, $\{\Omega_j,\widetilde{\Omega}_j\}$ 
do not necessarily correspond to a spatial bipartition of $j$.
      If  a subregion 
      $\tilde{A}\subseteq A$ exists  
      such that $\mathcal{H}_{\tilde{A}} \subseteq \mathcal{H}_{\Omega_A}$, 
      then we must have
     \begin{equation}
     \label{bound-CV-A}
         \mathsf{S}_{A}(\tilde{A}) \leqslant S_{\Omega_A} \,.
     \end{equation}
     Thus, if 
     $\Omega_A \cup \widetilde{\Omega}_A$ is 
     a spatial bipartition of $A$, then $\mathsf{S}_{A}(\Omega_A) = S_{\Omega_A}$ and $\mathsf{S}_{A}(\widetilde{\Omega}_A) = S_{\widetilde{\Omega}_A}$.

The expression \eqref{ent-contour-subregion} 
and the above requirements (a), (b) and (c) 
can be adapted to the contour function $\mathsf{S}_A^{(n)}(i)$ 
for the R\'enyi entropies with $n \geqslant2$ 
in a straightforward way.

In the case of the multimode bosonic Gaussian states that we are exploring, the entanglement entropies are obtained through 
(\ref{ee-from-sigma-k}), (\ref{Fn-entropy-def}) and (\ref{F1-entropy-def}).
A natural way \cite{Botero:2004vpl} to construct $\mathsf{S}_A^{(n)}(i)$ 
exploits the mode participation function $p_k(i)$ satisfying the following properties
\be
\label{EE-prob-distribution-def}
\sum_{i \,\in\, A} p_k(i)  \,=\, 1
\;\;\;\;\qquad\;\;\;\;
p_k(i)  \geqslant 0
\;\;\;\;\;\;\;\;\qquad\;\;\;\;\;\;\;\;
1 \leqslant k \leqslant L_A
\ee
where $i \in A$ and $k$ labels the symplectic spectrum,
which tell us that $p_k(i)$ assigns a probability distribution on $A$ to each symplectic eigenvalue. 
Combining (\ref{ee-from-sigma-k}) and (\ref{EE-prob-distribution-def}),
one realizes that
a proposal for the contour functions $\mathsf{S}_A(i)$ and $\mathsf{S}_A^{(n)}(i)$
can be written as 
\be
\label{contour_constr}
\mathsf{S}_A(i) = \sum_{k=1}^{L_A}  p_k(i) \, F_1(\sigma_k)
    \;\;\;\qquad\;\;\;
\mathsf{S}_A^{(n)}(i) = \sum_{k=1}^{L_A}  p_k(i)  \, F_n(\sigma_k)
    \;\;\;\;\;\qquad\;\;\;\;\;
    \forall \, i \in A
\ee
in terms of the mode participation function $p_k(i)$ 
and of the symplectic spectrum of $\gamma_A$.

For a system of free fermions on a lattice in generic spatial dimension, 
a mode participation $p_k(i)$ has been constructed in \cite{ChenVidal2014}
by employing both the eigenvalues and the eigenvectors 
of the correlation matrix reduced to the subsystem $A$.

In the context of multimode bosonic Gaussian states for free lattice models, 
a mode participation function $p_k(i)$ has been proposed in \cite{Coser:2017dtb}.
This construction exploits the Euler decomposition 
(also known as Bloch-Messiah decomposition) \cite{Arvind1995,Braunstein2005}
of the real symplectic matrix $W$ occurring 
in the Williamson decomposition of $\gamma_A$ 
(see (\ref{williamson-dec-gamma-A})), that is given by 
\be
\label{euler-dec-W-matrix}
W = K_{\textrm{\tiny L}} \, E\, K_{\textrm{\tiny R}}
\;\;\;\qquad \;\;\;
E =\e^{\chi} \oplus \e^{-\chi} 
\qquad
\chi = \textrm{diag}\big(\chi_{_1} , \dots , \chi_{_{L_A}}\big)
\ee
where $\chi_{_{j}} \geqslant 0$ and the real matrices 
$K_{\textrm{\tiny L}} $ and $K_{\textrm{\tiny R}} $ are 
symplectic and orthogonal.
The main proposal in \cite{Coser:2017dtb} is based on the $(2L_A)\times (2L_A)$ symplectic and orthogonal matrix $K$ defined as follows 
\be
\label{K-matrix-def}
K \equiv K_{\textrm{\tiny L}} \, K_{\textrm{\tiny R}}
=
 \bigg( \,
\begin{array}{cc}
U_K  \;& Y_K
\\
Z_K   \;& V_K
\end{array} 
\,\bigg)
\ee
where its decomposition in $L_A \times L_A$ blocks has been introduced.
The matrix (\ref{K-matrix-def}) provides the following mode participation function
\be
\label{mode-part-W}
p_k(i) = \frac{1}{2} \Big(
\big[\big(U_K\big)_{k,i}\big]^2
+ \big[\big(Y_K\big)_{k,i}\big]^2
+ \big[\big(Z_K\big)_{k,i}\big]^2
+ \big[\big(V_K\big)_{k,i}\big]^2
\Big)
\ee
which fulfills the requirements in 
(\ref{EE-prob-distribution-def}). 
It is worth remarking that $K$ in (\ref{K-matrix-def}) 
can be written in terms of $W$ in (\ref{williamson-dec-gamma-A})
as follows
\begin{equation}
\label{K-from-W}
    K 
    = \big( W \, W^{\textrm{t}}\big)^{-1/2} \; W 
    \,=\, 
    W \,\big(W^{\textrm{t}}\, W \big)^{-1/2} \,.
\end{equation}
A convenient way to express
$\mathsf{S}_A(i)$ and $\mathsf{S}_A^{(n)}(i)$ in (\ref{contour_constr}) 
 is based on (\ref{ee-from-O}) and reads 
 \cite{Coser:2017dtb}
\begin{equation}
\label{contour-function-EE-projector}
    \mathsf{S}_A(i)
    =
    \frac{1}{2} \, \textrm{Tr} 
    \big[ X^{(i)} F_1 (\Phi_A)\big]
        \;\;\qquad\;\;
    \mathsf{S}_A^{(n)}(i)
    =
    \frac{1}{2} \, \textrm{Tr} 
    \big[ X^{(i)} F_n (\Phi_A)\big]
    \;\;\;\qquad\;\;\;
    \Phi_A \equiv 
    K^{\textrm{t}} \big(  D \oplus D \big) K
\end{equation}
in terms of the diagonal matrix $D$ in (\ref{williamson-dec-gamma-A}), the orthogonal and symplectic matrix $K$ defined in (\ref{K-matrix-def})
and the set of orthogonal projectors 
$\big\{ X^{(i)} \, ; 1\leqslant i \leqslant L_A \big\}$,
made by  $(2L_A)\times (2L_A)$
real, symmetric and positive definite matrices 
satisfying  $X^{(i)}X^{(j)} = \delta_{i,j} \, X^{(i)}$ and $\sum_{i =1}^{L_A} X^{(i)} = \boldsymbol{1}$.
The projector $X^{(i)}$ can be written as 
$X^{(i)} = \delta^{(i)} \oplus \delta^{(i)}$,
where the generic element of the $L_A \times L_A$ matrix $\delta^{(i)}$ is $\delta^{(i)}_{r,s} = \delta_{r,i} \, \delta_{s,i} $.
The mode participation function (\ref{mode-part-W}) 
can be written also in terms of these projectors as follows
\begin{equation}
\label{mpf-proj-ent}
    p_k(i)=\frac{1}{2} 
    \Tr \! \big( X^{(i)} K^{\textrm{t}} X^{(k)} K \big)
\end{equation}
which is obtained by exploiting the explicit form 
of $\Phi_A$ in (\ref{contour-function-EE-projector}).
The expression (\ref{mpf-proj-ent}) is also employed 
in the discussion reported in the 
appendices\;\ref{app-contour-properties} and \ref{app-relation}.

Although the matrix $W$ in the Williamson decomposition \eqref{williamson-dec-gamma-A} 
is not unique \cite{de2006symplectic, Son_2021},
the contour construction 
\eqref{contour-function-EE-projector} is well defined. 
Indeed, it has been shown 
(see e.g. Proposition 8.12 in \cite{de2006symplectic}) that, 
if two different symplectic matrices $W$ and $W'$ 
provide the same Williamson decomposition 
(see (\ref{williamson-dec-gamma-A})) 
for an assigned symplectic spectrum, 
meaning that 
$W^{\textrm{t}}\,(D\,\oplus\,D)\,W = (W')^{\textrm{t}}\,(D\,\oplus\,D)\,W'$, 
then the matrices 
$\mathcal{K} \equiv W'\,W^{-1}$ and $\mathcal{O} \equiv \left( D\,\oplus\,D\right)^{\frac{1}{2}}\,\mathcal{K}\,\left( D\,\oplus\,D\right)^{-\frac{1}{2}}$ are symplectic and orthogonal. 
Combining this properties and \eqref{K-from-W}, 
one finds first that  
$K'=\mathcal{K}\,K$, and then,
by plugging this relations in \eqref{contour-function-EE-projector}, 
also that 
\begin{eqnarray}
\label{uniqueness-ent-contour}
         \mathsf{S}_A^{(n)}(i)'
         &\equiv & 
         \frac{1}{2}
         \Tr\!\big[\,X^{(i)}\, (K')^\textrm{t}\,
         F_n(D\,\oplus\,D)\,K'\, \big] 
         = 
         \frac{1}{2}
         \Tr\! \big[\,
         X^{(i)}K^{\textrm{t}} \,
         F_n\big(\mathcal{K} (D\,\oplus\,D)
         \mathcal{K}^{\textrm{t}} \big) \,K \, \big]
         \nonumber\\
         \rule{0pt}{.7cm}
         &=& 
         \frac{1}{2}
         \Tr\!\big[
         X^{(i)}\,K^\textrm{t}\,
         F_n\big(
         ( D\,\oplus\,D)^{\frac{1}{2}} \, 
         \mathcal{O}^\textrm{t} \mathcal{O} \,
         ( D\,\oplus\,D)^{\frac{1}{2}} 
         \big)
         \,K \, \big] 
         =
         \mathsf{S}_A^{(n)}(i)
\end{eqnarray}
where $S_A^{(n)}(i)$ and $S_A^{(n)}(i)'$ are the contour functions built 
from $W$ and $W'$ respectively.

In \cite{Coser:2017dtb} it has been shown that the contour functions for the entanglement entropies \eqref{contour-function-EE-projector} 
satisfy certain properties that are strictly included in
the above properties (a), (b) and (c). 
In the appendix\;\ref{C-V-Gaussian}, we show that 
\eqref{contour-function-EE-projector} does not satisfy the properties 
(a), (b) and (c) in the form given above. 

Numerical results for the contour functions obtained from  
(\ref{contour_constr}) and (\ref{mode-part-W}) 
in harmonic lattices have been reported in 
\cite{Coser:2017dtb,DiGiulio:2019lpb}.

\section{A contour function for the entanglement negativity}
\label{sec:contour-neg}

In this section we provide an explicit construction  
of contour functions for $\mathcal{E}$ and $\mathcal{E}^{(n)}$
that satisfy (\ref{neg-density-property-intro}) 
and (\ref{neg-density-positivity-property-intro}), 
for the multimode bosonic Gaussian states of free lattice models.
This is done by adapting the procedure followed in \cite{Coser:2017dtb}
for the contour functions of the entanglement entropies, 
reviewed in Sec.\,\ref{sec:contour-entropies}.

In the context of the multimode bosonic Gaussian states of free lattice models,
consider a mixed state characterized by the density matrix $\rho_A$ 
for the region $A$ containing $L_A$ sites.
Assuming that $A$ is bipartite into two spatial regions $A_1$ and $A_2$, 
made by $L_1$ and $L_2$ sites respectively,
the partial transpose 
$\rho_A^{\textrm{\tiny $\Gamma_2$}}$ of $\rho_A$ w.r.t. $A_2$
is given by (\ref{rhoA-T2-def-intro}).
In our analysis, it is crucial that both $\rho_A$ and 
$\rho_A^{\textrm{\tiny $\Gamma_2$}}$ are Gaussian states. 
Furthermore, in the setup that we are considering, 
the partial transposition in $A_2$ corresponds to change 
the signs of the momenta only in $A_2$ \cite{Simon:1999lfr}.
Thus, the covariance matrix of  $\rho_A^{\textrm{\tiny $\Gamma_2$}}$ reads
\be
\label{gamma-T2-from-gamma}
\gamma_A^{\textrm{\tiny $\Gamma_2$}} 
\equiv 
\mathcal{R}_{2} \, \gamma_A \, \mathcal{R}_{2} 
\ee
where we have introduced the following 
$(2L_A)\times (2L_A)$ real and symmetric matrix
\be
\label{R2-matrix-def}
\mathcal{R}_{2} 
\equiv 
 \bigg( \,
\begin{array}{cc}
\boldsymbol{1} \; & \boldsymbol{0} 
\\
\boldsymbol{0}  \; & R_{2}
\end{array} 
\bigg)
\ee
being $R_2$ defined as the $L_A \times L_A$ diagonal matrix 
whose elements on the diagonal are given by  
$(R_2)_{i,i} =1 $  for $ i \leqslant L_1$ and 
$(R_2)_{i,i} = - 1 $  for $ L_1 < i \leqslant L_A$.
The fact that $\mathcal{R}_{2} \notin \textrm{Sp}(L_A)$ in (\ref{gamma-T2-from-gamma})
makes the relation between 
$\gamma_A^{\textrm{\tiny $\Gamma_2$}}$ and $\gamma_A$ 
highly non trivial.

Since $\gamma_A$ is real, symmetric and positive definite, also $\gamma_A^{\textrm{\tiny $\Gamma_2$}}$ in \eqref{gamma-T2-from-gamma} have these features. Hence, 
the Williamson decomposition of $\gamma_A^{\textrm{\tiny $\Gamma_2$}}$ can be introduced, namely
\begin{equation}
\label{williamson-dec-gamma-A-transpose}
    \gamma_A^{\textrm{\tiny $\Gamma_2$}}
    = \widetilde{W}^{\textrm{t}} 
    \big(  \widetilde{D} \oplus \widetilde{D} \big) \,
    \widetilde{W}
\end{equation}
where $\widetilde{W} \in \textrm{Sp}(L_A)$
and the $L_A \times L_A$ diagonal matrix 
$\widetilde{D} = \textrm{diag}( \tilde{\sigma}_1 , 
\dots , \tilde{\sigma}_{L_A})$ contains 
the symplectic eigenvalues of $\gamma_A^{\textrm{\tiny $\Gamma_2$}}$.
In the following 
the symplectic eigenvalues of $\gamma_A^{\textrm{\tiny $\Gamma_2$}}$ 
are ordered in the increasing way, 
i.e. 
$\tilde{\sigma}_1 \leqslant \tilde{\sigma}_2 \leqslant \dots \leqslant \tilde{\sigma}_{L_A}$,
as in (\ref{williamson-dec-gamma-A})
for the symplectic eigenvalues of $\gamma_A$.
We remark that the symplectic eigenvalues of $\gamma_A^{\textrm{\tiny $\Gamma_2$}}$ 
are not constrained by the uncertainty principle; 
hence $0 <  \tilde{\sigma}_k < 1/2$ is allowed for some values of $k$.
Combining (\ref{gamma-T2-from-gamma}) and (\ref{williamson-dec-gamma-A}), 
one obtains
the relation between the symplectic spectra 
of $\gamma_A^{\textrm{\tiny $\Gamma_2$}}$ and of $\gamma_A$ given by 
\begin{equation}
   \widetilde{D} \oplus \widetilde{D} 
    \,=\,
    \big(W\, \mathcal{R}_{2}\, \widetilde{W}^{-1}\big)^{\textrm{t}}
    \,\big(  D \oplus D \big)
    \big(W\, \mathcal{R}_{2}\, \widetilde{W}^{-1}\big)
\end{equation}
in terms of (\ref{R2-matrix-def}) and the symplectic matrices 
introduced in the Williamson decompositions (\ref{williamson-dec-gamma-A}) 
and (\ref{williamson-dec-gamma-A-transpose}).

The logarithmic negativity (\ref{log-neg-def-intro}) 
and the corresponding quantity associated to the moments of $\gamma_A^{\textrm{\tiny $\Gamma_2$}}$ 
introduced in (\ref{neg-moments-def-intro})
can be found from the symplectic spectrum of 
$\gamma_A^{\textrm{\tiny $\Gamma_2$}}$ as follows 
\cite{Audenaert_2002}
\be
\label{neg-from-sigma-k}
\mathcal{E}\, 
= \sum_{k=1}^{L_A} 
\widetilde{F}_1 (\tilde{\sigma}_k)
\;\;\;\;\qquad\;\;\;
\mathcal{E}^{(n)} 
= 
\sum_{k=1}^{L_A} \widetilde{F}_n(\tilde{\sigma}_k)
\ee
being $\widetilde{F}_1(x)$ and $\widetilde{F}_n(x)$ 
defined respectively by 
\bea
\label{F1-neg-def}
\widetilde{F}_1(x) 
&\equiv& 
\, - \log \! \Big[  \big| x+1/2 \big| - \big| x -1/2 \big| \Big]
\,=\,
\log\!\bigg[
\textrm{max} \bigg( 1\, , \, \frac{1}{2x} \bigg)
\bigg]
\\
\label{Fn-neg-def}
\rule{0pt}{.6cm}
\widetilde{F}_n(x) 
&\equiv& 
 - \log \! \Big[  \big( x+1/2 \big)^n - \big( x -1/2 \big)^n \Big]
\eea
for a generic integer $n \geqslant 2$.
The functions in (\ref{Fn-neg-def}) and (\ref{Fn-entropy-def})
are related as $\widetilde{F}_n(x)  = (1-n) F_n(x) $ 
for $n > 1$.
Since $\widetilde{F}_1(x) >0$ for $0<x<1/2$ and 
vanishes for $x \geqslant 1/2$, 
only the symplectic eigenvalues satisfying $0<\tilde{\sigma}_k < 1/2$ 
provide a non vanishing contribution to the logarithmic negativity in (\ref{neg-from-sigma-k}).
Instead, all the eigenvalues $\tilde{\sigma}_k$ contribute to $\mathcal{E}^{(n)}$ in (\ref{neg-from-sigma-k}).
In particular, $\widetilde{F}_n(\tilde{\sigma}_k) > 0$ when $0<\tilde{\sigma}_k < 1/2$
and $\widetilde{F}_n(\tilde{\sigma}_k) \leqslant 0$ when $\tilde{\sigma}_k \geqslant 1/2$;
hence $\mathcal{E}^{(n)}$ do not have a definite sign. 
The expressions in (\ref{neg-from-sigma-k}) have been widely employed 
to explore the properties of $\mathcal{E}$ and $\mathcal{E}^{(n)}$
in harmonic lattices 
\cite{Audenaert_2002, Plenio:2004tsx, Marcovitch:2008sxc, Eisler:2014dze, Eisler_2016,
Calabrese:2012nk, Calabrese:2014yza, Coser:2014gsa, De_Nobili_2016, 
Klco:2020rga,Klco:2021cxq,Arias:2026bqh}.

The entanglement quantifiers in (\ref{neg-from-sigma-k}) can be written 
through a generic $(2L_A)\times (2L_A)$ orthogonal matrix $O$
respectively as 
\begin{equation}
    \label{neg-from-O}
\mathcal{E}\, = 
\frac{1}{2} \, 
\textrm{Tr} \big[ \widetilde{F}_1\big(\widetilde{\Upsilon}_A\big) \big]
\;\;\qquad\;\;
\mathcal{E}^{(n)} = \frac{1}{2} \, 
\textrm{Tr} \big[ \widetilde{F}_n\big(\widetilde{\Upsilon}_A\big) \big]
\;\;\;\;\quad\;\;\;\;
\widetilde{\Upsilon}_A \equiv 
O^{\textrm{t}} \big( \, \widetilde{D} \oplus \widetilde{D} \,\big) O
\end{equation}
in analogy with (\ref{ee-from-O}) 
for the entanglement entropies (\ref{ee-from-sigma-k}).

In order to adapt to (\ref{neg-from-sigma-k})
the procedure described in Sec.\,\ref{sec:contour-entropies}
for the entanglement entropies 
(see (\ref{EE-prob-distribution-def}) and (\ref{contour_constr})), 
we first consider a mode participation function $\tilde{p}_k(i)$ 
such that
\be
\label{NEG-prob-distribution-def}
\sum_{i \,\in\, A} \tilde{p}_k(i)  \,=\, 1
\;\;\;\;\qquad\;\;\;\;
\tilde{p}_k(i)  \geqslant 0
\;\;\;\;\;\;\;\;\qquad\;\;\;\;\;\;\;\;
1 \leqslant k \leqslant L_A
\ee
(like in (\ref{EE-prob-distribution-def}))
and then define the contour functions $\mathsf{E}(i) $ 
and $\mathsf{E}^{(n)}(i)$ 
for the entanglement quantifiers 
in (\ref{neg-from-sigma-k}) respectively as  
\be
\label{contour_constr_neg}
\mathsf{E}_A(i) = \sum_{k=1}^{L_A}  
\tilde{p}_k(i) \, \widetilde{F}_1(\tilde{\sigma}_k)
    \;\;\;\;\qquad\;\;\;\;\;
\mathsf{E}_A^{(n)}(i) = \sum_{k=1}^{L_A}  
\tilde{p}_k(i)  \, \widetilde{F}_n(\tilde{\sigma}_k)
    \;\;\;\;\;\qquad\;\;\;\;\;\;\;
    \forall \, i \in A \,.
\ee

Combining (\ref{NEG-prob-distribution-def}) 
with the sign properties of (\ref{F1-neg-def}) 
and (\ref{Fn-neg-def}), 
for the contour functions in (\ref{contour_constr_neg})
we observe that $\mathsf{E}_A(i) \geqslant 0$ 
for all $i \in A$, 
while $\mathsf{E}_A^{(n)}(i)$ do not have a definite sign
for $i \in A$.
This means that, 
while $\mathsf{E}_A(i)$ can be considered 
as a spatial density of the logarithmic negativity, 
the interpretation of $\mathsf{E}^{(n)}(i)$ as a spatial density of $\mathcal{E}^{(n)}$ is less straightforward. 

Following the procedure described in \cite{Coser:2017dtb} 
for the entanglement entropies 
(see (\ref{euler-dec-W-matrix})-(\ref{mode-part-W})),
we construct a mode participation function $\tilde{p}_k(i)$ 
by considering the Euler decomposition 
of the real symplectic matrix $\widetilde{W}$
occurring in the Williamson decomposition of 
$\gamma_A^{\textrm{\tiny $\Gamma_2$}}$ 
given in (\ref{williamson-dec-gamma-A-transpose}), 
which reads
\be
\label{tilde-W-euler-dec}
\widetilde{W} = 
\widetilde{K}_{\textrm{\tiny L}} \, 
\widetilde{E}\, 
\widetilde{K}_{\textrm{\tiny R}}
\;\;\;\qquad \;\;\;
\widetilde{E} =\e^{\widetilde{\chi}} \oplus \e^{-\widetilde{\chi}} 
\qquad
\widetilde{\chi} = 
\textrm{diag}\big(\widetilde{\chi}_{_1} , \dots , 
\widetilde{\chi}_{_{L_A}}\big)
\ee
where $\widetilde{\chi}_{_{j}} \geqslant 0$ and the real matrices 
$\widetilde{K}_{\textrm{\tiny L}} $ and $\widetilde{K}_{\textrm{\tiny R}} $ are 
symplectic and orthogonal.
Then, we introduce the following symplectic and orthogonal matrix 
\be
\label{tilde-K-def}
\widetilde{K} \equiv 
\widetilde{K}_{\textrm{\tiny L}} \, 
\widetilde{K}_{\textrm{\tiny R}}
=
 \Bigg( \,
\begin{array}{cc}
\widetilde{U}_K  \;& \widetilde{Y}_K
\\
\widetilde{Z}_K   \;& \widetilde{V}_K
\end{array} 
\Bigg)
\ee
where in the last step its decomposition in $L_A \times L_A$ blocks
has been considered. 
By adapting the analysis of \cite{Coser:2017dtb} 
to (\ref{tilde-K-def}), 
we find that (\ref{tilde-K-def}) provides 
the following mode participation function
\be
\label{tilde-pk-i-def}
\tilde{p}_k(i) = \frac{1}{2} \Big(
\big[\big(\widetilde{U}_K\big)_{k,i}\big]^2
+ \big[\big(\widetilde{Y}_K\big)_{k,i}\big]^2
+ \big[\big(\widetilde{Z}_K\big)_{k,i}\big]^2
+ \big[\big(\widetilde{V}_K\big)_{k,i}\big]^2
\Big)
\ee
which fulfills the requirements in (\ref{NEG-prob-distribution-def}). 
A relation similar to (\ref{K-from-W}) occurs also in this case
and it allows to express (\ref{tilde-K-def}) through the symplectic matrix in (\ref{williamson-dec-gamma-A-transpose}). It reads
\begin{equation}
\label{K-from-W-tilde}
    \widetilde{K} 
    = \big( \,\widetilde{W} \, \widetilde{W}^{\textrm{t}}\,\big)^{-1/2} \; 
    \widetilde{W} 
    \,=\, 
    \widetilde{W} \,\big(\,\widetilde{W}^{\textrm{t}}\, \widetilde{W} \,\big)^{-1/2}\,.
\end{equation}
From (\ref{neg-from-O}),  
the contour functions  $\mathsf{E}(i)$ and $\mathsf{E}^{(n)}(i)$ 
in (\ref{contour_constr_neg}) can be written as 
\begin{equation}
\label{contour-function-NEG-projector}
    \mathsf{E}_A(i)
    =
    \frac{1}{2} \, \textrm{Tr} 
    \big[ X^{(i)} \widetilde{F}_1 (\tilde{\Phi}_A)\big]
        \;\;\;\quad\;\;\;
    \mathsf{E}_A^{(n)}(i)
    =
    \frac{1}{2} \, \textrm{Tr} 
    \big[ X^{(i)} \widetilde{F}_n ( \tilde{\Phi}_A)\big]
    \;\;\;\;\quad\;\;\;\;
    \tilde{\Phi}_A \equiv 
    \widetilde{K}^{\textrm{t}} 
    \big(\,  \widetilde{D} \oplus \widetilde{D}\, \big) \widetilde{K}
\end{equation}
where $\widetilde{D}$ is the diagonal matrix 
occurring in (\ref{williamson-dec-gamma-A-transpose}), 
$\widetilde{K}$ is the orthogonal and symplectic matrix 
defined in (\ref{tilde-K-def})
and $\big\{ X^{(i)} \, ; 1\leqslant i \leqslant L_A \big\}$ 
is the set of orthogonal projectors 
introduced in (\ref{contour-function-EE-projector}).
By adapting the step performed to obtain (\ref{mpf-proj-ent}) to this case, 
we find that the mode participation function (\ref{tilde-pk-i-def}) 
can be written in terms of the orthogonal projectors introduced 
below (\ref{contour-function-EE-projector}) as follows
\begin{equation}
\label{mpf-proj-neg}
    \tilde{p}_k(i)=\frac{1}{2}
    \Tr \! \big( \,X^{(i)} \widetilde{K}^{\textrm{t}} X^{(k)} \widetilde{K} \,\big)
\end{equation}
which is amployed also in the 
appendices\;\ref{app-contour-properties} and \ref{app-relation}.

An important property of the contour functions (\ref{contour_constr_neg}), 
defined through (\ref{tilde-K-def}) and (\ref{tilde-pk-i-def}), 
is that both the partial transposition w.r.t. $A_1$ and w.r.t. $A_2$ lead to the same result. 
Indeed, considering $\rho_A^{\textrm{\tiny $\Gamma_1$}}$ 
(i.e. the partial transpose of $\rho_A$ w.r.t. $A_1$),
its covariance matrix $\gamma_A^{\textrm{\tiny $\Gamma_1$}} $ reads 
\be
\label{gamma-T1-from-gamma}
\gamma_A^{\textrm{\tiny $\Gamma_1$}} 
\equiv 
\mathcal{R}_{1} \, \gamma_A \, \mathcal{R}_{1} 
\;\;\;\;\qquad \;\;\;\;
\mathcal{R}_{1} 
\equiv 
 \bigg( \,
\begin{array}{cc}
\boldsymbol{1} \; & \boldsymbol{0} 
\\
\boldsymbol{0}  \; & R_{1}
\end{array} 
\bigg)
\ee
where the $(2L_A)\times (2L_A)$ real matrix
$\mathcal{R}_{1} $ is defined through the 
$L_A \times L_A$ diagonal matrix $R_1$,
whose non vanishing elements are  
$(R_1)_{i,i} =-1 $  for $ i \leqslant L_1$ and 
$(R_1)_{i,i} =  1 $  for $ L_1 < i \leqslant L_A$.
The matrices $\mathcal{R}_{j}$ with $j \in \{1,2\}$
(see (\ref{R2-matrix-def}) and (\ref{gamma-T1-from-gamma}))
 satisfy $\mathcal{R}_{j}= \mathcal{R}_{j}^{-1}$, 
 i.e. they correspond to involutions. 
From (\ref{gamma-T2-from-gamma}) and (\ref{gamma-T1-from-gamma}),
we  observe that 
\begin{equation}
    \gamma_A^{\textrm{\tiny $\Gamma_1$}} 
    = 
    \mathcal{R}_1\mathcal{R}_2 \,
    \gamma_A^{\textrm{\tiny $\Gamma_2$}} 
    \,\mathcal{R}_2\mathcal{R}_1 
    \;\; \;\qquad\;\;\; 
    \mathcal{R}_1\mathcal{R}_2
    =
     \bigg( \,
\begin{array}{cc}
\boldsymbol{1} \; & \boldsymbol{0} 
\\
\boldsymbol{0}  \; & -\boldsymbol{1}
\end{array} 
\bigg)
\end{equation}
where $\mathcal{R}_1\mathcal{R}_2$ is symplectic;
hence the symplectic spectra of $\gamma_A^{\textrm{\tiny $\Gamma_1$}} $ 
and $\gamma_A^{\textrm{\tiny $\Gamma_2$}} $ coincide. 
Furthermore, denoting by $\widetilde{W}_j$ the symplectic matrix occurring in the Williamson decomposition of $\gamma_A^{\textrm{\tiny $\Gamma_j$}} $,  
it is straightforward to find that $\widetilde{W}_1= \widetilde{W}_2\,\mathcal{R}_2\mathcal{R}_1$.
Combining this result with 
the Euler decompositions of $\widetilde{W}_j$,
namely $\widetilde{W}_j= \widetilde{K}_{\textrm{\tiny L},j} \,
\widetilde{E}_j\, 
\widetilde{K}_{\textrm{\tiny R},j}$,
we arrive to 
$\widetilde{K}_{\textrm{\tiny L},1}=\widetilde{K}_{\textrm{\tiny L},2}$, $\widetilde{E}_1=\widetilde{E}_2$ and
$\widetilde{K}_{\textrm{\tiny R},1}=\widetilde{K}_{\textrm{\tiny R},2}\,\mathcal{R}_2\mathcal{R}_1$.
Hence, denoting by $\widetilde{K}_j$ the orthogonal and symplectic matrices (\ref{tilde-K-def}) corresponding to $\gamma_A^{\textrm{\tiny $\Gamma_j$}} $, 
we have that $\widetilde{K}_1 = \widetilde{K}_2\,\mathcal{R}_2\mathcal{R}_1$.
From this relation and (\ref{tilde-pk-i-def}), where the square of the matrix elements occur and therefore their signs are irrelevant, 
we conclude that $\widetilde{K}_1$ and $\widetilde{K}_2$ 
provide the same mode participation function $\tilde{p}_k(i)$.
Finally, this observation and the fact that $\gamma_A^{\textrm{\tiny $\Gamma_1$}} $ 
and $\gamma_A^{\textrm{\tiny $\Gamma_2$}} $ have the same symplectic spectrum 
tell us that they also provide the same contour functions (\ref{contour_constr_neg}).

By adapting the observations made in the final part of Sec.\,\ref{sec:contour-entropies} 
in a straightforward way, we find that the contour functions for the entanglement negativity defined between (\ref{contour_constr_neg}) 
and (\ref{contour-function-NEG-projector}) are well defined, 
despite the non uniqueness of the matrix $\widetilde{W}$ occurring in the Williamson decomposition (\ref{williamson-dec-gamma-A-transpose}).

An important qualitative difference between 
the contour functions (\ref{contour_constr_neg})
and the contour functions for the entanglement entropies in (\ref{contour_constr}) 
is given by the domains supporting them.
Indeed, while 
for the entanglement entropy,
that measures the bipartite entanglement between $A$ and $B$ when $A \cup B$ is in a pure state, the contour function is defined either in $A$ or in $B$,
in the case of the logarithmic negativity,
that measures the bipartite entanglement between $A_1$ and $A_2$ even when $A_1 \cup A_2$ is in a mixed state, the contour function is naturally defined on the entire domain $A_1 \cup A_2$.
This could be due to the mixed nature of the state in $\rho_{A_1 \cup A_2}$.
Indeed, 
when $\rho_{A_1 \cup A_2}$ is a pure state, 
we show below (see \eqref{neg_ent_loc_logneg}) that 
the contour function for the logarithmic negativity becomes 
the sum of two functions supported in $A_1$ and $A_2$.

By adapting the properties of the contour functions 
for the entanglement entropies 
discussed in \cite{ChenVidal2014, Coser:2017dtb} 
(see also the appendix\;\ref{C-V-Gaussian}),
in the appendix\;\ref{app-CV-neg-contour}
we introduce some reasonable constraints for generic contour functions (\ref{contour_constr_neg}), showing also which of them are satisfied by
by the specific contour functions in \eqref{contour-function-NEG-projector}.
This analysis requires to introduce 
\begin{equation}
\label{ent-contour-subregion-neg}
    \mathsf{E}_A(\tilde{A}) \equiv 
    \sum_{i\in\tilde{A}} \, \mathsf{E}_A(i)
    \;\;\;\;\qquad\;\;\;\;
    \tilde{A} \subseteq A
\end{equation}
in analogy with (\ref{ent-contour-subregion}).

When the bipartite system $A = A_1 \cup A_2$ is in a pure state, 
i.e. $\rho_A = |\Psi\rangle \langle \Psi|$,
the moments of $\rho_{A}^{\textrm{\tiny $\Gamma_j$}}$
are related to the moments of $\rho_{A_r}$, with $j,r \in \{1,2\}$,
as follows \cite{Calabrese:2012ew, Calabrese:2012nk}
\begin{equation}
\label{pure-states-moments-relation}
    \Tr \! \big( \rho_A^{\textrm{\tiny $\Gamma_j$}} \big)^n = 
    \left\{
    \begin{array}{ll}
        \; \Tr \rho_{A_r}^{n_{\textrm{\tiny o}}} 
        \hspace{2cm} & 
        n=n_{\textrm{\tiny o}} \;\;\textrm{odd}
        \\
        \rule{0pt}{.7cm}
        \left( \Tr \rho_{A_r}^{n_{\textrm{\tiny e}} / 2}\right)^2 
        & 
        n=n_{\textrm{\tiny e}} \;\;\textrm{even}
    \end{array}
    \right.
\end{equation}
where the parity of $n$ plays a crucial role
and we remind that $\Tr \rho_{A_1}^{n}  = \Tr \rho_{A_2}^{n}$ 
for any $n$.
The latter identity allows us to write (\ref{pure-states-moments-relation}) 
as follows
\begin{equation}
\label{neg_ent}
    \mathcal{E}^{(n)} = 
    \left\{
    \begin{array}{ll}
    \displaystyle
        \frac{1- n_{\textrm{\tiny o}}}{2}  
        \left( 
        S_{A_1}^{(n_{\textrm{\tiny o}})}
        +
        S_{A_2}^{(n_{\textrm{\tiny o}})}\right) 
        \hspace{2cm} & 
        n=n_{\textrm{\tiny o}} \;\;\textrm{odd}
        \\
        \rule{0pt}{.9cm}
    \displaystyle
        \bigg( 
        1-\frac{n_{\textrm{\tiny e}}}{2}
        \bigg) 
        \left(
        S_{A_1}^{(n_{\textrm{\tiny e}}/2)} 
        + 
        S_{A_2}^{(n_{\textrm{\tiny e}}/2)} 
        \right)
        & 
        n=n_{\textrm{\tiny e}} \;\;\textrm{even}\,.
    \end{array}
    \right.
\end{equation}
As first observed in \cite{DeNobiliThesis},
from (\ref{entropy-density-property}) and (\ref{neg-density-property-intro}), 
a sufficient condition for (\ref{neg_ent}) is given by its local version, namely
\begin{equation}
\label{neg_ent_loc}
    \mathsf{E}_A^{(n)}(i) 
    = 
        \left\{
    \begin{array}{ll}
        \displaystyle
        \frac{1-n_{\textrm{\tiny o}}}{2} 
        \left( 
        \Theta_{1}(i)\,\mathsf{S}_{A_1}^{(n_{\textrm{\tiny o}})}(i)
        +
        \Theta_{2}(i)\,\mathsf{S}_{A_2}^{(n_{\textrm{\tiny o}})}(i)
        \right) 
        \hspace{2cm} & 
        n=n_{\textrm{\tiny o}} \;\;\textrm{odd}
        \\
        \rule{0pt}{.9cm}
    \displaystyle
        \bigg( 1-\frac{n_{\textrm{\tiny e}}}{2}\bigg) \left( \Theta_{1}(i)\, 
        \mathsf{S}_{A_1}^{(n_{\textrm{\tiny e}}/2)}(i) 
        + 
        \Theta_{2}(i)\,  \mathsf{S}_{A_2}^{(n_{\textrm{\tiny e}}/2)}(i) 
        \right)
        & 
        n=n_{\textrm{\tiny e}} \;\;\textrm{even}
      \end{array}
    \right.
\end{equation}
where $\Theta_{j}(i)$ denotes the characteristic function for the block $A_j$, with $j \in \{1,2\}$, 
namely $\Theta_{j}(i)=1$ when $i \in A_j$ 
and $\Theta_{j}(i)=0$ otherwise.
The replica limit $n_{\textrm{\tiny e}} \to 1$ of (\ref{neg_ent_loc}) provides the following relation 
\begin{equation}
\label{neg_ent_loc_logneg}
    \mathsf{E}_A(i) = \frac{1}{2} \left( 
    \Theta_{1}(i)\, 
    \mathsf{S}_{A_1}^{(1/2)}(i)
    +
    \Theta_{2}(i)\, 
    \mathsf{S}_{A_2}^{(1/2)}(i)\right) .
\end{equation}
While (\ref{neg_ent}) holds for a generic system in a pure state, 
the relation (\ref{neg_ent_loc}) is only a sufficient condition for 
(\ref{neg_ent}); hence its validity, 
and therefore the validity of (\ref{neg_ent_loc_logneg}) 
that is derived from it as well, 
is not guaranteed in general. 
However, the contour functions given by 
(\ref{contour_constr_neg}) and (\ref{tilde-pk-i-def}) 
for the entanglement negativity and 
by (\ref{contour_constr}) and (\ref{mode-part-W}) 
for the entanglement entropies
satisfy the relations 
(\ref{neg_ent_loc}) and (\ref{neg_ent_loc_logneg}).
The proof of this non trivial result is reported in the Appendix\;\ref{app-relation}.

Since the relations (\ref{neg_ent_loc}) and (\ref{neg_ent_loc_logneg}) hold for the contour functions that we are exploring, 
it is not worth considering the pure states 
because the corresponding contour functions can be obtained from these relations and the results obtained in \cite{Coser:2017dtb} 
for the contour functions of the entanglement entropies. 

We focus our numerical investigations 
only on bipartite mixed states. 
In particular, we consider the mixed states  that are 
either some reduced density matrices coming from the ground state
of a harmonic chain in the infinite line
(see Sec.\,\ref{sec-GS-line-adjacent} and Sec.\,\ref{sec-GS-line-disjoint})
or the thermal state of a finite harmonic chain on the circle
(see Sec.\,\ref{sec-thermal-state})
\cite{Bombelli1986,Srednicki1993,Peschel1999,Audenaert_2002,Plenio_2005,Cramer_2006,Botero:2004vpl,Amico2008,Eisert2010,de2006symplectic}.
The contour functions 
given by (\ref{contour_constr_neg}) and (\ref{tilde-pk-i-def}) 
for the entanglement negativity and 
by (\ref{contour_constr}) and (\ref{mode-part-W}) 
for the entanglement entropies
are based on the reduced covariance matrix $\gamma_A$ 
(see Sec.\,\ref{sec:contour-entropies}).
The covariance matrices for these static scenarios 
are characterized by the corresponding correlators
$\langle \hat{q}_i \,\hat{q}_j \rangle $ 
and $\langle \hat{p}_i \,\hat{p}_j \rangle $.

In Sec.\,\ref{sec-GS-line-adjacent} and Sec.\,\ref{sec-GS-line-disjoint}
we study the contour functions for $\mathcal{E}$ and $\mathcal{E}^{(n)}$,
given by (\ref{contour_constr_neg}) and (\ref{tilde-pk-i-def}),
in the harmonic chain on the infinite line and in its ground state
(hence $d=1$ and $i \in \mathbb{Z}$ in (\ref{hamiltonian})).
The correlators characterizing this setup read  
\cite{Botero:2004vpl}
\begin{eqnarray}
\label{qq-corr-gs-line}
        \langle \hat{q}_i \,\hat{q}_j \rangle 
        &=& 
        \frac{\mu^{|i-j|+1/2}}{2} 
    \binom{|i-j|-1/2}{|i-j|} \;
    {}_2F_1 \!\left( 
    \frac{1}{2}\, , |i-j|+\frac{1}{2}\, ; |i-j|+1\,; \mu^2 
    \right)
    \\
    \label{pp-corr-gs-line}
    \rule{0pt}{.9cm}
        \langle \hat{p}_i \,\hat{p}_j \rangle 
        &=& 
        \frac{\mu^{|i-j|-1/2}}{2} 
    \binom{|i-j|-3/2}{|i-j|} \;
    {}_2F_1\!\left( 
    -\frac{1}{2}\, , |i-j|-\frac{1}{2}\, ; |i-j|+1\, ; \mu^2 
    \right) 
\end{eqnarray}
where 
\begin{equation}
    \mu \equiv \frac{1}{4} \left( \sqrt{\frac{m\omega_0^2}{\kappa}+4} - \sqrt{\frac{m\omega_0^2}{\kappa}} \; \right)^2 .
\end{equation}
When $\omega \to 0$, the correlator (\ref{qq-corr-gs-line}) diverges \cite{Botero:2004vpl}, as a consequence of the occurrence of the zero mode associated to the invariance under translations
 of the model. 

In Sec.\,\ref{sec-thermal-state} we consider the harmonic chain on the circle 
made by $L$ sites (hence $d=1$ and $N=L$ in (\ref{hamiltonian}))
and in the thermal state characterized by the 
finite inverse temperature $\beta$.
The correlators characterizing this state are
(see e.g. \cite{DiGiulio:2020hlz})
\begin{eqnarray}
\label{qq-corr-temp}
        \langle \hat{q}_i \,\hat{q}_j \rangle 
        &=& 
        \frac{1}{2L} \sum_{k=1}^{L} 
        \frac{1}{\Omega_k} \,
        \coth\!\left(\frac{\sqrt{\kappa}\,\Omega_k}{2\sqrt{m}\,T} \right) 
        \cos \! \big[ 2 \pi k (i-j) / L \big] 
        \\
        \label{pp-corr-temp}
        \rule{0pt}{.9cm}
        \langle \hat{p}_i \,\hat{p}_j \rangle 
        &=& 
        \frac{1}{2L} \sum_{k=1}^{L} 
        \Omega_k \,
        \coth\!\left(\frac{\sqrt{\kappa}\,\Omega_k}{2\sqrt{m}\,T} \right) 
        \cos \!\big[ 2 \pi k (i-j) / L \big]
\end{eqnarray}
where
\begin{equation}
     \Omega_k \equiv 
     \sqrt{\frac{m\omega_0^2}{\kappa} 
     + 4\big( \sin \left[ \pi k / L \right]\big)^2}  \;.
\end{equation}
Notice that the zero mode corresponds to $k=L$ 
and it makes (\ref{qq-corr-temp}) divergent in the massless regime.

Since the correlators in 
(\ref{qq-corr-gs-line})-(\ref{pp-corr-gs-line}) 
and in (\ref{qq-corr-temp})-(\ref{pp-corr-temp}) 
depend only the dimensionless parameter $\omega$,
in our numerical analyses we set $m_0 = \kappa =1$  
without loss of generality. 
We are not allowed to set $\omega =0$ in our numerical analysis because of the occurrence of the zero mode;
hence we explore the massless regime by taking $\omega \ll 1$.
Some technical details about the procedure to determine the Williamson 
decomposition of the covariance matrices of interest 
have been reported in appendix\;\ref{app-W-matrix}.

\begin{figure}[t!]
\vspace{-.5cm}
\hspace{-.5cm}
\includegraphics[width=1.\textwidth]{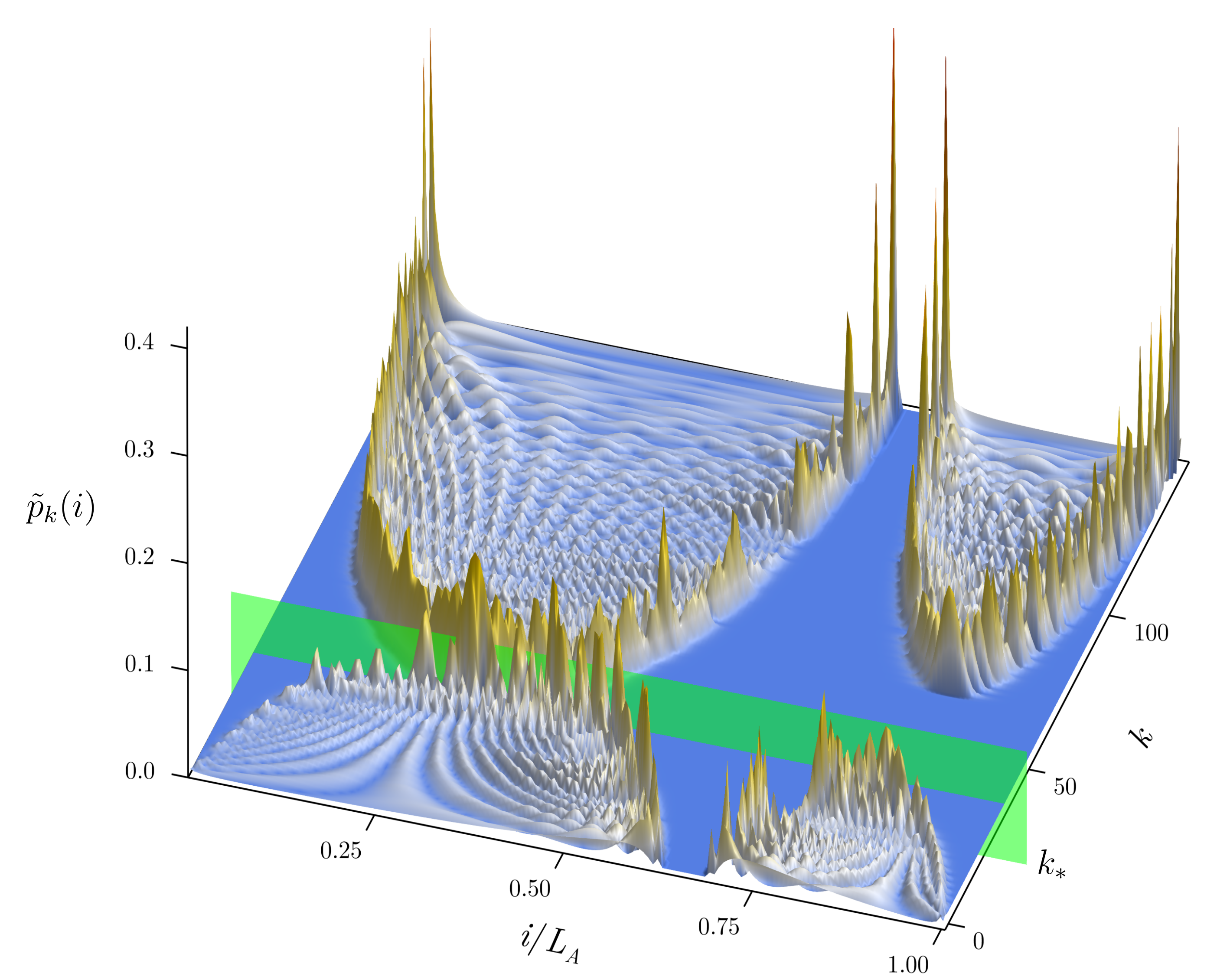}
\caption{\small
Mode participation function \eqref{tilde-pk-i-def} 
for an infinite chain on the line and in its ground state, 
when $A$ is the union of two disjoint blocks made by 
$L_1 =100$ and $L_2 =50$ consecutive sites,
separated by $d = 10$ consecutive sites. 
Here $\omega L_1 = 10^{-10}$. 
The green plane highlights the threshold mode at  $k = k_\ast$,
defined in (\ref{threshold-mode-def}).
}
\label{fig-3D-MPF-example}
\end{figure}

In Fig.\,\ref{fig-3D-MPF-example} we show the 
mode participation function $\tilde{p}_k(i)$ 
defined in \eqref{tilde-pk-i-def} 
in the massless regime of the harmonic chain in the ground state
(the correlators (\ref{qq-corr-gs-line}) and (\ref{pp-corr-gs-line}) 
have been employed)
and for a typical case where $A$ is made by two disjoint blocks.
We recall that increasing values of $k$ 
correspond to increasing values of the symplectic eigenvalue $\tilde{\sigma}_k$,
and therefore to a decreasing value of $\widetilde{F}_1(\tilde{\sigma}_k)$,
that provides the contribution of the $k$-th term 
to $\mathcal{E}$ in (\ref{neg-from-sigma-k}).
A similar plot has been made in \cite{DeNobiliThesis} 
for a different mode participation function
that does not satisfy the positivity condition 
in (\ref{NEG-prob-distribution-def}).
Instead, analogous plots for $p_k(i)$ 
are shown in \cite{Botero:2004vpl, Coser:2017dtb}.
In the rectangular domain parameterized by 
$1 \leqslant i \leqslant L_A$ and $1 \leqslant k \leqslant L_A$,
characteristic fronts occur 
along which $\tilde{p}_k(i)$ takes large values.
These fronts separate 
domains where $\tilde{p}_k(i)$ takes non vanishing but almost negligible values 
(see the flat blue region in Fig.\,\ref{fig-3D-MPF-example})
from domains where $\tilde{p}_k(i)$ oscillates about small values. 
The occurrence of these kind of fronts have been observed also 
for $p_k(i)$ in  \cite{Botero:2004vpl, Coser:2017dtb}.

A crucial feature of $\tilde{p}_k(i)$
(already highlighted in \cite{DeNobiliThesis} for a different mode participation function) is the occurrence of a threshold mode at $k=k_\ast$ characterizing the symplectic eigenvalues $\tilde{\sigma}_k$ that provide a non vanishing contribution 
to the logarithmic negativity in (\ref{neg-from-sigma-k}). In particular, 
we introduce $k_\ast$ such that
\begin{equation}
\label{threshold-mode-def}
    \tilde{\sigma}_k \geqslant \frac{1}{2} 
    \;\;\;\;\qquad\;\;\;\;
    k_\ast \leqslant k \leqslant L_A
\end{equation}
where $\widetilde{F}_1 (\tilde{\sigma}_k) = 0$.
The occurrence of this threshold mode is due to fact that $\widetilde{F}_1(x)$ in (\ref{F1-neg-def}) vanishes at $x=1/2$
in a sharp and non smooth way.
Thus, from (\ref{neg-from-sigma-k}) and (\ref{threshold-mode-def}), 
we have that $\mathcal{E} = \sum_{k=1}^{k_\ast} \widetilde{F}_1 (\tilde{\sigma}_k)$.
In Fig.\,\ref{fig-3D-MPF-example}, 
the threshold mode $k_\ast$ corresponds to the green plane 
and $k_\ast \simeq 40$.
It is important to remark that, 
although the modes satisfying (\ref{threshold-mode-def}) 
do not contribute to $\mathcal{E}$,
they provide a non vanishing contribution to 
$\mathcal{E}^{(n)}$ in (\ref{neg-from-sigma-k}).
In particular, 
$\widetilde{F}_n (\tilde{\sigma}_k) < 0$ 
for integer values $n \geqslant 2$
and when (\ref{threshold-mode-def}) holds.

\begin{figure}[t!]
\includegraphics[width=1.\textwidth]{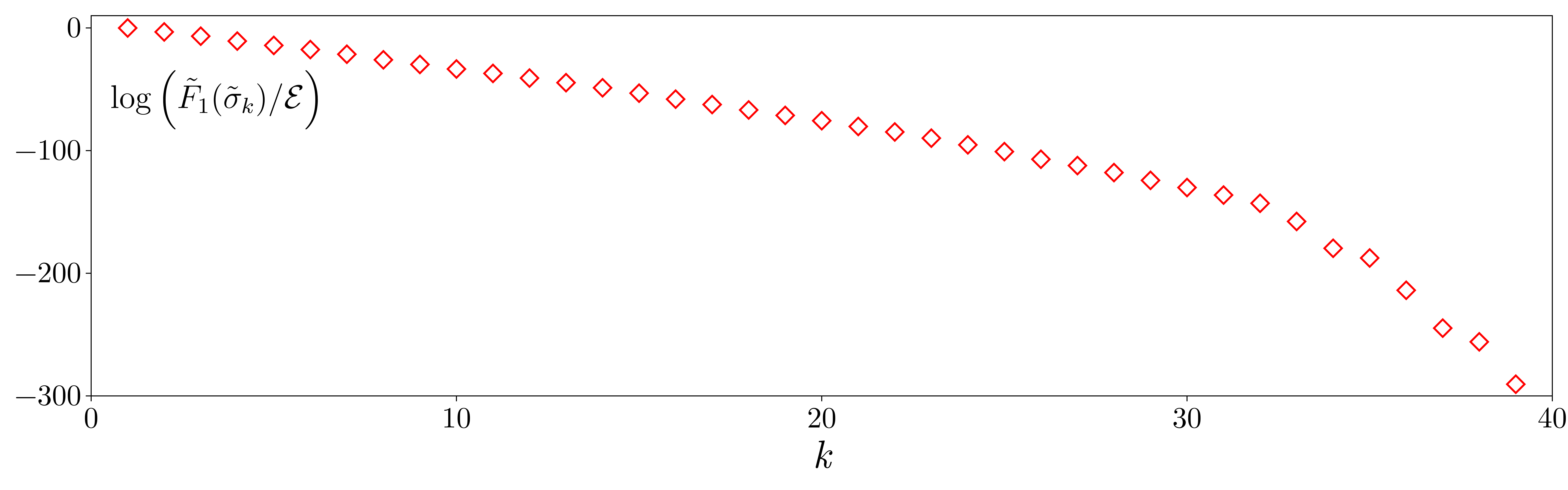} 
\caption{\small
Contribution of the $k$-mode to the logarithmic negativity 
for $ 1 \leqslant k < k_\ast$ 
(see \eqref{neg-from-sigma-k} and \eqref{threshold-mode-def}),
in the same setup of Fig.\,\ref{fig-3D-MPF-example}.}
\label{fig-log-F1-neg}
\end{figure}

In Fig.\,\ref{fig-log-F1-neg}, 
considering the same setup of  Fig.\,\ref{fig-3D-MPF-example},
we show the contribution of various modes to the logarithmic negativity 
(see the first expression in (\ref{neg-from-sigma-k})).
This plot highlights that the largest contribution to the logarithmic negativity
is provided by the modes corresponding to small values of $k$.
From Fig.\,\ref{fig-3D-MPF-example}, we observe that 
the mode participation function 
$\tilde{p}_k(i)$ corresponding to the modes at small values of $k$ 
is peaked around the endpoints of $A_1 \cup A_2$
that are closest to each other, 
which become the entangling point in the limit of adjacent blocks.

\section{Insights for the contour functions in CFT$_2$}
\label{sec:cft}

In this section we discuss the construction of functions 
describing some universal features of the contour functions for 
$\mathcal{E}$ and $\mathcal{E}^{(n)}$ in CFT$_2$, when these quantities are UV divergent,
considering how analogous properties are captured by the contour functions for 
$S_A$ and $S_A^{(n)}$.

Consider a CFT$_2$ on the line and in its ground state, 
with  the spatial line bipartited by
the union of $N$ disjoint intervals
$A = \bigcup_{j=i}^N A_j$, where $A_j \equiv (a_j, b_j)$,
and its complement $B$. 
In the initial part of our analysis, 
it is convenient to assume that 
all the pairs of intervals in $A$ are disjoint, 
namely 
$b_{j} < a_{j+1}$ for $1\leqslant j \leqslant N-1$. 
From the perspective of the entanglement entropy, 
which measures the bipartite entanglement between $A$ and $B$,
all the endpoints of the intervals are also 
entangling points of the bipartition.  

Considering the R\'enyi entropies, 
it has been argued \cite{Calabrese:2009qy} that 
a specific divergence can be associated to  each entangling point
and it is given by 
$ R_n(\epsilon) \equiv - \tfrac{\Delta_n}{n-1} \log \epsilon$ 
as $\epsilon \to 0^+$, 
where $\epsilon $ is the UV cutoff 
and $\Delta_n$ is the conformal dimension of the twist fields 
$\mathcal{T}_n$ and $\overline{\mathcal{T}}_n$ which reads
\begin{equation}
\label{conf-dimensions-Tn}
    \Delta_n \equiv \frac{c}{12} \left( n - \frac{1}{n}\,\right)
\end{equation}
in terms of the central charge $c$ of the CFT$_2$. 
The argument of \cite{Calabrese:2009qy} relies on the expected form of the entanglement entropy in the massive case, 
when the dominant scale is the correlation length, 
and on the fact that this divergence is a UV feature
and therefore it is independent of the mass.

In order to evaluate the entanglement entropies from their contour functions, 
we find it convenient to adopt the regularization procedure 
employed in \cite{Ohmori:2014eia,Cardy:2016fqc},
where an infinitesimal interval of width $2\epsilon$ 
centered in each endpoint is removed, 
and this defines $A_\epsilon \subsetneq A$. 
It is reasonable to expect that the behavior of a contour function in the neighborhood of an entangling point provides 
the corresponding UV divergence in $S_A^{(n)}$.
This requirement implies that the asymptotic behaviour  
of the contour function close to the entangling points 
is 
$- \,\partial_{\epsilon} R_n(\epsilon)
\big|_{\epsilon = b_j - x} +O(1)$ as $x \to b_j$
and 
$- \,\partial_{\epsilon} R_n(\epsilon)
\big|_{\epsilon = x- a_j}+O(1)$ as $x \to a_j$,
with $1 \leqslant j \leqslant N$.
This asymptotic behaviour has been observed in all the contour functions considered in \cite{Coser:2017dtb}.
To construct a reasonable and simple candidate for the contour function, 
one can just superpose functions that guarantee this asymptotic behaviour. 
In the case mentioned above, where $A = \bigcup_{j=i}^N A_j$, 
this criterion gives the following contour function 
for the entanglement entropies \cite{Coser:2017dtb}
\begin{equation}
\label{contour-entropies-cft-N-int}
    \mathsf{s}^{(n)}_{_A}(x) 
    \,\equiv\,
    \tilde{\mathsf{s}}^{(n)}_{_A}(x)
    + \textrm{const}
\;\qquad\;
    \tilde{\mathsf{s}}^{(n)}_{_A}(x) 
    \,\equiv\, 
    \frac{1}{n-1} \,\sum_{j=1}^N
    \left(
    \frac{\Delta_n }{x-a_j} + \frac{\Delta_n }{b_j - x}
    \right) 
\;\qquad\;
x\in A
\end{equation}
where the constant additive term is independent of $x$ 
and can be exploited to guarantee that 
the integral of $\mathsf{s}^{(n)}_{_A}(x) $ 
over $A_\epsilon$ gives $S_A$.
In \cite{Coser:2017dtb},
the proposal (\ref{contour-entropies-cft-N-int}) has been checked for $N=2$ against the lattice computation in the harmonic chain based on the contour function discussed in Sec.\,\ref{sec:contour-entropies},
finding a good agreement close the endpoints and a discrepancy in the central parts of the two intervals.

The expression (\ref{contour-entropies-cft-N-int}) 
has been obtained in \cite{Coser:2017dtb} 
by considering the special CFT$_2$ given by the massless Dirac field on the line. 
Indeed, for the ground state of this model, 
the entanglement Hamiltonian of $A = \bigcup_{j=i}^N A_j$ 
is known \cite{Casini:2009vk} 
and it is written as the sum of a local term and a bilocal term. 
The weight function of the local term is proportional to the function $\beta_{N}(x)$ defined as 
\begin{equation}
\label{beta-local-N-int-def}
\beta_{N}(x) \equiv \frac{1}{\partial_x\Omega_N(x)}
    \;\;\;\;\qquad\;\;\;
    \Omega_N(x) \equiv \,
    \log \!
    \left(
    \frac{ \prod_{j=1}^N (x-a_j) }{ \prod_{j=1}^N (b_j - x) }
    \right) .
\end{equation}
The function $\tilde{\mathsf{s}}^{(n)}_{_A}(x)$ in (\ref{contour-entropies-cft-N-int}) is proportional to $1/\beta_{N}(x)$ through a coefficient that encodes the dependence on $n$ as follows \cite{Coser:2017dtb}
\begin{equation}
\label{contour-from-Omega}
    \tilde{\mathsf{s}}^{(n)}_{_A}(x)
    =
    \frac{\Delta_n}{(n-1)\, \beta_{N}(x)}  
    =
    \frac{\Delta_n}{n-1} \, \partial_x\Omega_N(x) \,.
\end{equation}

The expression (\ref{contour-entropies-cft-N-int}),
which has been found by considering the special case of the massless Dirac field,
captures the behaviour of the contour function 
providing the universal divergences of the entanglement entropies.
However, it does not describe an important model dependent contribution 
that determines the behaviour of the contour function 
in the part of $A$ far from the entangling points of the bipartition. 
Indeed, considering a generic CFT$_2$ on the line 
and in its ground state, for the simplest case of $N=2$
the moments of the reduced density matrix read
\cite{Caraglio:2008pk,Furukawa:2008uk, Calabrese:2009ez,Calabrese:2010he}
\begin{equation}
\label{tr-rhoA-n-cft-2int}
    \Tr \rho_A^n
    \,=\,
    \langle 
    \,\mathcal{T}_n(a_1)
    \,\overline{\mathcal{T}}_n(b_1)
    \,\mathcal{T}_n(a_2)
    \,\overline{\mathcal{T}}_n(b_2)
    \, \rangle
\,=\,
    C_n^2 \, \mathcal{P}_A^{ 2\Delta_n }\, \mathcal{F}_n(\eta)
\end{equation}
where $a_1 < b_1 < a_2 < b_2$ are the entangling points that support the twist fields, $C_n$ is the normalization constant occurring in the two-point function of the twist fields and 
\begin{equation}\label{eta}
    \mathcal{P}_A 
    \equiv 
    \frac{\epsilon^2\,(a_2-a_1)\,(b_2-b_1)}{(b_1 - a_1)\,(b_2 - a_2)\,(b_2 - a_1)\,(a_2 - b_1)}
    \;\;\;\;\qquad\;\;\;\;
        \eta \equiv \frac{(a_1 - b_1)\, (a_2 - b_2)}{(a_1 - a_2)\, (b_1 - b_2)} \in (0,1)
\end{equation}
being $\eta$ defined as the cross ratio of the four entangling points. 
The function $\mathcal{F}_n(\eta)$ is model dependent 
and very few explicit examples are available in the literature
\cite{Furukawa:2008uk, Calabrese:2009ez,Calabrese:2010he,Arias:2018tmw}. 
From (\ref{tr-rhoA-n-cft-2int}), the R\'enyi entropies 
(\ref{EntEnt-def-intro}) in this case become
\begin{equation}
\label{renyi-log-form-CFT-2int}
    S_A^{(n)} =\,
    \frac{2\Delta_n}{1-n} \, \log \mathcal{P}_A
    +
    \frac{2 \log(C_n)}{1-n}
    +
    \frac{1}{1-n}\, \log \! \big[\mathcal{F}_n(\eta)\big] \,.
\end{equation}
Considering (\ref{contour-entropies-cft-N-int}) specialized to the $N=2$ case,
we have that \cite{Coser:2017dtb}
\begin{equation}
\label{integral-contour-cft-2int}
    \int_{A_\epsilon} 
    \tilde{\mathsf{s}}^{(n)}_{_A}(x) \, \rd x
    \,=\,-
    \frac{2\Delta_n}{n-1}\, \log \mathcal{P}_A  
\end{equation}
where $A_\epsilon \equiv (a_1 + \epsilon,\, b_1 - \epsilon) \cup 
(a_2 + \epsilon, \,b_2 - \epsilon)$ is the integration domain 
after the implementation of the regularization procedure of 
\cite{Ohmori:2014eia,Cardy:2016fqc}.
Comparing (\ref{renyi-log-form-CFT-2int}) and (\ref{integral-contour-cft-2int}), it is straightforward to realize that 
(\ref{contour-entropies-cft-N-int}) for $N=2$ cannot be a contour function for (\ref{renyi-log-form-CFT-2int}) because
it does not capture the contribution to $S_A^{(n)}$ coming from $\mathcal{F}_n(\eta)$, which depends on the specific CFT$_2$ model.  
This has been already observed in \cite{Coser:2017dtb},
where (\ref{contour-entropies-cft-N-int}) for $N=2$
has been compared with numerical results 
for the contour function for the entanglement entropy in the harmonic chain,
obtained from (\ref{contour_constr}) and (\ref{mode-part-W}). 

In order to find reasonable candidates for the contour functions 
$\mathcal{E}$ and $\mathcal{E}^{(n)}$,
in the following we adapt to these quantities 
the criterion employed to construct (\ref{contour-entropies-cft-N-int}).
In particular, we employ a heuristic procedure 
to assign a proper divergent term to each boundary and entangling point.
In the case of an entangling point associated to two adjacent intervals, 
the corresponding divergent contribution 
is divided equally  between the two intervals. 
Considering the CFT$_2$ setup described above,  
in this case we also admit the possibility that some intervals are adjacent, 
namely that $b_j = a_{j+1}$ holds for some values of $j$. 
When $N=1$, the logarithmic negativity 
is obtained from the R\'enyi entropy with index $n=1/2$ 
through the relation (\ref{pure-states-moments-relation}); 
hence this case will not be explored.

The simplest setup providing non trivial results for 
$\mathcal{E}$ and $\mathcal{E}^{(n)}$
corresponds to $N=2$ in the limiting regime where the two intervals are adjacent, i.e. $A_1 = (a,p)$ and $A_2 = (p,b)$ with $a < p < b$.
In this case, since $\mathcal{E}$ and $\mathcal{E}^{(n)}$
explore the bipartite entanglement between $A_1$ and $A_2$ 
in the mixed state given by $\rho_A$,
only $x=p$ is a proper entangling point, 
while $x=a$ and $x=b$ are just endpoints.
Indeed, in higher dimensions and for adjacent spatial regions, 
only the area of the shared hypersurface provides the area law 
of the logarithmic negativity 
\cite{Eisler_2016,De_Nobili_2016}.

When $A$ is made by two adjacent intervals, 
the moments of $\rho_A^{\textrm{\tiny $\Gamma_2$}}$
are provided by the following three-point function 
\cite{Calabrese:2012ew, Calabrese:2012nk}
\begin{equation}
\label{tr-rhoA-T2-n-cft-2int-adj}
    \Tr \! \big( \rho_A^{\textrm{\tiny $\Gamma_2$}} \big)^n
    \,=\,
    \langle 
    \,\mathcal{T}_n(a)
    \,\overline{\mathcal{T}}^2_n(p)
    \,\mathcal{T}_n(b)
    \, \rangle
\,=\,
    \frac{\mathcal{C}_n\, \epsilon^{2\Delta_n+\Delta^{(2)}_n}}{ 
    \big(b-a\big)^{2\Delta_n-\Delta^{(2)}_n } 
    \big[\big(b-p\big)\big(p-a\big)\big]^{\Delta^{(2)}_n}  }
\end{equation}
where 
$\Delta^{(2)}_n$ is the scaling dimension of $\mathcal{T}^2_n$ and $\overline{\mathcal{T}}^2_n$, 
that depends on the parity of $n$ as follows
\begin{equation}
\label{Delta-Tn-squared-def}
        \Delta^{(2)}_n \equiv 
        \left\{\begin{array}{ll}
        \Delta_{n_{\textrm{\tiny o}}} \hspace{1.5cm} 
        & n=n_{\textrm{\tiny o}} \;\;\textrm{odd}
        \\
        \rule{0pt}{.5cm}
        2\,\Delta_{n_{\textrm{\tiny e}}/2}  
        & n=n_{\textrm{\tiny e}} \;\;\textrm{even}
        \end{array}\right.
\end{equation}
in terms of (\ref{conf-dimensions-Tn}),
and $\mathcal{C}_{n}$ is the structure constant of this three-point function, which depends on the underlying CFT$_2$ model \cite{Ginsparg:1988ui,DiFrancesco:1997nk}. 
It is worth remarking that, in the special case given by $n=2$, 
we have that $\mathcal{T}^2_2 = \overline{\mathcal{T}}^2_2$ 
is the identity operator; 
hence $\overline{\mathcal{T}}_n = \mathcal{T}_n$
and $\Delta^{(2)}_2 = 0$. 
The logarithmic negativity for a CFT$_2$ on the line and in its ground state when $A$ is the union of two adjacent intervals is obtained by applying 
the replica limit (\ref{neg-replica-limit-intro}) 
to (\ref{tr-rhoA-T2-n-cft-2int-adj}) and the result reads
(see also (\ref{logneg-CFT-adj-intro}))
\begin{equation}
\label{logneg-CFT-adj}
    \mathcal{E} = \widetilde{\mathcal{E}}
    + \textrm{const}
    \;\;\;\qquad\;\;\;
    \widetilde{\mathcal{E}}
    \equiv 
    \frac{c}{4} \, 
    \log \! \left( 
    \frac{(b-p)(p-a)}{(b-a)\, \epsilon}
    \right) 
\end{equation}
where the constant term in the r.h.s. of the first expression 
originates from the structure constant $\mathcal{C}_n$ 
in \eqref{tr-rhoA-T2-n-cft-2int-adj}, 
hence, although it is independent of $a$, $b$, $p$ and $\epsilon$,
it depends on the underlying CFT$_2$ model.

The structure of the UV divergence in 
(\ref{tr-rhoA-T2-n-cft-2int-adj}) leads us to introduce 
the following candidate 
for the contour function for the moments of $\rho_A^{\textrm{\tiny $\Gamma_2$}}$ 
\begin{equation}
\label{neg-contour-adjacent-cft}
    \mathsf{e}^{(n)}_{_A}(x)  
    \,\equiv\,
    \tilde{\mathsf{e}}^{(n)}_{_A}(x)
    + \textrm{const}
    \;\;\;\qquad\;\;\;
    \tilde{\mathsf{e}}^{(n)}_{_A}(x)  
    \equiv
    - \frac{\Delta_n}{x-a} 
    - \frac{\Delta^{(2)}_n}{2\,\big|x-p\big|} 
    - \frac{\Delta_n}{b-x} 
    \;\qquad\;
    x \in A
\end{equation}
where the constant additive term is independent of $x$ 
and can be exploited to guarantee that 
the integral of $\mathsf{e}^{(n)}_{_A}(x) $ 
over $A_\epsilon$ gives $\mathcal{E}^{(n)}$.
The expression (\ref{neg-contour-adjacent-cft}) has been constructed by imposing the same kind of divergence in all the three points 
characterizing the configuration of two adjacent intervals
with different multiplicative constants, 
which are determined by the conformal weights of the corresponding twist field.
Moreover, at the entangling point $x=p$, 
the factor $1/2$ has been introduced because
we have assumed that 
the same contribution to the divergence in $x=p$ 
is provided by $A_1$ and $A_2$, independently of their lengths.
This applies only at the location of the composite twist fields
$\mathcal{T}^2_n$ and $\overline{\mathcal{T}}^2_n$, i.e. 
where two intervals share an endpoint 
and the partial transposition is performed 
only in one of these two adjacent intervals. 

In the case of two adjacent intervals, 
the regularization procedure of \cite{Ohmori:2014eia,Cardy:2016fqc}
leads us to introduce the domain $A_\epsilon \equiv (a+\epsilon\, , p- \epsilon) \cup (p+\epsilon\, , b- \epsilon) $.
Integrating $\tilde{\mathsf{e}}^{(n)}_{_A}(x)$ in (\ref{neg-contour-adjacent-cft})
over $A_\epsilon$, one finds 
\begin{eqnarray}
\label{integ-tilde-en}
    \int_{A_\epsilon} 
    \tilde{\mathsf{e}}^{(n)}_{_A}(x) \, \rd x\,
    &=&
    -\,\log \! \Bigg(
    \bigg[\frac{b-a}{\epsilon}\bigg]^{2\Delta_n} 
    \bigg[\frac{(b-p)(p-a)}{\epsilon^2}\bigg]^{\Delta^{(2)}_n/2} 
    \,\Bigg)
    \\
    \rule{0pt}{.9cm}
    &=&
        -\,\log \! \Bigg(
    \bigg[\frac{b-a}{\epsilon}\bigg]^{2\Delta_n-\Delta^{(2)}_n } 
    \bigg[\frac{(b-p)(p-a)}{\epsilon^2}\bigg]^{\Delta^{(2)}_n} 
    \Bigg)
    + \frac{\Delta^{(2)}_n}{2} \,
    \log \! \bigg( \frac{(b-p)(p-a)}{(b-a)^2}\bigg)
    \nonumber
\end{eqnarray}
where the first term in the last expression gives 
the logarithm of the three-point function (\ref{tr-rhoA-T2-n-cft-2int-adj}), 
except for the term coming from the structure constant $\mathcal{C}_n$.

The replica limit $n_{\textrm{\tiny e}} \to 1$ 
of  (\ref{neg-contour-adjacent-cft}) provides 
the following candidate contour function for the logarithmic negativity
\begin{equation}
\label{neg-adj-proposal}
    \mathsf{e}_{_A}(x)  
    \,\equiv\,
    \tilde{\mathsf{e}}_{_A}(x)
    + \textrm{const}
    \;\;\;\qquad\;\;\;
    \tilde{\mathsf{e}}_{_A}(x)  
    \equiv
    \frac{c}{8\,\big|x-p\big|} 
    \;\;\;\;\;\qquad\;\;\;\;\;
    x \in A_1 \cup A_2
\end{equation}
which is positive in $A$  
and displays a divergence only at the entangling point
because $\Delta_{n_{\textrm{\tiny e}}} \to 0$ while  
$\Delta^{(2)}_{n_{\textrm{\tiny e}}} \to -c/4$
in the replica limit.

When $A= A_1 \cup A_2$ is the union of two disjoint intervals
$A_1 = (a_1,b_1)$ and $A_2 = (a_2,b_2)$, with $a_1 < b_1 < a_2 < b_2$, 
from the perspective of the logarithmic negativity, 
which measures the bipartite entanglement between $A_1$ and $A_2$
for the bipartite mixed state characterised by the reduced density matrix $\rho_A$,
the four endpoints of the intervals are not entangling points.
This simple observation already suggests 
that the logarithmic negativity is UV finite. 
The moments of $\rho_A^{\textrm{\tiny $\Gamma_2$}}$ can be written as the following four-point function 
\cite{Calabrese:2012ew, Calabrese:2012nk}
\begin{equation}
\label{tr-rhoA-T2-n-cft-2int}
    \Tr \! \big( \rho_A^{\textrm{\tiny $\Gamma_2$}} \big)^n
    \,=\,
    \langle 
    \,\mathcal{T}_n(a_1)
    \,\overline{\mathcal{T}}_n(b_1)
    \,\overline{\mathcal{T}}_n(a_2)
    \,\mathcal{T}_n(b_2)
    \, \rangle
    \,=\,
    C_n^2 \, \mathcal{P}_A^{ 2\Delta_n }\, \mathcal{G}_n(\eta)
\end{equation}
in terms of the quantities introduced in (\ref{tr-rhoA-n-cft-2int}).
The expression (\ref{tr-rhoA-T2-n-cft-2int}) is obtained 
by exchanging $ a_2 \leftrightarrow b_2$ in (\ref{tr-rhoA-n-cft-2int}); hence $\mathcal{G}_n(\eta) = (1-\eta)^{4\Delta_n} \mathcal{F}_n\big(\eta/(\eta-1)\big) $.
We remark that the correlator (\ref{tr-rhoA-T2-n-cft-2int}) 
depends on all the conformal data of the specific CFT$_2$ model, 
in contrast with (\ref{tr-rhoA-T2-n-cft-2int-adj}), 
which instead depends only on the central charge 
and on the structure constant $\mathcal{C}_n$.
Taking the replica limit (\ref{neg-replica-limit-intro}) 
in (\ref{tr-rhoA-T2-n-cft-2int}), 
one finds  $\mathcal{E} = \lim_{n_{\textrm{\tiny e}} \to 1} \mathcal{G}_n(\eta)$
for the logarithmic negativity,
which  is UV finite because the dependence on $\epsilon$ 
has been removed by the replica limit. 
This is in contrast with the case of adjacent intervals,
where the logarithmic negativity (\ref{logneg-CFT-adj}) is UV divergent 
and its divergent term depends only on the central charge
of the underlying CFT$_2$ model.
The crucial difference between (\ref{tr-rhoA-n-cft-2int}) and (\ref{tr-rhoA-T2-n-cft-2int}) is the sequence of the fields within the correlator,
which is ordered by the positions of the endpoints $a_1 < b_1 < a_2 < b_2$.
In particular, the limit $b_1 \to a_2$ of adjacent intervals in (\ref{tr-rhoA-T2-n-cft-2int}) leads to consider the operator product expansion of $\overline{\mathcal{T}}_n$ with itself, whose leading term defines the field $\overline{\mathcal{T}}^2_n$.

As for the contour function for the moments of $\rho_A^{\textrm{\tiny $\Gamma_2$}}$ 
when $A$ is the union of two disjoint intervals,
since the structure of the divergence in (\ref{tr-rhoA-T2-n-cft-2int}) 
is the same of (\ref{tr-rhoA-n-cft-2int}), 
a candidate is obtained from (\ref{contour-entropies-cft-N-int}) for $N=2$.
It reads
\begin{equation}
\label{contour-neg-CFT-2int-disjoint}
    \mathsf{e}^{(n)}_{_A}(x)  
    \,\equiv\,
    \tilde{\mathsf{e}}^{(n)}_{_A}(x)
    + \textrm{const}
    \;\;\;\;\;\qquad\;\;\;\;
    x \in A
\end{equation}
with 

\begin{equation}
\label{contour-neg-CFT-2int-disjoint-tilde}
    \tilde{\mathsf{e}}^{(n)}_{_A}(x)  
    \,=\, 
    (1-n)\,\tilde{\mathsf{s}}^{(n)}_{_A}(x) 
    \,=\,
    -
    \left(
    \frac{\Delta_n }{x-a_1} + \frac{\Delta_n }{b_1 - x}
    +
    \frac{\Delta_n }{x-a_2} + \frac{\Delta_n }{b_2 - x}
    \right) .
\end{equation}
By employing (\ref{integral-contour-cft-2int}), it is straightforward to observe that, for the massless scalar on the line, 
the contour function in (\ref{contour-neg-CFT-2int-disjoint}) 
cannot capture the contribution coming from $\mathcal{G}_n(\eta)$ 
in (\ref{tr-rhoA-T2-n-cft-2int}), which is the only term that provides a non vanishing logarithmic negativity, 
as highlighted in the text below (\ref{tr-rhoA-T2-n-cft-2int}).
This can be realized also by observing that 
(\ref{contour-neg-CFT-2int-disjoint})
vanishes identically in the replica limit 
$n_{\textrm{\tiny e}} \to 1$.
A contour function for the logarithmic negativity in the setup where the two intervals are disjoint is highly model dependent
and we do not have  any proposal for this quantity 
in the case of the massless scalar.

Since the first candidate for the contour of the entanglement entropies of disjoint intervals (\ref{contour-entropies-cft-N-int}) is related to the weight function of the local term of the entanglement Hamiltonian of the massless Dirac field of \cite{Casini:2009vk} as reported in (\ref{contour-from-Omega}), 
it is worth considering the possibility that a similar relation 
could occur also for the contour of the logarithmic negativity. 
For the massless Dirac on the line and in the ground state, 
when $A$ is the union of two disjoint intervals 
and the partial transposition w.r.t. $A_2$ is considered,
it has been proposed \cite{Murciano:2022vhe} 
to explore the operator obtained 
by exchanging $a_2$ and $b_2$ in the corresponding entanglement Hamiltonian
(see also \cite{Rottoli:2022plr} for further results).  
In this operator, the weight function of the local term is 
$\beta_{2}^{\textrm{\tiny $\Gamma_2$}}(x) \equiv \beta_{2}(x)\big|_{a_2 \leftrightarrow b_2}$, 
that is obtained by specializing (\ref{beta-local-N-int-def}) to $N=2$.
We remark that $\beta_{2}^{\textrm{\tiny $\Gamma_2$}}(x)$ vanishes only 
at the four endpoints of $A$.  
The case of adjacent interval is found by taking the limit $b_1 \to a_2$,
and the resulting weight function of the local term vanishes
both at the entangling point and at the endpoints. 
Since $\beta_{2}^{\textrm{\tiny $\Gamma_2$}}(x)$ is positive 
for  $x\in A_1$ and negative for $x\in A_2$, 
it is natural to introduce 
$\tilde{\mathsf{w}}^{(n)}_{_A}(x)  \equiv 
d_n /|\beta_{2}^{\textrm{\tiny $\Gamma_2$}}(x)|$ with $x\in A$
as a possible candidate for the contour function for $\mathcal{E}^{(n)}$,
where the factor $d_n0$ encodes all the dependence on $n$. 
The corresponding contour function 
for the logarithmic negativity is obtained through 
the replica limit $n_{\textrm{\tiny e}} \to 1$,
that gives $\tilde{\mathsf{w}}_{_A}(x)  \equiv 
\tilde{d} /|\beta_{2}^{\textrm{\tiny $\Gamma_2$}}(x)|$, 
where $\tilde{d}  \equiv \lim_{n_{\textrm{\tiny e}} \to 1} d_{n_{\textrm{\tiny e}}}$ 
and $\tilde{\mathsf{w}}_{_A}(x) > 0 $ for $x \in A$ 
is guaranteed by imposing $\tilde{d} >0$.

We find it worth anticipating here 
that $\tilde{\mathsf{w}}_{_A}(x)$  
does not agree with the qualitative behaviour 
of the numerical data for the 
contour function for the logarithmic negativity 
in the harmonic chain,
reported in Sec.\,\ref{sec-GS-line-adjacent} and 
Sec.\,\ref{sec-GS-line-disjoint}.
Indeed, in the case of two disjoint intervals,
$\tilde{\mathsf{w}}_{_A}(x)$ diverges as $x$ approaches any endpoint of $A$, 
while the numerical data points 
discussed in Sec.\,\ref{sec-GS-line-disjoint}
tell us that the contour function for the logarithmic negativity
does not diverge in $A$ (see Fig.\,\ref{fig-disjoint}).
Moreover, still in the case of two disjoint intervals,
we observe that  the integral of the positive function 
$\tilde{\mathsf{w}}_{_A}(x)$ over $A$ is divergent and therefore it should be regularized by considering $A_\epsilon$ as integration domain; 
hence the UV finite quantity $\mathcal{E}$ cannot be recovered.
Instead, 
considering $\tilde{\mathsf{w}}^{(n)}_{_A}(x)$ with  $d_n = -\Delta_n$, 
we observe that it is very similar to (\ref{contour-neg-CFT-2int-disjoint-tilde})
and its integral over $A_\epsilon$
provides the divergent term occurring in (\ref{tr-rhoA-T2-n-cft-2int}).
However, 
since $\Delta_{n_{\textrm{\tiny e}}} \to 0$ as $n_{\textrm{\tiny e}} \to 1$,
both this expression and (\ref{contour-neg-CFT-2int-disjoint-tilde})
vanish identically in the replica limit.

In the case of adjacent intervals, 
similar observations can be made. 
In particular, 
at the endpoints $x=a$ and $x=b$,
we have that $\tilde{\mathsf{w}}_{_A}(x)$ diverges
and this is in contrast with the numerical data points 
discussed in Sec.\,\ref{sec-GS-line-adjacent} 
(see the top panel of Fig.\,\ref{fig-adj}).
Instead, considering the behaviour close to the entangling point at $x=p$, we observe that $\tilde{\mathsf{w}}_{_A}(x)$ 
with $\tilde{d} \equiv c/8$
displays the same behaviour of $\tilde{\mathsf{e}}_{_A}(x)$ 
in (\ref{neg-adj-proposal}),
which is in agreement with the numerical data points 
shown in Fig.\,\ref{fig-adj}. 
As for the contour function for $\mathcal{E}^{(n)}$,
the divergencies displayed by $\tilde{\mathsf{w}}^{(n)}_{_A}(x)$ depend on $n$ only through the coefficient $d_n$ and this is in contrast with
the numerical results reported in Fig.\,\ref{fig-adj-ren}, 
especially for $n=2$,
where the contour function does not diverge around the entangling point.

\section{Ground state: Two adjacent blocks in the line}
\label{sec-GS-line-adjacent}

In this section, considering  
the harmonic chain on the line and in its ground state
(hence the correlators (\ref{qq-corr-gs-line}) 
and (\ref{pp-corr-gs-line}) are employed),
we explore the entanglement negativity when
$A_1$ and $A_2$ are adjacent blocks made by $L_1$ and $L_2$ consecutive sites respectively. 
We can set
$A_1 = \{ 1 \leqslant i \leqslant L_1\}$ 
and $A_2 = \{ L_1+1 \leqslant i \leqslant L_1+L_2\}$ without loss of generality.
In Sec.\,\ref{subsec-GS-line-adjacent-contour}
we discuss our numerical results about
the contour functions for $\mathcal{E}$ and $\mathcal{E}^{(n)}$
defined in Sec.\,\ref{sec:contour-neg}.
Instead, in Sec.\,\ref{subsec-GS-line-adjacent-logneg},
focusing on the massive regime, 
we employ the logarithmic negativity 
to introduce a  UV finite quantity that might be useful 
to explore RG flows.

\subsection{Contour functions}
\label{subsec-GS-line-adjacent-contour}

\begin{figure}[t!]
\vspace{-.5cm}
\includegraphics[width=1\textwidth]{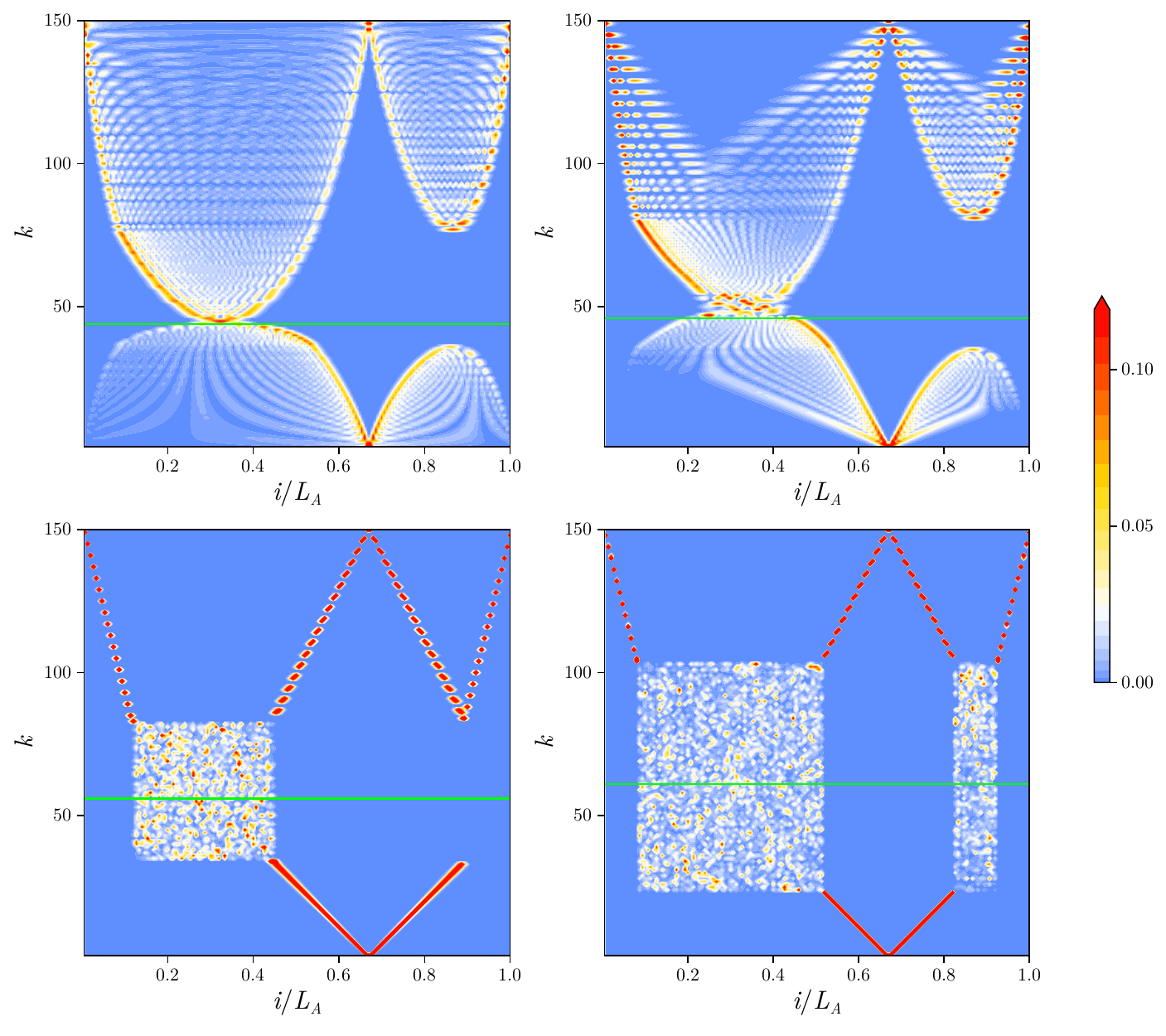}
\caption{\small
Mode participation function \eqref{tilde-pk-i-def} 
for two adjacent blocks with $L_1=2L_2 = 100$
in the infinite harmonic chain in its ground state. 
The parameter $\omega$ changes
in the different panels as follows:
$\omega L_1 = 10^{-10}$ (top left), 
$\omega L_1 = 40$ (top right), 
$\omega L_1 = 500$ (bottom left) 
and $\omega L_1 = 2000$ (bottom right). 
The green straight line corresponds to the threshold mode at $k=k_\ast$.
}
\label{mpf-adj}
\end{figure}

In Fig.\,\ref{mpf-adj} we report the numerical data points 
for the mode participation function $\tilde{p}_k(i)$ 
defined in (\ref{tilde-pk-i-def}) 
in a configuration where $L_1=2L_2$, 
for different values of the mass parameter $\omega$. 
Also very large values for $\omega$ are considered 
because 
interesting features have been observed for the entanglement Hamiltonians 
in harmonic chains in this regime 
\cite{Eisler:2020lyn,Gentile:2025koe}.
The horizontal green line corresponds to the 
threshold index $k_\ast$ introduced in (\ref{threshold-mode-def})
(see also the green plane in Fig.\,\ref{fig-3D-MPF-example}).
Since we are ordering the symplectic eigenvalues of 
$\gamma_A^{\textrm{\tiny $\Gamma_2$}}$ in the increasing way
(see the text below (\ref{williamson-dec-gamma-A-transpose})),
the modes labeled by $ k_\ast \leqslant k \leqslant L_A$ 
do not contribute to the $\mathcal{E}$, 
but they provide non vanishing contributions to $\mathcal{E}^{(n)}$
(see (\ref{neg-from-sigma-k})-(\ref{Fn-neg-def})).

In the massless regime 
(see the top left panel of Fig.\,\ref{mpf-adj}), 
characteristic profiles in $\tilde{p}_k(i)$ are observed
that are qualitatively very different from the ones 
shown in \cite{Coser:2017dtb}
for the mode participation function $p_k(i)$,
defined in (\ref{mode-part-W}),
obtained from $\gamma_A$.
This qualitative difference is mainly due to 
the occurrence of the threshold mode $k_\ast$. 
Also the behaviour of these profiles 
as $\omega$ increases is very different in 
$\tilde{p}_k(i)$ w.r.t. the one observed in $p_k(i)$. 
However, $\tilde{p}_k(i)$ and $p_k(i)$ 
share interesting common features.
Indeed, at any value of $\omega$, 
the modes providing the largest contributions 
to $\mathcal{E}$ and $S_A$ display local maxima in
$\tilde{p}_k(i)$ and $p_k(i)$, respectively, 
at the entangling points. 
Moreover, as $\omega$ increases, 
both $\tilde{p}_k(i)$ and $p_k(i)$ localize along insightful curves. 
These limiting curves in the regime of large $\omega$ 
can be qualitatively understood by assuming that, as $\omega$  increases, 
the correlations that are more relevant for the bipartite entanglement 
get localized close to the entangling points.

It is instructive to identify the positions of the local maxima of 
$\tilde{p}_{k}(i)$ as a function of $i$.
Indeed, the contribution to the contour functions
$\mathsf{E}_A(i)$ and $\mathsf{E}_A^{(n)}(i)$ (see (\ref{contour_constr_neg}))
of the $k$-th mode is determined by $\tilde{p}_{k}(i)$ 
as $i \in A$, weighted by 
either $\widetilde{F}_1(\tilde{\sigma}_k)$ or $\widetilde{F}_n(\tilde{\sigma}_k)$ respectively. 
In particular, for $k \gtrsim 0$,
the largest value of $\tilde{p}_{k}(i)$ 
corresponds to the entangling point at $x=p$, 
for all the values of $\omega$ that we have explored. 
This feature is observed also for 
the mode participation function $p_k(i)$ when $k \gtrsim 0$
in the case of two disjoint intervals
(see Fig.\,10 of \cite{Coser:2017dtb}), 
where the entangling points are given by 
all the four endpoints, 
even for very close intervals.
As for the endpoints at $x=a$ and $x=b$,
the mode participation function $\tilde{p}_{k}(i)$ 
displays local maxima at these points for $k \simeq L_A$;
hence, while they provide an important contribution to $\mathcal{E}^{(n)}$,
they do not contribute to $\mathcal{E}$.

In the massless regime, $\tilde{p}_k(i)$ at $k=k_\ast$ 
takes relevant values
only in a domain $A_{1,\ast} \subsetneq A_1$ properly contained in the largest interval 
(see the top left panel of Fig.\,\ref{mpf-adj} and Fig.\,\ref{fig-3D-MPF-example},
for a configuration where $A$ is made by two either adjacent or disjoint blocks respectively).
When $\omega$ increases, 
the size of $A_{1,\ast}$ increases as well, 
as shown in 
the top right and bottom left panel of Fig.\,\ref{mpf-adj}.
If $\omega$ further increases, 
a similar proper subset $A_{2,\ast} \subsetneq A_2$ occurs 
(see the bottom right panel of Fig.\,\ref{mpf-adj}).
In the regime of large $\omega$,
which is considered e.g. in the bottom panels of 
Fig.\,\ref{mpf-adj},
we can identify some rectangular domains in the plane parameterized by $(i,k)$, with 
$1\leqslant k \leqslant L_A$ 
and $1\leqslant i \leqslant L_A$, 
where  $\tilde{p}_k(i)$ is relevant and seems to display a random behavior. 
For the values of $k$ defining these rectangular regions, 
we observe that the corresponding symplectic eigenvalues 
are almost all degenerate to the threshold value 
$\tilde{\sigma}_k \simeq 1/2$.
This explains the randomness of  $\tilde{p}_k(i)$ in this region; 
indeed, the algorithm determining $W$ 
(see the appendix\;\ref{app-W-matrix}) 
is unstable when the symplectic eigenvalues are almost degenerate. 
In this regime, the threshold mode $k_\ast$ is not well defined anymore and it is replaced by threshold rectangles in the domain parameterized by $(i,k)$.
In the bottom panels of 
Fig.\,\ref{mpf-adj} the green line is shown anyway, 
but it indicates that the threshold mode could be anywhere in the range corresponding to the rectangle; hence but it is not meaningful like in the other panels.
When $\omega$ becomes very large, 
the state represented by $\rho_A$ is approximated by a product state, 
whose symplectic spectrum is degenerate and equal to $1/2$. As a consequence, the number of threshold modes (i.e. the size of these rectangular regions along the $k$-axis) increases as the mass parameter $\omega$ increases. 
In the asymptotic limit $\omega \to +\infty$,
we expect that these rectangular regions cover 
the whole domain parameterized by $(i,k)$ since all the symplectic eigenvalues 
of $\gamma_A^{\textrm{\tiny $\Gamma_2$}}$ 
become equal to $1/2$ in this limit.
This also means that the bipartite entanglement between $A_1$ and $A_2$ 
in the mixed state given by $\rho_A$ vanishes in this limit.

Another interesting feature of the numerical results 
for the large mass regime reported in the bottom panels Fig.\,\ref{mpf-adj} is that,
besides the rectangular domains discussed above, 
the domains in $(i,k)$ where $\tilde{p}_{k}(i)$ 
is relevant are given by well defined segments 
with simple relation occurring among their slopes. 
In particular, from the entangling point and at $k=0$
two segments depart with opposite slopes, 
one in $A_1$ and the other one in $A_2$.
These two segments are important because they entirely determine the logarithmic negativity in the large mass regime.
When both rectangles associated to $\tilde{\sigma}_k \simeq 1/2$ are formed, 
each segment reaches the rectangle 
within the corresponding interval (see the bottom right panel of Fig.\,\ref{mpf-adj}).
However, these segments are well defined 
even for lower values of $\omega$, 
when only one of these rectangles occurs 
(see e.g. the bottom left panel of Fig.\,\ref{mpf-adj}).

Four other segments where $\tilde{p}_{k}(i)$ 
is relevant depart from points along the line corresponding to $k=L_A$:
two of them from the entangling point at $x=p$,
that corresponds to $i=L_1$,
and the remaining ones from the endpoints at $x=a$ and $x=b$,
corresponding to $i=1$ and $i=L_A$ respectively. 
Although they do not contribute to $\mathcal{E}$, 
they provide non vanishing contributions to $\mathcal{E}^{(n)}$.
Simple relations occur among the slopes of these four segments.
Indeed, the two segments departing from the entangling point have opposite slopes and the same feature is observed for the two segments attached to the endpoints. 
Considering the absolute values of the slopes of these four segments
evaluated with respect to the vertical line at a fixed site of the chain, 
we observe that the slope of the segments departing from the entangling point 
is twice the one of the segments departing from the endpoints. 
This relative factor of $2$ between these slopes could be related to 
the fact that, in the continuum limit, 
the computation of (\ref{neg-moments-def-intro})
is based on the three-point function given by the first equality in 
(\ref{tr-rhoA-T2-n-cft-2int-adj}), which tells us that,
while $\mathcal{T}_n$ and $\overline{\mathcal{T}}_n$ occur at the endpoints of $A$,
either $\mathcal{T}_n^{2}$ or $\overline{\mathcal{T}}_n^{2}$ 
occurs at the entangling point.

Considering the two slopes (in absolute value) of the four segments departing at $k=L_A$,
we also observe numerically that the smallest one multiplied by $4$ 
gives the absolute value of the slopes of the two segments at $k=1$.

\begin{figure}[t!]
\vspace{-.5cm}
\includegraphics[width=1\textwidth]{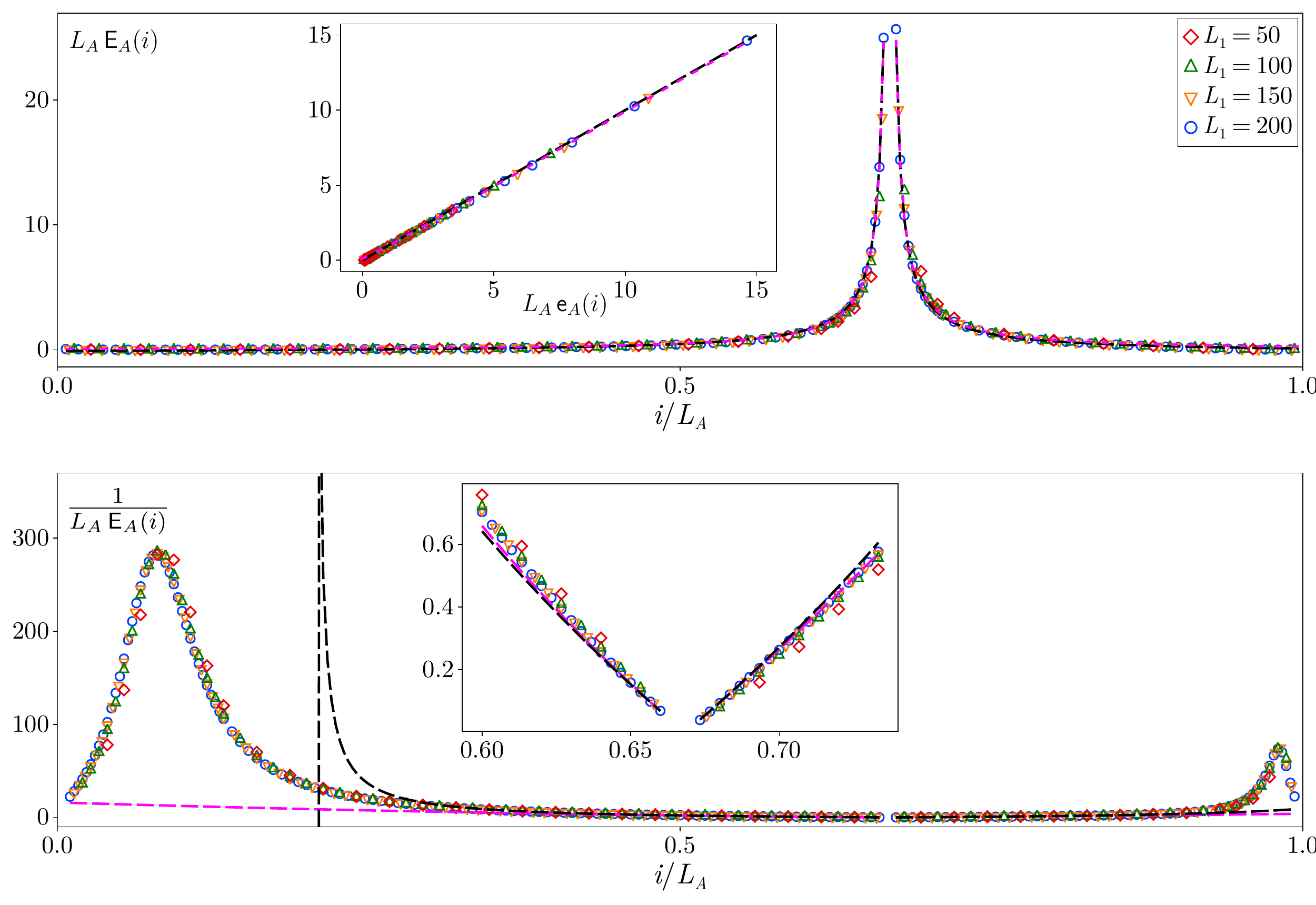}
\caption{\small
Contour function $\mathsf{E}_A(i)$ (see (\ref{contour-function-NEG-projector})) 
for the logarithmic negativity of two adjacent blocks 
with $L_1 = 2L_2$
in the infinite harmonic chain with $\omega L_1 = 10^{-10}$ 
in its ground state. 
The black and magenta dashed lines correspond to the analytical proposals 
given by  \eqref{neg-adj-proposal} and \eqref{Ryu-contour-adj-on-A} respectively.
The inset in the top panel highlights the quality of the fit, while the one in the bottom panel zooms in on the region around the entangling point. 
}
\label{fig-adj}
\end{figure}

In Fig.\,\ref{fig-adj}, 
we report some numerical results for $\mathsf{E}(i)$,
defined in (\ref{contour-function-NEG-projector}),
in the case of adjacent blocks with $L_1 = 2 L_2$
and in the massless regime.
The dimensionless quantity $L_A \, \mathsf{E}(i)$ is shown 
in terms of the ratio $i/L_A$.
A remarkable collapse of the data points corresponding to different values of $L_1$ is observed,
which strongly indicates that 
an analytic expression in the continuum could be obtained through CFT$_2$ 
methods. 

The top panel of Fig.\,\ref{fig-adj} 
shows that $L_A \, \mathsf{E}(i)$
diverges at the entangling point, while 
it is finite at the endpoints. 
This qualitative behavior is in agreement the analytic expression $\mathsf{e}_{_A}(x)$,
defined in (\ref{neg-adj-proposal})
as the replica limit of (\ref{neg-contour-adjacent-cft}).

A different contour function for the logarithmic negativity 
of adjacent intervals in CFT$_2$
has been proposed in \cite{Kudler-Flam:2019nhr}.
This construction is not based on the replica limit,
the integration is performed only on one of the two adjacent intervals 
in $A = A_1 \cup A_2$
and the case corresponding to a generic value of the index $n$
has not been considered.

In order to obtain a contour function  for the logarithmic negativity 
of adjacent intervals that is  defined on $A = A_1 \cup A_2$,
and therefore it is suitable for a comparison 
with (\ref{neg-adj-proposal}), 
we slightly modify the construction of \cite{Kudler-Flam:2019nhr}.
In particular, from (\ref{logneg-CFT-adj}), we introduce
\begin{equation}
\label{Ryu-contour-adj-on-A-tilde}
    \tilde{\mathsf{r}}_{_A}(x)  
    \,\equiv\, 
    \frac{1}{2} 
    \Big[\,
    \Theta_{2}(x)\,
    \big(\partial_b \widetilde{\mathcal{E}}\,\big)\big|_{b=x}
    -
    \Theta_{1}(x)\,
    \big(\partial_a \widetilde{\mathcal{E}}\,\big)\big|_{a=x}
    \,\Big]
\end{equation}
where
(see also Eq.\,(A62) in \cite{Kudler-Flam:2019nhr})
\bea
\big(\partial_b \widetilde{\mathcal{E}}\,\big)\big|_{b=x}
&=&
\frac{c}{4} \left(\frac{1}{x-p} -  \frac{1}{x-a}\right)
=
\frac{c}{4}\; \frac{p-a}{(x-p)\,(x-a)}
\\
\rule{0pt}{.8cm}
\big(\partial_a \widetilde{\mathcal{E}}\,\big)\big|_{a=x}
&=&
\frac{c}{4} \left(\frac{1}{b-x} -  \frac{1}{p-x}\right)
=
\frac{c}{4}\; \frac{b-p}{(b-x)\,(x-p)}
\eea
the function $\Theta_{j}(x)$ is the characteristic function  
for the interval $A_j$, with $j \in \{1,2\}$, 
and 
\begin{equation}
\label{Ryu-contour-adj-on-A}
    \mathsf{r}_{_A}(x)  
    \,\equiv\,
    \tilde{\mathsf{r}}_{_A}(x)
    + \textrm{const}
    \;\;\;\;\;\qquad\;\;\;\;\;
    x \in A_1 \cup A_2 \,.
\end{equation}

\begin{figure}[t!]
\vspace{-.5cm}
\includegraphics[width=1\textwidth]{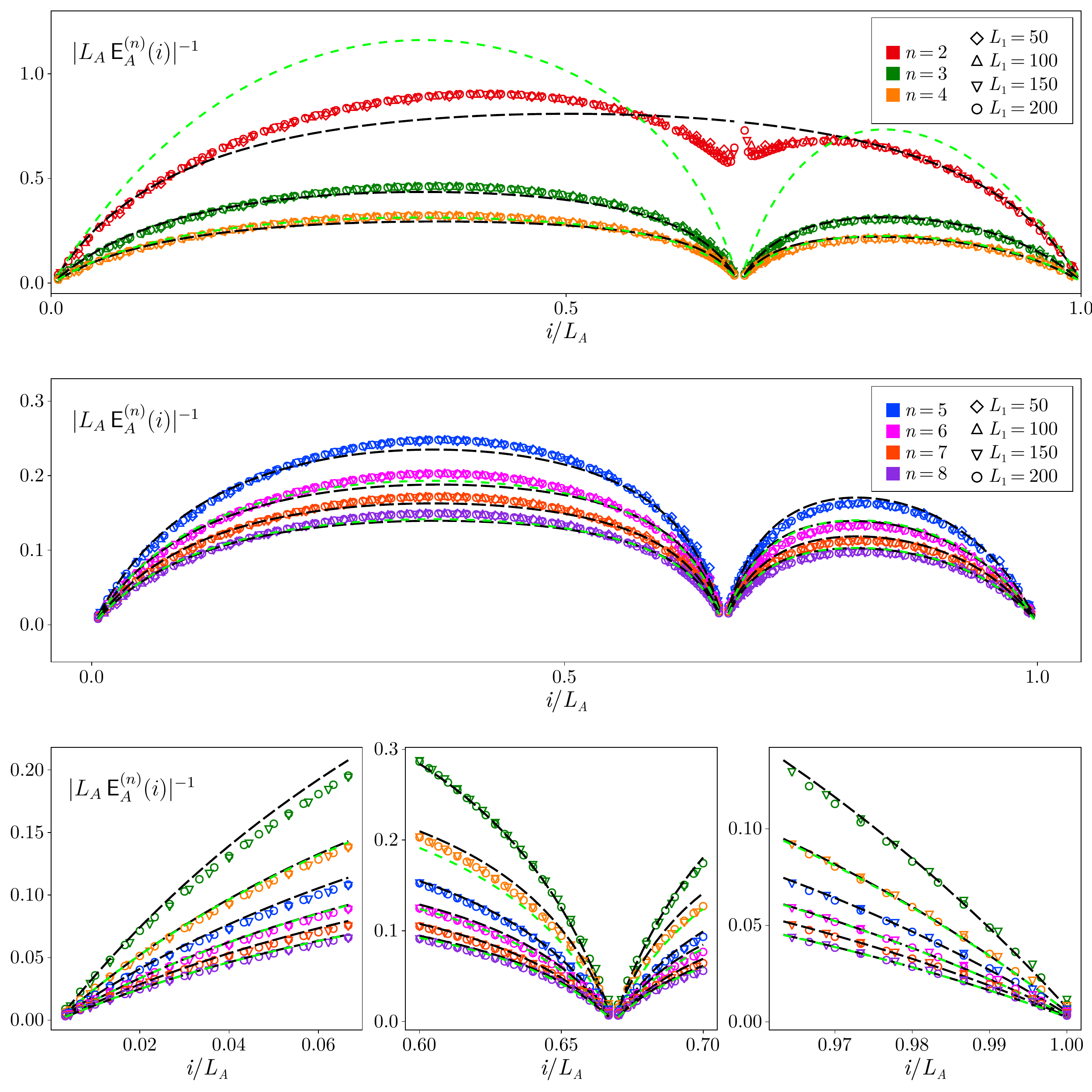}
\caption{\small
Contour function $\mathsf{E}_A^{(n)}(i)$ for the logarithm of the moments of 
$\gamma_A^{\textrm{\tiny $\Gamma_2$}}$
 in (\ref{contour-function-NEG-projector}) in the massless regime and
 for various values of $n$,
 in the same setup of Fig.\,\ref{fig-adj}.
 The black and green dashed lines correspond respectively to \eqref{neg-contour-adjacent-cft} and to its modified version, obtained by replacing $\Delta^{(2)}_n$ with $\Delta_n$. 
}
\label{fig-adj-ren}
\end{figure}

We remark that, because of the characteristic functions of $A_1$ and $A_2$, 
the expression in (\ref{Ryu-contour-adj-on-A}) diverges 
only at the entangling point $x=p$.
The expression (\ref{logneg-CFT-adj}) for the logarithmic negativity 
is recovered up to the constant term $\mathcal{C}_n$;
indeed
\begin{equation}
\widetilde{\mathcal{E}} 
= 
\int_{\tilde{A}_{\epsilon}} \!\! 
\tilde{\mathsf{r}}_{_A}(x)\, \rd x
\,= \,
-\int_{a}^{p-\epsilon} \!\! 
\big(\partial_a \widetilde{\mathcal{E}}\,\big)\big|_{a=x} 
 \,\rd x
\,= 
\int_{p+\epsilon}^{b} \!\! 
\big(\partial_b \widetilde{\mathcal{E}}\,\big)\big|_{b=x}
\, \rd x
\end{equation}
where $\tilde{A}_\epsilon \equiv 
(a\, , p- \epsilon) \cup (p+\epsilon\, , b) $,
which satisfies $ A_\epsilon  \subsetneq \tilde{A}_\epsilon $
(see the text above (\ref{integ-tilde-en})). 
By taking $A_\epsilon$ as the domain of integration, 
we get the same result up to a $O(\epsilon)$ term.

%

In Fig.\,\ref{fig-adj}, we compare our numerical results 
for the contour function $\mathsf{E}_A(i)$ of two adjacent blocks 
(see (\ref{contour-function-NEG-projector})) in the massless regime, 
with the analytic expressions 
in (\ref{neg-adj-proposal}) and (\ref{Ryu-contour-adj-on-A}).
As for the additive constant term, 
let us introduce 
\begin{equation}
\label{lattice-shift}
    \mathfrak{C}_{\textrm{\tiny lat}}^{(n)}
    \equiv
    \frac{1}{L_A} \sum_{i \in A} \Big(
    \mathsf{E}_A^{(n)}(i)-\mathsf{e}^{(n)}_A(i-1/2) 
    \Big)
\end{equation}
which is obtained by adapting the corresponding expression employed in 
\cite{Coser:2017dtb} to study the contour function for the entanglement entropies,
as a reasonable expression suggested by the normalization condition. 
The numerical data points obtained from the lattice display 
a very nice collapse on a well defined curve
and an overlap between this curve and the analytic expressions in the continuum
has been obtained by setting the additive constant 
equal to (\ref{lattice-shift}) for (\ref{neg-adj-proposal})
and equal to zero for (\ref{Ryu-contour-adj-on-A}).
Both these expressions nicely captures the divergent behaviour of the numerical data points around the entangling point at $x=p$. 
However, a slight disagreement is observed 
around the two endpoints at $x=a$ and $x=b$, 
as highlighted in the bottom panel of Fig.\,\ref{fig-adj}, 
where the $(L_A \, \mathsf{E}_A(i))^{-1}$ is shown, in order to highlight the part of $L_A \, \mathsf{E}_A(i)$ 
that cannot be seen in the top panel 
because of the divergence at the entangling point. 
In particular, the curve obtained from the numerical data points 
displays an interesting non monotonic behaviour 
that would be worth understanding analytically. 
%

In Fig.\,\ref{fig-adj-ren} we report some numerical results 
for the contour functions $\mathsf{E}_A^{(n)}(i)$, 
defined in (\ref{contour-function-NEG-projector}), 
for various values of $n$ 
and for the same setup of Fig.\,\ref{fig-adj}.
In order to highlight both the divergences and the finite part 
of $L_A \, \mathsf{E}^{(n)}(i)$,
we show $(L_A \, \mathsf{E}_A^{(n)}(i))^{-1}$.
Also for this quantity the numerical data sets 
obtained for different values of $L_1$ 
nicely collapse, providing well defined curves.
We remark that the curves for $L_A \, \mathsf{E}_A^{(n)}(i)$ obtained 
from the numerical analysis display divergences both at the entangling point and at the endpoints,
while the curve for $L_A \, \mathsf{E}_A(i)$ in Fig.\,\ref{fig-adj} 
diverges only at the entangling point. 
This is compatible with the fact that $\tilde{p}_k(i)$ 
have peaks both at the entangling point and at the endpoints
when $k$ is close to $L_A$,
which label modes that provide 
non vanishing contributions only to $\mathcal{E}_n$
(see the discussions of 
Fig.\,\ref{fig-3D-MPF-example} and Fig.\,\ref{mpf-adj}).

The curves for $L_A \, \mathsf{E}_A^{(n)}(i)$
obtained from the collapse of the data points 
are compared with the ones 
(see the dashed black lines in Fig.\,\ref{fig-adj-ren})
corresponding to 
$\tilde{\mathsf{e}}^{(n)}_{_A}(x) + \mathfrak{C}_{\textrm{\tiny lat}}^{(n)}$,
from (\ref{neg-contour-adjacent-cft}) and (\ref{lattice-shift}),
and with the ones 
(see the dashed green lines in Fig.\,\ref{fig-adj-ren})
corresponding to a slightly modified expression obtained by 
replacing $\Delta^{(2)}_n$ with $\Delta_n$ 
in \eqref{neg-contour-adjacent-cft} 
and using (\ref{lattice-shift}) for the additive constant,
which therefore coincide with 
$\tilde{\mathsf{e}}^{(n)}_{_A}(x) + \mathfrak{C}_{\textrm{\tiny lat}}^{(n)}$
when $n$ is odd. 
For $n \geqslant 3$, both these analytic proposals nicely describe
$L_A \, \mathsf{E}_A^{(n)}(i)$ around its divergencies,
although a slight discrepancy occurs 
in the internal parts of the blocks $A_1$ and $A_2$.
Instead, for $n=2$ the dashed black and dashed green curves are drastically different. In particular (\ref{neg-contour-adjacent-cft}) is not divergent at the entangling point because $\Delta^{(2)}_2 = 0$ (see (\ref{Delta-Tn-squared-def})), while its modified version (see the top dashed green line in the top panel of Fig.\,\ref{fig-adj-ren}) diverges at this point. 
Since the corresponding numerical data points 
(see the red markers in the top panel of Fig.\,\ref{fig-adj-ren})
do not display a divergence at the entangling point
in agreement with \eqref{neg-contour-adjacent-cft}, 
its modified version can  be discarded.

When $n=2$ and the intervals $A_1$ and $A_2$ are adjacent, 
the identity 
$\Tr \!\big( \rho_{A}^{\textrm{\tiny $\Gamma_2$}}\big)^{2} = \Tr \rho_{A}^{2}$ 
holds (see Eq.\,(36) of \cite{Calabrese:2012nk}).
However, 
by comparing the numerical data points 
in the top panel of Fig.\,\ref{fig-adj-ren} for $n=2$ 
and the results for the contour function of the $n=2$ R\'enyi entropy
shown in Fig.\,5 of \cite{Coser:2017dtb},
it is straightforward to realize that 
$\mathsf{E}_A^{(2)}(i)$ and $\mathsf{S}_A^{(2)}(i)$ 
do not coincide. 
The most visible difference occurs around the shared endpoint at $x=p$,
where both these contour functions are finite, 
but, while $\mathsf{S}_A^{(2)}(i)$ is smooth, 
$\mathsf{E}_A^{(2)}(i)$ displays a singular behaviour
which is not described by the analytic expressions considered above.

In almost all the cases that we have explored numerically, 
$\mathsf{E}_A^{(n)}(i)$ 
has the same sign of the corresponding $\mathcal{E}^{(n)}$.
We have observed only one exception
for  $n=2$, where $\mathsf{E}_A^{(2)}(i) < 0$, like $\mathcal{E}^{(2)}$,
except for few data points close to the entangling point,
where it becomes positive.

We find it worth trying to construct analytical expressions 
for the contour functions for the logarithm of the moments of 
$\gamma_A^{\textrm{\tiny $\Gamma_2$}}$ 
by adapting the procedure proposed in \cite{Kudler-Flam:2019nhr}. 
As first attempt, let us consider \eqref{Ryu-contour-adj-on-A-tilde} 
with $\widetilde{\mathcal{E}}$ replaced by 
$\widetilde{\mathcal{E}}^{(n)}$, 
reported in \eqref{tr-rhoA-T2-n-cft-2int-adj}.
Since the resulting function 
is finite at the endpoints $x=a$ and $x=b$, 
it is incompatible with the numerical results shown in Fig.\,\ref{fig-adj-ren}.
Another ansatz, inspired by Eq.\,(9) of \cite{Kudler-Flam:2019nhr},
reads 
\begin{equation} 
\label{Ryu-deriv-ren-neg-E}
\tilde{\mathsf{r}}^{(n)}_{_A}(x)  
    \equiv
    \frac{1}{2} \left[ \,\frac{\Theta_1(x)}{2} 
    \Big( \partial_{p}\mathcal{E}^{(n)}\big|_{p=x}
    -\partial_{a}\mathcal{E}^{(n)}\big|_{a=x}
    \Big) + \frac{\Theta_2(x)}{2} 
    \Big( \partial_{b}\mathcal{E}^{(n)}\big|_{b=x}
    -\partial_{p}\mathcal{E}^{(n)}\big|_{p=x} 
    \Big) \right]\,.
\end{equation}
However, also this expression should be discarded 
because its integral over $A_\epsilon$
(see the text above (\ref{integ-tilde-en}))
does not reproduce the dependence on $\epsilon$ 
of \eqref{tr-rhoA-T2-n-cft-2int-adj}.

\begin{figure}[t!]
\vspace{-.5cm}
\includegraphics[width=1\textwidth]{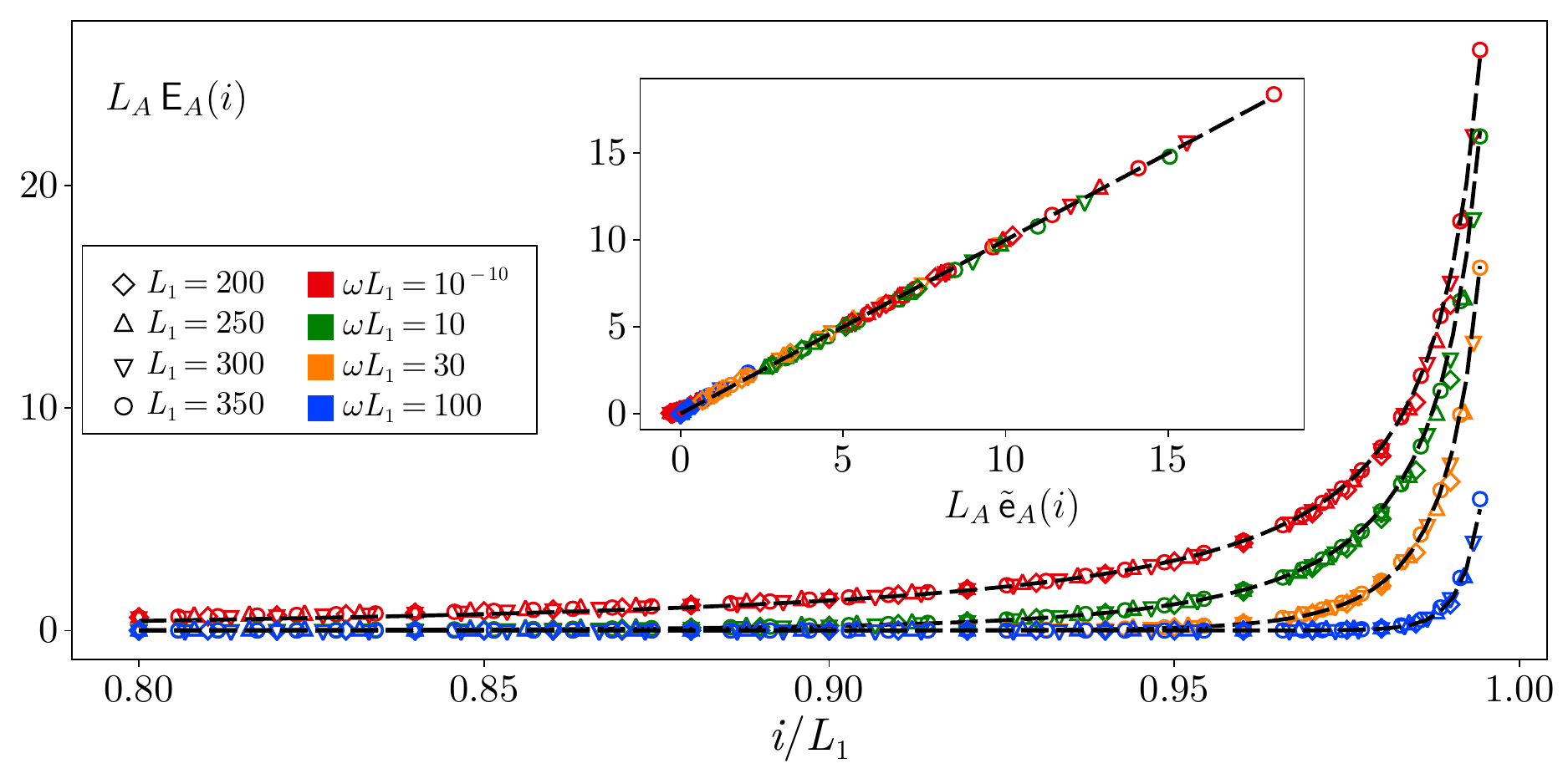}
\caption{\small
Contour function $\mathsf{E}_A(i)$ (see (\ref{contour-function-NEG-projector})) 
for the logarithmic negativity of two adjacent blocks in the ground state of the infinite harmonic chain
with $L_1 = 2L_2$ and 
for various $\omega L_1$. 
The part of $\mathsf{E}_A(i)$ in $A_1$ is reported 
and the one in $A_2$ is analogous. 
The dashed curve correspond to \eqref{neg-cont-massive-analytic}.
}
\label{fig-adj-mass}
\end{figure}

\begin{figure}[t!]
\vspace{-.5cm}
\includegraphics[width=1\textwidth]{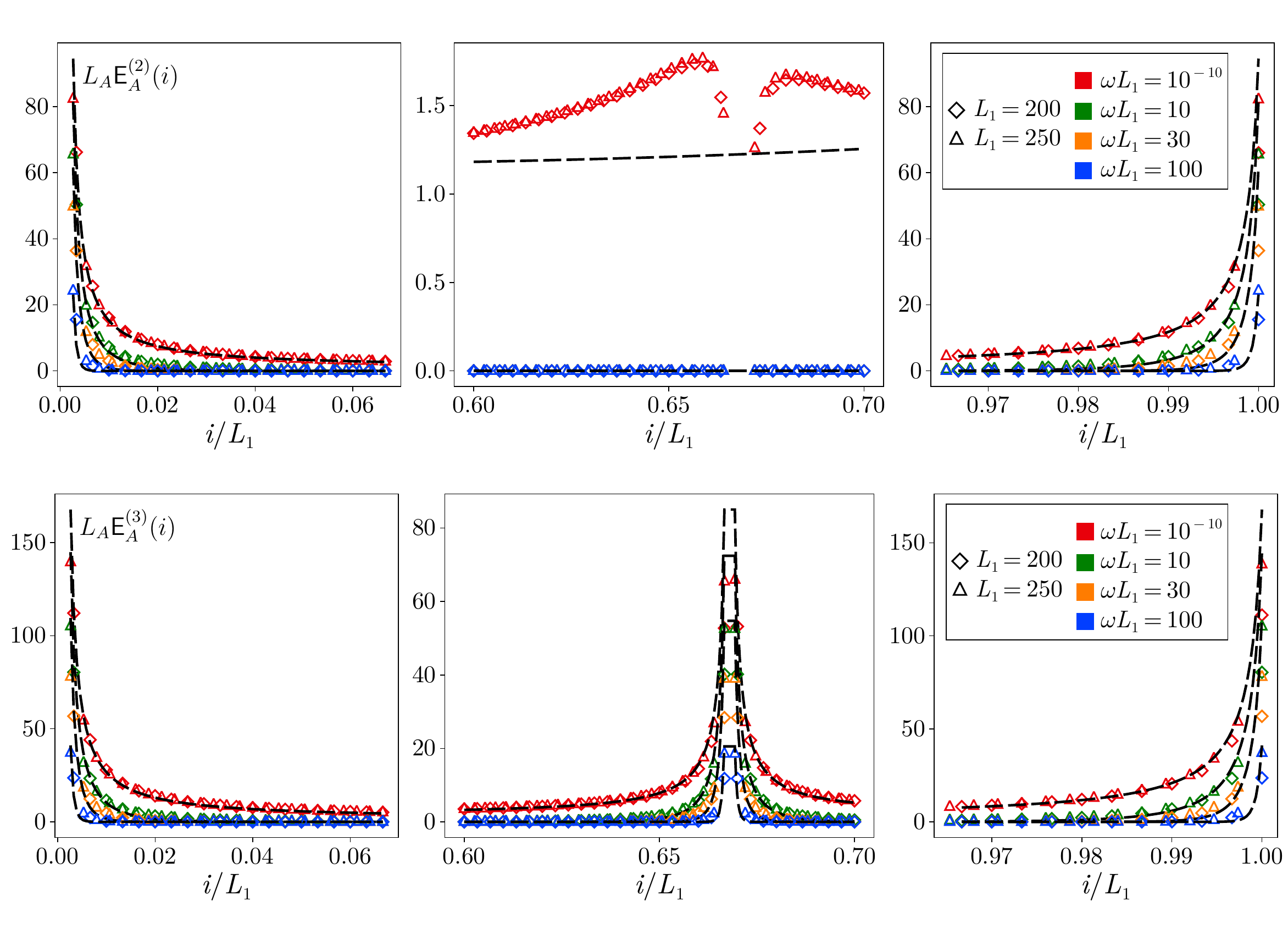}
\caption{\small 
Contour function $\mathsf{E}^{(n)}(i)$ 
(see \ref{contour-function-NEG-projector}))
for the logarithm of the moments of $\gamma_A^{\textrm{\tiny $\Gamma_2$}}$
of two adjacent blocks with $L_1 = 2L_2$
and in the massive regime, 
for $n=2$ (top) and $n=3$ (bottom).
The regions of $A_1 \cup A_2$ close to 
the left endpoint, the entangling point and the right endpoint 
are considered in the left, central and right panel respectively. 
The black dashed lines correspond to 
\eqref{renyi-neg-cont-massive-analytic}.
In the top middle panel, the data points corresponding to the green and orange markers coincide with the ones denoted by the blue markers. 
}
\label{fig-renyi-adj-mass}
\end{figure}

In Fig.\,\ref{fig-adj-mass} and Fig.\,\ref{fig-renyi-adj-mass}
we extend to the massive regime, 
where a finite correlation length $\xi = \omega^{-1}$ occurs \cite{Botero:2004vpl},
the analysis performed in Fig.\,\ref{fig-adj} and Fig.\,\ref{fig-adj-ren} 
respectively.
A configuration with $L_1 = 2L_2$ is considered again,
with increasing values of $\omega L_1$ 
and for different sizes.
The dimensionless quantity $L_A \, \mathsf{E}^{(n)}(i)$
diverges only at the entangling point also in the massive case, as expected.
Moreover, as the distance from the entangling point increases,
the rate of the decay depends on $\omega L_1$.  
In the massive regime, an area law occurs \cite{Calabrese:2012nk} 
and it is governed by the entangling region \cite{Eisler_2016,De_Nobili_2016}, 
that given by the single point at $x=p$ 
in the case of adjacent blocks that we are considering.

In Fig.\,\ref{fig-adj-mass},
the data sets corresponding to the same value of $\omega L_1$ 
nicely collapse on a well defined curve.
In order to explore this curve, let us introduce
\begin{equation}
\label{neg-cont-massive-analytic}
    \tilde{\mathsf{e}}_{_A}(x)
    = 
    \frac{\textrm{e}^{-2m |x-p|}}{8\, \big|x - p \big|}
\end{equation}
which is obtained by 
first specialising \eqref{neg-adj-proposal} to $c=1$ 
and then introducing a proper exponential damping factor, 
whose exponent is proportional to the dimensionless quantity 
$m |x-p|$.
For $ x\to p$, this provides the same kind of divergence 
observed in the massless case, as expected.
The factor $2$ in the exponent of the damping factor 
has been fixed through a numerical fit of the data points. 
The dashed black curves in Fig.\,\ref{fig-adj-mass},
which have been obtained from (\ref{neg-cont-massive-analytic})
by using the explicit relation $\omega = m\,\mathsf{a}$ 
(see Sec.\ref{sec:contour-entropies})
and without introducing any additive constant,
nicely capture the curves provided by the 
corresponding numerical data points.

As for $\mathsf{E}_A^{(n)}(i)$ 
(see (\ref{contour-function-NEG-projector})) in the massive regime, 
our numerical results for $n=2$ and $n=3$
are shown in the top and bottom panels of
Fig.\,\ref{fig-renyi-adj-mass} respectively.
The relevant domains close to $x=a$, $x=p$ and $x=b$
are shown in the left, central and right panels respectively. 
The massless regime (see Fig.\,\ref{fig-adj-ren}) 
corresponds to the red markers in Fig.\,\ref{fig-renyi-adj-mass}.
The data points obtained for the same value of $\omega L_1$ collapse on a well defined curve 
parameterized by $n$.
Notice that $\mathsf{E}_A^{(2)}(i)$
does not diverge at the entangling point,
as shown in the top middle panel of 
Fig.\,\ref{fig-renyi-adj-mass}. 

In order to describe the curves provided by the collapses of the data points, 
we introduce proper exponential damping factors 
in (\ref{neg-contour-adjacent-cft}) as follows
\begin{equation} 
\label{renyi-neg-cont-massive-analytic}
    \tilde{\mathsf{e}}^{(n)}_{_A}(x)  
    \equiv
    - \,\frac{\Delta_n  \,\textrm{e}^{-\alpha_a \,m |x-a|}}{x-a} 
    - \frac{\Delta^{(2)}_n\, \textrm{e}^{-\alpha_p \,m |x-p|}}{2\,\big|x-p\big|} 
    - \frac{\Delta_n\, \textrm{e}^{-\alpha_b \,m |b-x|}}{b-x} 
\end{equation}
where $c=1$ in $\Delta_n$ and $\Delta^{(2)}_n$,
and the parameters $\alpha_j$ with $j \in \{a,p,b\}$ 
are obtained through a fitting procedure 
and might depend on $n$.
The dashed black curves in Fig.\,\ref{fig-renyi-adj-mass}
correspond to (\ref{renyi-neg-cont-massive-analytic}), 
with the simplifying assumption that the three parameters $\alpha_j$ 
are all equal to $\alpha$ and independent of $n$.
Our best fit gives $\alpha =7$.
This numerical result for the parameters $\alpha_j$ seems to be in contrast with 
(\ref{neg-cont-massive-analytic}), which is not recovered 
through the replica limit of (\ref{renyi-neg-cont-massive-analytic}).
However, we remark that this independence of $n$ for $\alpha_j$ is due to the roughness of our fitting procedure; indeed, 
we expect that an improved analysis would lead to 
a non trivial dependence on $n$.

\subsection{Logarithmic negativity in the massive regime}
\label{subsec-GS-line-adjacent-logneg}

In the massive regime, the logarithmic negativity of two adjacent blocks
displays an interesting feature. 
In order to describe this property, 
we find it instructive to briefly recall  
an important result for the entanglement entropy of a single block.

Considering the entanglement entropy $S_A$ 
of an interval in a relativistic quantum field theory 
on the line and in its ground state, 
Casini and Huerta \cite{Casini:2004bw,Casini:2006es}
have introduced the following $C$-function
\begin{equation} 
\label{c-function-CH}
    C_S \,\equiv\, \ell\; \frac{\partial S_A}{\partial \ell}
\end{equation}
where $\ell$ is the length of the interval,
and showed that it decreases along the RG flow \cite{Casini:2004bw}.
The crucial property employed in the proof is the 
the strong subadditivity inequality for the entanglement entropy. 
At the fixed points of the RG flow, 
where the model becomes a CFT$_2$ and 
$S_A = \tfrac{c}{3} \log(\ell/\epsilon)+ \dots$ holds, where $c$ is the central charge, 
the entropic $C$-function (\ref{c-function-CH}) becomes $C_S = c/3$.
Thus, this leads to $c_{\textrm{\tiny UV}} \geqslant c_{\textrm{\tiny IR}}$ along an RG flow, 
as originally found by Zamolodchikov \cite{Zamolodchikov:1986gt}
through a different method. 
In the case of a free massive scalar, 
$C_S$ is a function of the dimensionless 
quantity $m\ell$  that can be written through the solution of a certain Painlev\'e equation \cite{Casini:2005zv}
(see also \cite{Castro-Alvaredo:2009yqb} for an analysis based on the form factors).

\begin{figure}[t!]
\vspace{-.5cm}
\includegraphics[width=1\textwidth]{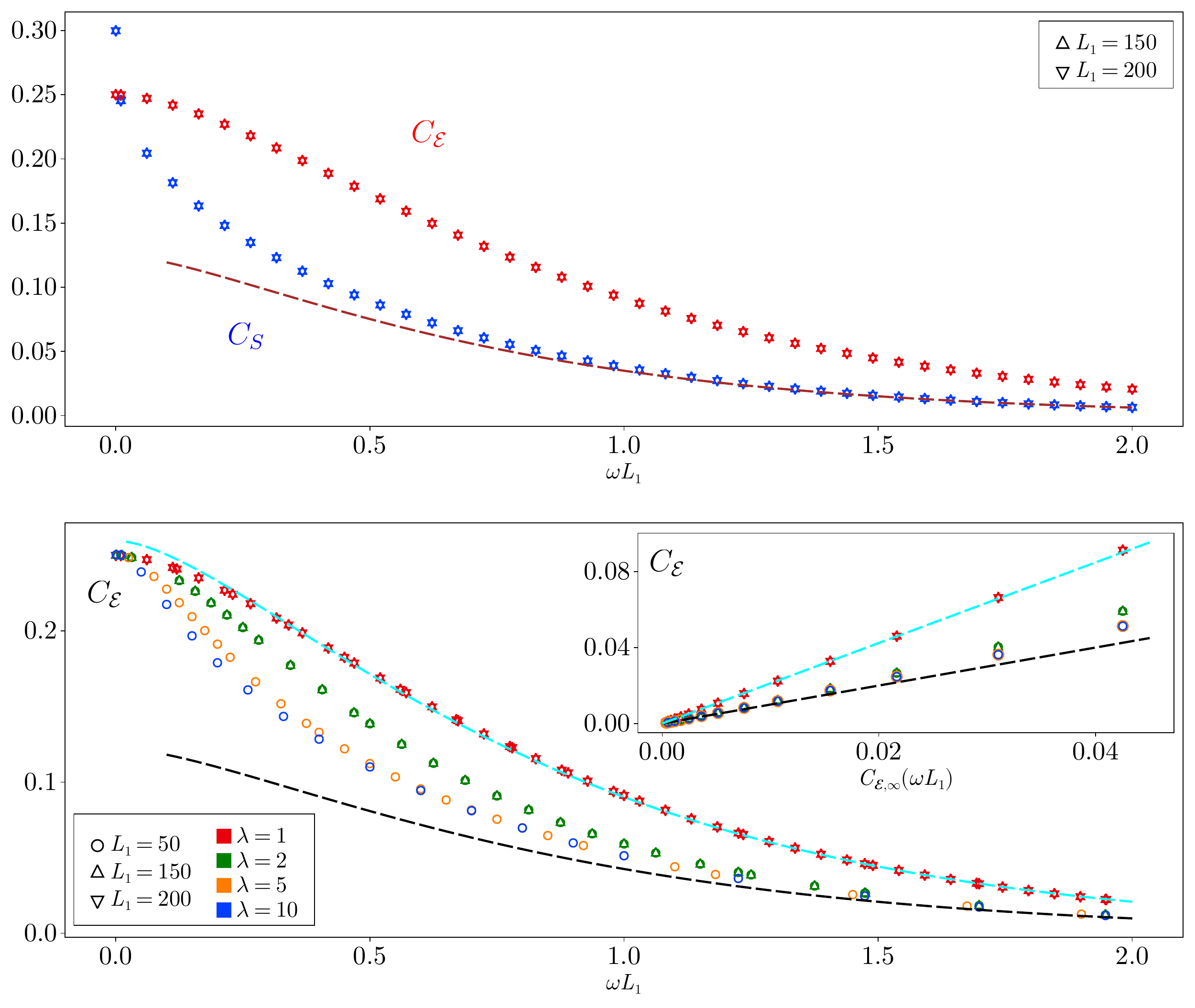}
\caption{\small
Top: The entropic $C$-function $C_S$ in \eqref{c-function-CH} 
(blue markers)
and $C_\mathcal{E}$ in \eqref{c-neg} (red markers) 
for the massive harmonic chain, when $\lambda=1$. 
The brown dashed curve corresponds to \eqref{asymp-C-entropy}.
Bottom: $C_\mathcal{E}$ in \eqref{c-neg} for various values of 
$\lambda \geqslant 1$. The black dashed curve
corresponds to \eqref{asymp-C-neg}, 
while the cyan dashed curve is give by the same formula 
multiplied by the factor $2.12$, obtained through a fitting procedure. 
}
\label{c_theo}
\end{figure}

Our numerical results for (\ref{c-function-CH}), 
computed for a block $A$ made by $L_1$ consecutive sites 
in the harmonic chain in its ground state,
are represented in the top panel of Fig.\,\ref{c_theo}
by the blue markers. 
This numerical analysis has been already carried 
out in \cite{Casini:2005zv} (see also \cite{Arias:2023kni});
hence we tested our algorithm by reproducing 
the dotted curve in Fig.\,2 of \cite{Casini:2005zv},
as shown by the blue marker in the top panel of Fig.\,\ref{c_theo}.

In this procedure, one has to numerically evaluate a derivative 
through the discrete results obtained from the lattice model. 
We perform this analysis by considering the functional defined as 
$\mathsf{D}[f](\mathcal{L},\delta) \equiv \frac{f(\mathcal{L}+\delta) - f(\mathcal{L}-\delta)}{2\delta}$.
%
Since $S_A$ depends on dimensionless arguments, 
for the massive scalar in the continuum
we can consider $S_A$ as a function of 
either $\ell/ \mathsf{a}$ and $m \ell$
or $m \ell$ and $ m \mathsf{a}$, equivalently.
In the latter case, 
the functional introduced above for  $f=S_A(m\ell, m\mathsf{a})$,
with $\mathcal{L}=\ell$ and $\delta = \mathsf{a}$,
gives
$\mathsf{D}[S_A](\ell,\mathsf{a}) = \frac{S_A(m(\ell+\mathsf{a}),m\mathsf{a})-S_A(m(\ell-\mathsf{a}),m\mathsf{a})}{2\mathsf{a}}$.
As $\mathsf{a} \to 0$, we have that 
$\ell\,\mathsf{D}[S_A](\ell,\mathsf{a}) = 
\ell \, \partial_\ell S_A + o(1)$, where we used the fact that 
the leading term is UV finite
and $\big[ \partial_x^p \partial_y^q S_A(x,y) \big]\big|_{x=m\ell\, , \,y = 0}$ 
with  $p> 0$ and $q\geqslant 0$
are UV finite as well \cite{Casini:2005zv}.
Hence, 
by employing the relations between the parameters in the lattice 
and in the continuum (see Sec.\,\ref{sec:intro})
and using that $\omega \ll 1$,
a reliable approximation for the entropic $C$-function 
\eqref{c-function-CH}
is given by 
$L_1 \,\mathsf{D}[S_A](L_1,1)= L_1 (S_A|_{L_1+1} - S_A|_{L_1-1})/2$,
whose evaluation has provided the points 
corresponding to the blue markers 
in the top panel of Fig.\,\ref{c_theo}.

We remark that $C_S \to 1/3$ as $\omega L \to 0$ 
in the top panel of Fig.\,\ref{c_theo}, 
as expected for the fact that the massless scalar is a CFT$_2$ with $c=1$. 
The numerical data points of $C_S$
display a divergence in the slope of the curve as $\omega L \to 0$. 
This is due to the first subleading logarithmic term 
in the expansion of $C_S$ as $m \ell \to 0$,
found in \cite{Casini:2005zv} 
through the solution of a certain Painlev\'e equation, 
which reads 
$C_S =  \tfrac{1}{3} + \tfrac{1}{2\log(m\ell)}
+ O\big(1/[\log(m\ell)]^{2}\big)$.
In the opposite regime, where $\omega L \gg 1$, 
one can specialize to the massive scalar the general expansion given in Eq.\,(1.4) of \cite{Cardy:2007mb} 
for any integrable quantum field theory
and obtained through the form factors of the twist fields.
This leads to introduce 
(see also Eq.\,(88) of \cite{Casini:2005zv})
\begin{equation}
\label{asymp-C-entropy}
C_{S,\infty}(s) 
\equiv 
s \, \partial_s
\left(
- \frac{1}{8} K_0(2s)
\right)
=
    \frac{s}{4} \, K_{1}(2s)
\end{equation}
in terms of the modified Bessel functions of the second kind $K_0(z)$ and $K_1(z)$.
The brown dashed curve in the top panel of Fig.\,\ref{c_theo}
corresponds to $C_{S,\infty}(\omega L_1)$.

Within the setup of the previous discussion, 
given by a $1+1$ dimensional relativistic quantum field theory 
on the line 
and in its ground state, let us consider
$\mathcal{E}$ and $\mathcal{E}^{(n)}$
of two adjacent intervals $A_1$ and $A_2$.

Denoting by $\ell_1$ and $\ell_2$ 
the lengths of the adjacent intervals $A_1$ and $A_2$ respectively,
their harmonic mean $\ell_{1,2}$ and their ratio $\lambda > 0$ 
are defined respectively as 
\begin{equation}
\label{ell12-rho-defs}
        \ell_{1,2} \equiv \frac{2\,\ell_1 \, \ell_2}{\ell_1 + \ell_2}
            \;\;\;\;\;\;\qquad\;\;\;\;\;
            \lambda \equiv \frac{\ell_2}{\ell_1}
\end{equation}
where the harmonic mean 
is invariant under the exchange $\ell_1 \leftrightarrow \ell_2$
and it can be written also as 
$\ell_{1,2} = \tfrac{2\lambda}{1+\lambda}\,\ell_1
= \tfrac{2}{1+\lambda}\,\ell_2$;
hence $\ell_{1,2} = \ell_1 =\ell_2$
in the case of equal intervals.
Thus, we have that
$\ell_1 = \tfrac{1+\lambda}{2\lambda} \, \ell_{1,2}$
and $\ell_2 = \tfrac{1+\lambda}{2} \, \ell_{1,2}$,
which give
$\ell_1+\ell_2 = \tfrac{(1+\lambda)^2}{2\lambda} \, \ell_{1,2}\,$.
Let us also introduce the function of the ratio $\lambda >0 $ given by 
\begin{equation}
\label{rho-tilde-def}
    \tilde{\lambda} \,\equiv \sqrt{\lambda} + \frac{1}{\sqrt{\lambda}}
\end{equation}
which is invariant under the exchange $\ell_1 \leftrightarrow \ell_2$,
has only one minimum at $\lambda=1$ 
and is a bijective function 
of either $\lambda \geqslant 1$ or $\lambda \leqslant 1$.
Hence, $\tilde{\lambda}$ is equivalent to $\lambda$, 
when one of the interval is kept longer than the other one.   
Notice that $\ell_{1,2}$ has the dimension of a length, 
while both $\lambda$ and $\tilde{\lambda}$ are dimensionless.
Thus, the two dimensional domain parameterized by $\ell_1$ and $\ell_2$
where one interval is larger than the other one
is equivalently parameterized by $\ell_{1,2}$ and $\tilde{\lambda}$.

The result (\ref{logneg-CFT-adj}), 
for the logarithmic negativity of two adjacent intervals in the line
when the underlying CFT$_2$ is in its ground state, 
suggests to introduce
\begin{equation} 
\label{c-neg}
    C_{\mathcal{E}} \,\equiv\, 
    \ell_{1,2}\; 
    \frac{\partial \mathcal{E}}{\partial \ell_{1,2}}
\end{equation}
where the partial derivative w.r.t. $\ell_{1,2}$ is performed 
while $\lambda$ is kept fixed. 
At a critical point, 
from (\ref{logneg-CFT-adj}) we find that 
(\ref{c-neg}) becomes $C_{\mathcal{E}} = c/4$,
which is UV finite and proportional to the central charge.

As for the moments of the partial transpose in (\ref{neg-moments-def-intro}), at the critical point we have that (\ref{tr-rhoA-T2-n-cft-2int-adj}) holds.
By employing the following identity
(where $x_{ij}\equiv x_i - x_j$)
\begin{eqnarray}
    \big[\,
x_{12}^{h_1+h_2-h_3} \,
x_{23}^{h_2+h_3-h_1} \,
x_{13}^{h_1+h_3-h_2} \,    
\big] \big|_{h_1 =h_3 \,\equiv\, h}
&=&
\left( \frac{x_{12}\, x_{23}}{x_{13}} \right)^{h_2} 
x_{13}^{2h}
\\
\rule{0pt}{.8cm}
&=&
\left( \frac{x_{12}\, x_{23}}{x_{13}} \right)^{h_2+2h} 
\left( \sqrt{\frac{x_{12}}{x_{23}}} + \sqrt{\frac{x_{23}}{x_{12}}}\;\right)^{4h}
\hspace{1cm}
\end{eqnarray}
that holds for real values of $h$, $h_2$ and $x_1 < x_2 < x_3$,  
one finds that (\ref{tr-rhoA-T2-n-cft-2int-adj}) can be written as 
\begin{equation}
\label{En-CFT-adjacent-rhotilde-form}
    \mathcal{E}^{(n)}
    \,=\,
    -\big(\Delta^{(2)}_n+2\Delta_n \big) \,
    \log \! \left( \frac{\ell_{1,2}}{\epsilon} \right)
    - 4\Delta_n \log \tilde{\lambda}
    + \log \! \big( 2^{\Delta^{(2)}_n+2\Delta_n} \, \mathcal{C}_n\big)
\end{equation}
in terms of (\ref{conf-dimensions-Tn}), (\ref{Delta-Tn-squared-def}) 
and of the geometric variables introduced in 
(\ref{ell12-rho-defs}) and (\ref{rho-tilde-def}).
From (\ref{En-CFT-adjacent-rhotilde-form}),
it is straightforward to realise that 
the dependence on $\lambda$ disappears
in the replica limit $n_{\textrm{\tiny e}} \to 1$
while the dependence on $\ell_{1,2}$ remains non trivial,
providing (\ref{logneg-CFT-adj}) for 
the logarithmic negativity at the critical point.

The expression (\ref{En-CFT-adjacent-rhotilde-form})  
suggests to introduce also the following UV finite quantities
\begin{equation}
\label{c-neg-moments}
    C^{(n)}_{\mathcal{E}} \equiv
    \ell_{1,2}\; 
    \frac{\partial \mathcal{E}^{(n)}}{\partial \ell_{1,2}}
    \;\;\;\;\;\;\qquad\;\;\;\;\;
    \widetilde{C}^{(n)}_{\mathcal{E}}
    \equiv
    -\,\tilde{\lambda}\; 
    \frac{\partial \mathcal{E}^{(n)}}{\partial \tilde{\lambda}}
\end{equation}
where the partial derivative w.r.t. either $\ell_{1,2}$ 
or $\tilde{\lambda}$ is computed 
with either $\tilde{\lambda}$ or $\ell_{1,2}$ kept fixed, 
respectively.
At the critical point, 
by using (\ref{En-CFT-adjacent-rhotilde-form}), 
the quantities in (\ref{c-neg-moments}) become
$C^{(n)}_{\mathcal{E}}= -\big(\Delta^{(2)}_n+2\Delta_n \big) $ 
and $\widetilde{C}^{(n)}_{\mathcal{E}} = 4\Delta_n  $,
in terms of (\ref{Delta-Tn-squared-def}) 
and (\ref{conf-dimensions-Tn}), 
and therefore they are both proportional to the central charge. 
In the replica limit $n_{\textrm{\tiny e}} \to 1$,
these expressions become
$C^{(n_{\textrm{\tiny e}})}_{\mathcal{E}} \to c/4$ 
and $\widetilde{C}^{(n_{\textrm{\tiny e}})}_{\mathcal{E}} \to 0$ 
respectively.
Finally, 
taking the replica limit of (\ref{c-neg-moments}),
we observe that $C^{(n)}_{\mathcal{E}}$ gives (\ref{c-neg}),
while $\widetilde{C}^{(n)}_{\mathcal{E}}$ becomes 
a function that vanishes identically at the critical point.

In Fig.\,\ref{c_theo}, 
we report our numerical results for (\ref{c-neg}) 
in the case of the harmonic chain in its ground state 
and two adjacent blocks $A_1$ and $A_2$,
made by $L_1$ and $L_2$ consecutive sites respectively. 
Setting $L_2 = \lambda L_1$ with fixed $\lambda>0$ 
(see (\ref{ell12-rho-defs})), 
the numerical data have been obtained by 
varying $\omega$, for given values of $L_1$ and $\lambda$; 
hence both the lengths of the adjacent blocks do not change in a set of data points.  
Because of the invariance under the exchange $L_1 \leftrightarrow L_2$, 
we focus on $\lambda \geqslant 1$.
In the top panel of Fig.\,\ref{c_theo}, the data points 
correspond to $\lambda=1$ (see the red markers),
while in the bottom panel 
also other values of $\lambda > 1$ have been considered. 
The derivative occurring in (\ref{c-neg}) 
has been evaluated through its discrete approximation 
and the corresponding lattice data,
following the procedure 
explained above for (\ref{c-function-CH}).
In particular, the points in the bottom panel of Fig.\,\ref{c_theo} 
have been obtained by evaluating 
$L_1 (\mathcal{E}\,|_{L_1+1} - \mathcal{E}\,|_{L_1-1})/2$, 
for assigned values of $\lambda$ and $\omega$.
For  $\lambda =1 $ and $\lambda=2$, the data sets corresponding to different values of $L_1$  collapse in a remarkable way.
Our numerical results confirm that 
the dimensionless quantity defined in (\ref{c-neg}) depends only on 
$\omega \ell_{1,2}$ and $\lambda$
(or $\tilde{\lambda}$ equivalently). 
In the massless limit, 
all the curves for $C_{\mathcal{E}}$ corresponding 
to different values of $\lambda$ converge to $1/4$,
as expected from (\ref{logneg-CFT-adj}) with $c=1$.
Furthermore their approach to the critical point at $\omega L_1 = 0$
occurs through a finite derivative 
whose dependence on $\lambda$ seems very mild, according to the 
numerical data points. 
The latter feature provides an interesting qualitative difference 
between $C_{\mathcal{E}}$ and $C_S$ in their way to approach
the massless regime. 
This might be a general property of these UV finite quantities. 
From the bottom panel of Fig.\,\ref{c_theo}, 
it seems that $C_{\mathcal{E}}$ is a decreasing function of $\lambda$,
at fixed $\omega L_1$.
This feature is not observed anymore 
when the data points are shown in terms of 
$\omega L_{1,2} = \omega L_1 L_2 /(L_1 + L_2) $, while the monotonicity is preserved since
$\frac{\partial C_{\mathcal{E}}}{\partial (\omega L_{1,2})}  = 
\frac{1+\lambda}{2\lambda}\, \frac{\partial C_{\mathcal{E}}}{\partial(\omega L_1)}$
(see the relations between the various lengths in the continuum reported
in text below (\ref{ell12-rho-defs})).

The data points in Fig.\,\ref{c_theo} provide 
some insights about the ground state of the massive scalar 
in the continuum limit.
In particular, they tell us that $C_{\mathcal{E}}$ could be 
a function of the two dimensionless variables 
$\Omega_{1,2} \equiv \omega \ell_{1,2}$ and $\tilde{\lambda}$,
such that 
$\tfrac{\partial C_{\mathcal{E}} }{\partial \Omega_{1,2}}  \leqslant 0$
for a fixed value of $\tilde{\lambda}$.
This specific example suggests us that this interesting behaviour of $C_{\mathcal{E}}$
for the ground state of a relativistic quantum field theory on the line
could be a generic property of any RG flow, 
namely that the following property holds
\begin{equation}
\label{c-neg-decreasing-conjecture}
    \frac{\partial C_{\mathcal{E}} }{\partial \ell_{1,2}}  \leqslant 0
\end{equation}
when $\lambda$ and all the other parameters determining $C_{\mathcal{E}} $ are kept fixed. 
The validity of (\ref{c-neg-decreasing-conjecture}) 
would mean that (\ref{c-neg}) provides a new tool to explore RG flows. 
Thus, 
in order to check this monotonic behaviour along different RG flows.
it would be very interesting to study (\ref{c-neg}) for other models, 
on the lattice or in the continuum.

An analytic expression for the logarithmic negativity of two adjacent intervals 
with finite lengths in an integrable quantum field theory 
is not available in the literature. 
However, in \cite{Blondeau-Fournier:2015yoa} some limiting regimes have been explored, by employing the form factors of the twist fields.
In particular, 
considering the limiting regime where the length of the largest interval diverges, 
the expansion given in Eq.\,(8) of 
\cite{Blondeau-Fournier:2015yoa} specialised to the massive scalar
suggests us to introduce
\begin{equation}
\label{asymp-C-neg}
C_{\mathcal{E},\infty}(s) 
\,\equiv\, 
s \, \partial_s
\left(
-\frac{2}{3\sqrt{3}\, \pi}\, 
K_{0}\big(\sqrt{3}\,s\big)
\right)
=
\frac{2\,s}{3 \pi} \, K_{1}\big(\sqrt{3}\,s\big) \,.
\end{equation}
The dashed black curve in the bottom panel of Fig.\,\ref{c_theo}
corresponds to $C_{\mathcal{E},\infty}(\omega L_1)$.
Instead, the dashed cyan one is $\alpha\,C_{\mathcal{E},\infty}(\omega L_1)$
with $\alpha \simeq 2.12$, which has been found by looking for the best overlap with the lattice data points having $\lambda = 1$.
This is highlighted in the inset of this figure, 
where the black dashed curve (corresponding to \eqref{asymp-C-neg})
is a straight line with slope equal to one. 
For large values of the mass parameter, 
we expect that the entanglement between $A_1$ and $A_2$ is mainly provided by the region around the entangling point. 
Hence, the curves for $C_\mathcal{E}$ associated to different values of $\lambda$
should approach (\ref{asymp-C-neg}) in this limiting regime. 
The data sets displayed in the bottom panel of Fig.\,\ref{c_theo} confirm this intuition for $\lambda \neq 1$. As for the data corresponding to $\lambda = 1$,
larger values of $\omega L_1$ should be considered, 
in order to check this feature.

Given the features of $C_{\mathcal{E}}$ highlighted above, 
it would be interesting to explore also the quantities 
$C^{(n)}_{\mathcal{E}}$ and $\widetilde{C}^{(n)}_{\mathcal{E}}$
introduced in (\ref{c-neg-moments}) for integers $n \geqslant 2$,
both numerically and analytically.
We leave these analyses for future studies.

\section{Ground state: Two disjoint blocks in the line}
\label{sec-GS-line-disjoint}

In this section we explore the contour functions 
for $\mathcal{E}$ and $\mathcal{E}^{(n)}$
when the subsystem $A = A_1 \cup A_2$ is made by  two disjoint blocks 
$A_1 = \{ 1 \leqslant i \leqslant L_1\}$ 
and $A_2 = \{ L_1+d+1 \leqslant i \leqslant L_1+L_2+d\}$
separated by $d$ consecutive sites
in the infinite harmonic chain in its ground state,
for setups analogue to the ones considered in 
Sec.\,\ref{sec-GS-line-adjacent} for adjacent blocks.
Thus, in the harmonic chain the endpoints $a_1$, $b_1$, $a_2$ and $b_2$ 
correspond to $1$, $L_1$, $L_1 + d$ and $L_1 + d +L_2$ respectively.  

\begin{figure}[t!]
\vspace{-.5cm}
\includegraphics[width=1\textwidth]{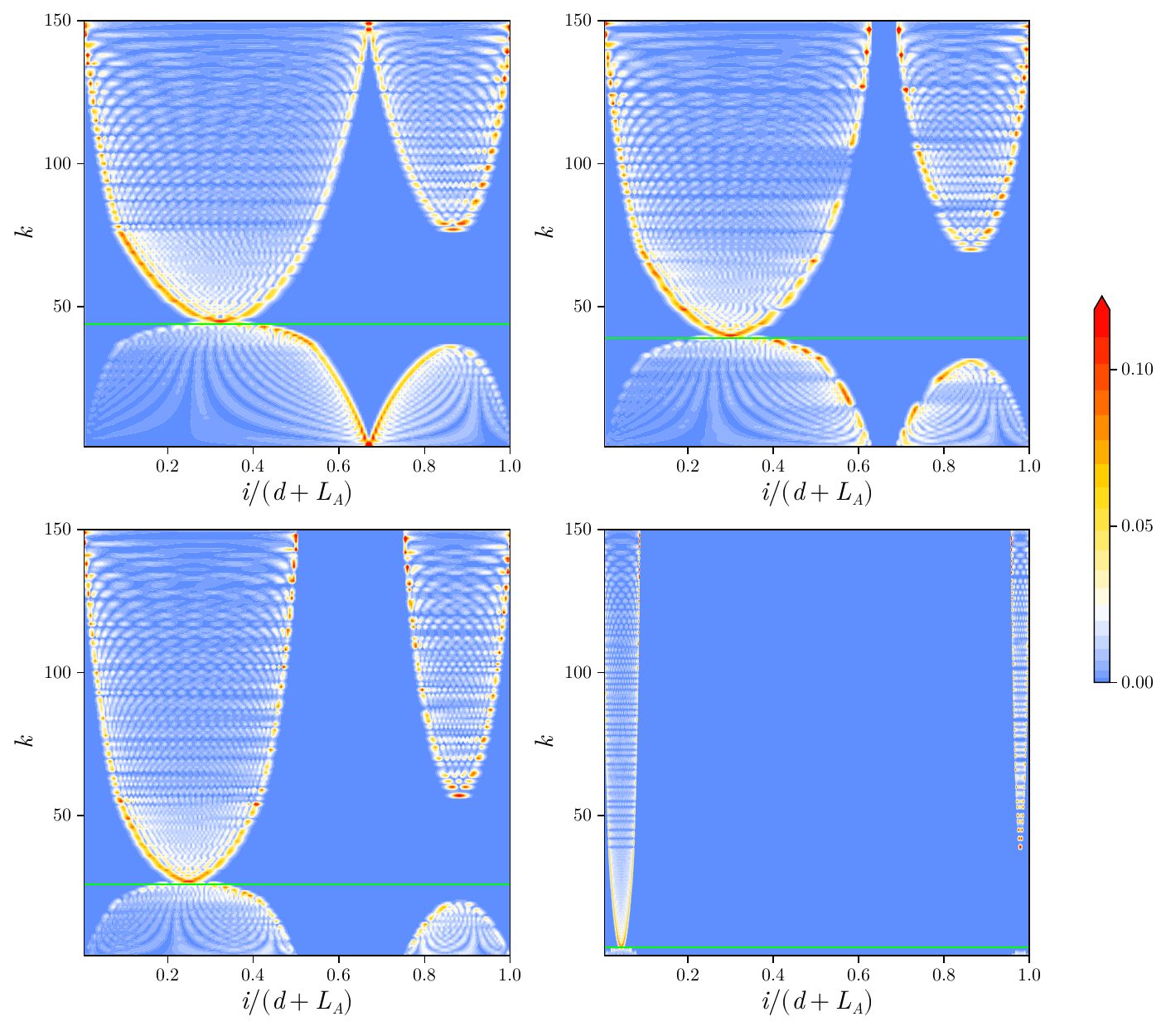}
\caption{\small
Mode participation function \eqref{tilde-pk-i-def} 
for two disjoint blocks with $L_1=2L_2 = 100$
in the infinite harmonic chain in its ground state
and in the massless regime ($\omega L_1 = 10^{-10}$).
The distance between the blocks changes as follows:
$d = 0$ (top left, which coincides with the top left panel of Fig.\,\ref{mpf-adj}), 
$d = L_1/10$ (top right), 
$d = L_1/2$ (bottom left) and $d = 10 \, L_1$ (bottom right). 
The green straight line corresponds to the threshold mode at $k=k_\ast$.
}
\label{fig-disjoint-MPF}
\end{figure}

\begin{figure}[t!]
\vspace{-.5cm}

\includegraphics[width=1\textwidth]{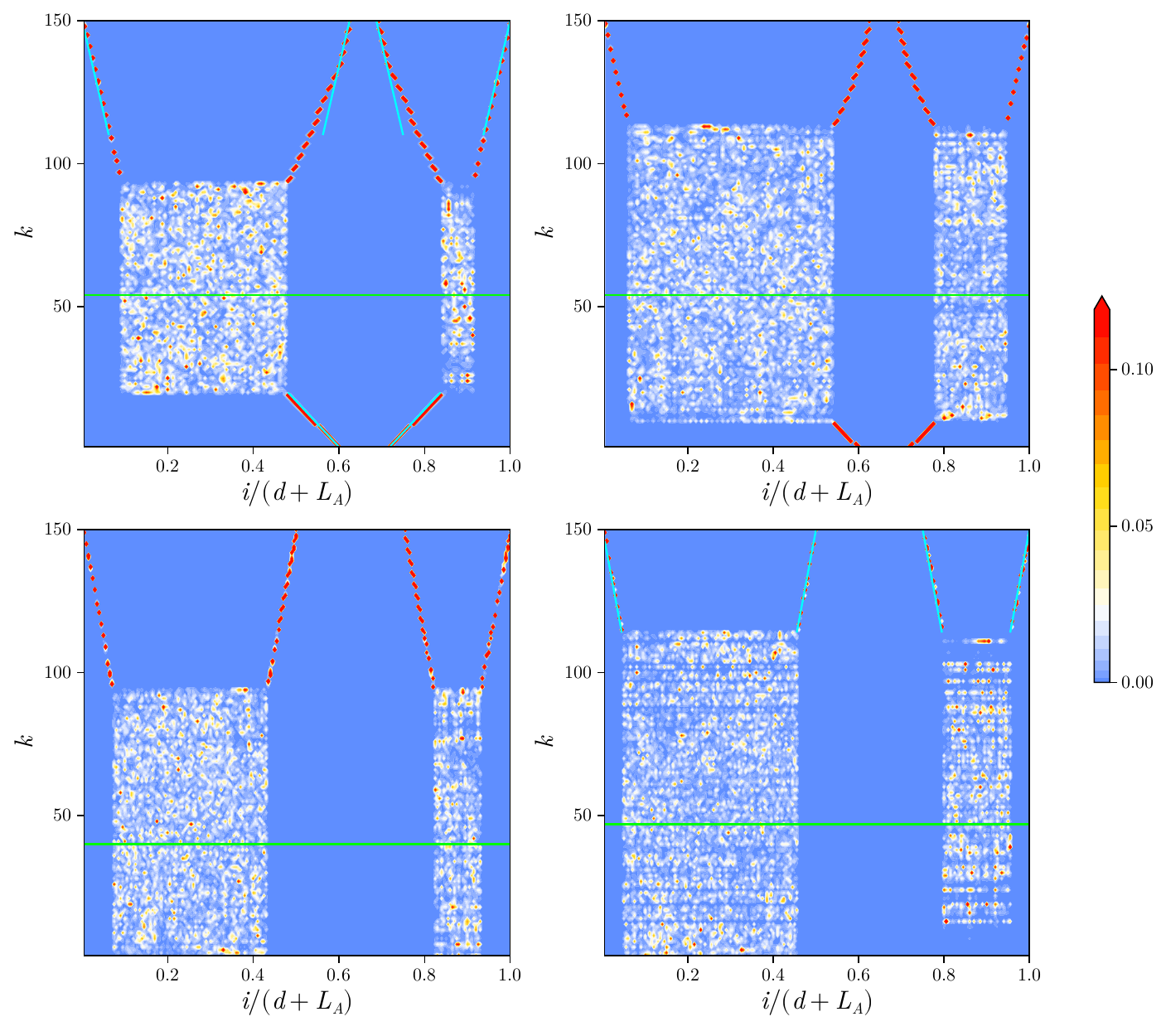}

\caption{\small
Mode participation function \eqref{tilde-pk-i-def} 
for two disjoint blocks with $L_1=2L_2 = 100$
in the infinite harmonic chain in its ground state,
in the regime of large mass. In particular, 
we set $d = L_1/10$ and $d = L_1/2$ 
in the top and bottom panels respectively, 
while $\omega L_1 = 1000$ and $\omega L_1 = 5000$  
in the left and right panels respectively.
}
\label{fig-mpf-disj-mass}
\end{figure}

In Fig.\,\ref{fig-disjoint-MPF} and Fig.\,\ref{fig-mpf-disj-mass} 
we show some numerical results for the mode participation function
$\tilde{p}_k(i)$ in (\ref{tilde-pk-i-def}) 
for  $L_1 = 2L_2=100$, 
in the regimes of small and large mass respectively. 

As for the massless regime, Fig.\,\ref{fig-disjoint-MPF} displays 
the behaviour of $\tilde{p}_k(i)$ 
as the separation distance between the blocks increases. 
The straight green line denotes the threshold index $k=k_*$, 
defined in (\ref{threshold-mode-def}). 
In particular, $d=0$
in  the top left panel of this figure  
and therefore it coincides with 
the top left panel of Fig.\,\ref{mpf-adj}.
As for the case where $d>0$,
it is useful to compare the top right panel in this figure with
Fig.\,\ref{fig-3D-MPF-example}.
Because of the lack of the entangling points when $d>0$,
the peaks of $\tilde{p}_k(i)$ at the endpoints are less prominent.
In the case of adjacent blocks,
the general features highlighted in the discussion of Fig.\,\ref{mpf-adj} 
are confirmed also in this setup. 
It is very instructive also to compare Fig.\,\ref{fig-disjoint-MPF} 
with the numerical data for the mode participation function 
$p_k(i)$ in (\ref{mode-part-W}), shown in Fig.\,10 of \cite{Coser:2017dtb}.
We remark that the value of the threshold mode $k_\ast$ approaches 
its smallest allowed value as $d$ increases
(see the bottom right panel of Fig.\,\ref{fig-disjoint-MPF}).
This is consistent with the fact that the logarithmic negativity 
between the two blocks should vanish as their separation distance diverges
while their lengths are kept fixed. 
However, the number of modes providing a non vanishing contribution 
to the moments of the partial transpose increases with $d$,
although also these quantities vanish for large separation distance. 

The large massive regime is explored in Fig.\,\ref{fig-mpf-disj-mass},
where $\omega L_1 = 1000$ and $\omega L_1 = 5000$
in the left and right panels respectively,
for the same bipartition of Fig.\,\ref{fig-disjoint-MPF}.
Two different separation distances are considered:
$d= L_1 /10 $ and $d= L_1 /2 $ in the top and bottom panels
respectively. 
The qualitative features highlighted for two adjacent blocks
in the discussion of the bottom panels of Fig.\,\ref{mpf-adj} 
are observed also for disjoint blocks.
However, additional observations can be made, due to 
the occurrence of a non vanishing separation distance. 
In particular, considering the rectangular domains 
where $\tilde{p}_k(i)$ seems to take random values, 
its size depends also on the separation distance. 
As for the horizontal green line, the same considerations made in the discussion of Fig.\,\ref{mpf-adj} can be repeated here. 

The remaining values of $(i,k)$ for which $\tilde{p}_k(i)$ is relevant provide well defined finite lines that are approximately straight
and whose slopes  are related to those occurring in Fig.\,\ref{mpf-adj},
once the lattice coordinate $i$ is shown along the horizontal axis.
The cyan segments in the top left and bottom right panels 
of Fig.\,\ref{fig-mpf-disj-mass}
(the same segments could have been shown also in the 
top right and bottom left panels, respectively)
provide a good approximation of these curves.
The two cyan segments corresponding to small values of $k$
in the top right panel have the same slope 
(with respect to a vertical line at a fixed value of $i$)
as their analogue ones occurring for the case of adjacent blocks, 
in the bottom panels of Fig.\,\ref{mpf-adj}.
Similarly, the four cyan segments corresponding to the highest values of $k$ 
in the top left panel of  Fig.\,\ref{fig-mpf-disj-mass}
have the same slopes (in absolute value) 
of the two straight lines corresponding to the highest values of $k$
and departing from endpoints at $x=a_1$ and $x=b_2$,
in the bottom panels of Fig.\,\ref{mpf-adj}.
This numerical observation 
could be interpreted as another consequence of the fact that, 
in the case of two disjoint blocks, 
all the four endpoints play the same role
(in contrast with the case of two adjacent blocks, considered in Fig.\,\ref{mpf-adj},
where the entangling point displays various characteristic features).
Thus, in the top left panel of Fig.\,\ref{fig-mpf-disj-mass},
the absolute value of the slopes of the two segments departing at $k=1$
is four times 
the absolute value of the slopes of the four segments departing at $k=L_A$,
consistently with the corresponding observation made for 
the bottom panels of Fig.\,\ref{mpf-adj}.
As $\omega$ increases, the curves associated with the highest values of $k$ 
tend to straight lines, as shown in the bottom right panel of 
Fig.\,\ref{fig-mpf-disj-mass}.

\begin{figure}[t!]
\vspace{-.5cm}
\includegraphics[width=1\textwidth]{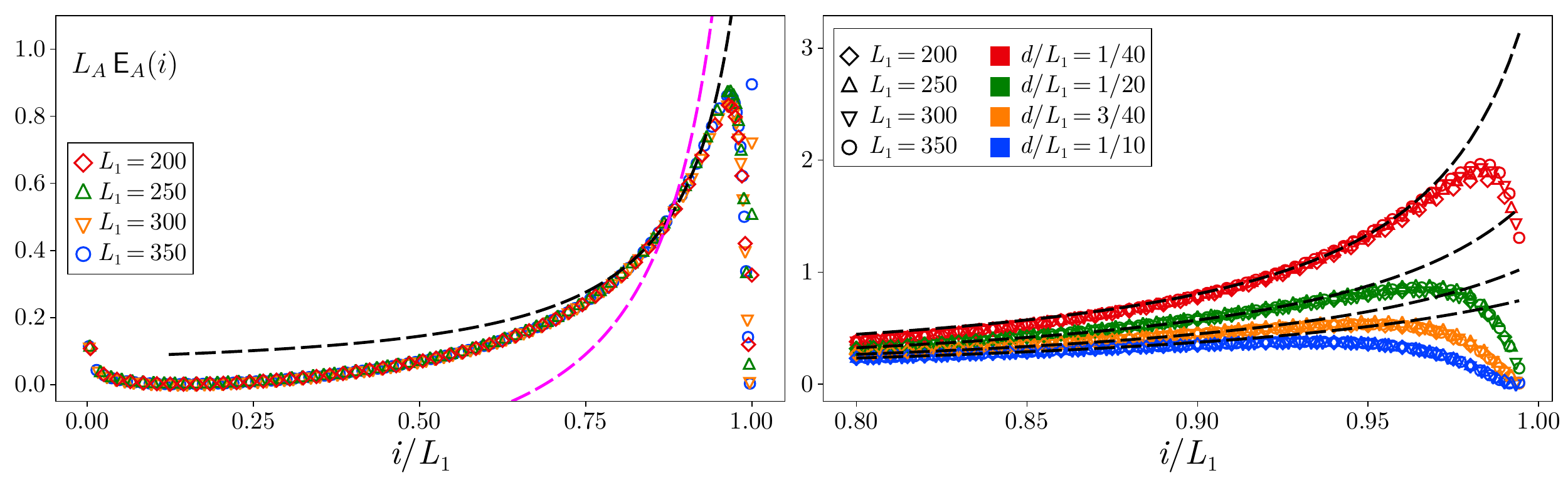}
\caption{\small 
Contour function $\mathsf{E}_A(i)$ for the logarithmic negativity 
of two disjoint and close blocks in the harmonic chain, 
from \eqref{contour-function-NEG-projector}, 
with $L_1=2 L_2$ and $\omega L_1 = 10^{-10}$, for various values of $L_1$ and $d$. 
In the left panel, 
the data points correspond to $d/L_1 = 1/20$, for various values of $L_1$,
and only the part corresponding to $i \in A_1$ is shown. 
In the right panel, 
where only a subregion of the block $A_1$ containing its right endpoint is considered,
the data obtained for various values of $d /L_1$ are reported. 
The dashed magenta and black lines correspond 
to \eqref{disjoint-fermions} and \eqref{disjoint-bosons} respectively. 
}
\label{fig-disjoint}
\end{figure}

In the left panel of Fig.\,\ref{fig-disjoint}, 
we show the contour function $\mathsf{E}_A(i)$ 
for the logarithmic negativity 
of two disjoint blocks in the harmonic chain with $L_1 = 2L_2$, 
in the regime of small separation distance 
(i.e. with $d/L_1 \ll 1$ for fixed $L_1$),
obtained from \eqref{contour-function-NEG-projector}.
In the figure we have reported only the domain $i \in A_1$ 
because $\mathsf{E}_A(i)$ for $i \in A_2$ is qualitatively analogous.  
The data sets corresponding to different values of $L_1$ 
and a fixed ratio $d/L_1$ nicely collapse and  
it would be very instructive to find an analytic expression for this curve. 
The most important feature to highlight in Fig.\,\ref{fig-disjoint}
is the fact that $\mathsf{E}_A(i)$ does not diverge as $i \in A_1 \cup A_2$.
This is related to the fact 
the logarithmic negativity of two disjoint blocks is UV finite in the continuum limit. 
This finiteness of $\mathsf{E}_A(i)$
is in contrast with $\mathsf{S}_A(i)$ of two disjoint blocks 
considered e.g. in \cite{Coser:2017dtb}, 
that diverges close to the four endpoints of the intervals
(which are the entangling points associated to the bipartition of the pure state
from the perspective of the entanglement entropy).

We remark that the replica limit $n_{\textrm{\tiny e}} \to 1$ 
of the analytic proposal (\ref{contour-neg-CFT-2int-disjoint})
gives a constant, and therefore gives a UV finite logarithmic negativity, 
as expected. 
However, (\ref{contour-neg-CFT-2int-disjoint}) 
is not a useful analytic prediction 
for the contour function of the logarithmic negativity
because it is a bounded function having a non trivial dependence on the position. 
This is due to the fact that (\ref{contour-neg-CFT-2int-disjoint}) 
cannot capture the contribution 
associated to the function $\mathcal{G}_n(\eta)$ 
in (\ref{tr-rhoA-T2-n-cft-2int}),
which is the only one surviving in the replica limit 
$n_{\textrm{\tiny e}} \to 1$ and therefore provides the logarithmic negativity,
as discussed in the text below (\ref{tr-rhoA-T2-n-cft-2int}).

In order to construct analytic expressions
to compare with the data points 
for the contour function of the logarithmic negativity
obtained from the lattice, 
in the following we employ the construction proposed in 
\cite{Kudler-Flam:2019nhr}.
This is based on the linear functional defined as 
\begin{equation} 
\label{Ryu-deriv-neg-E-generic}
    \tilde{\mathsf{r}}_{A}\big[\mathcal{E}\big](x) 
    \equiv
    \frac{1}{2} \sum_{i=1,2}\frac{\Theta_i(x)}{2} 
    \Big( 
    \partial_{b_i}\mathcal{E}\big|_{b_i=x}
    -
    \partial_{a_i}\mathcal{E}\big|_{a_i=x} 
    \Big) 
\end{equation}
which has been obtained by considering the expression provided in 
\cite{Kudler-Flam:2019nhr} for each interval 
and introducing a global factor $1/2$.
Notice the similarity between 
(\ref{Ryu-deriv-neg-E-generic}) and the expression 
given by (\ref{Ryu-deriv-ren-neg-E})
where $\mathcal{E}^{(n)}$ is replaced by $\mathcal{E}$.

An analytic expression involves 
the logarithmic negativity $\mathcal{E}_{\textrm{\tiny F}}$  in the ground state 
associated to the partial time-reversal of the reduced density matrix for free fermionic systems \cite{Shapourian:2016cqu}.
It is obtained from the functional (\ref{Ryu-deriv-neg-E-generic}) 
as follows
\begin{equation} 
\label{Ryu-deriv-neg}
    \mathsf{r}_{A}(x) 
    \equiv
    \tilde{\mathsf{r}}_{A}
    \big[\mathcal{E}_{\textrm{\tiny F}}\big](x)
    + 
    \textrm{const}
\end{equation}
where, for two disjoint intervals, we have that
\cite{Herzog:2016ohd,Shapourian:2019xfi,Arias:2026bqh}
\begin{equation} 
\label{disj-fermions-neg}
    \mathcal{E}_{\textrm{\tiny F}} 
    = -\frac{1}{4}\log (1-\eta)
\end{equation}
in terms of the cross ratio $\eta$ 
introduced in \eqref{eta}.
More explicitly, (\ref{Ryu-deriv-neg}) reads
\begin{equation} 
\label{disjoint-fermions}
    \mathsf{r}_{A}(x) = 
    \frac{1}{2} \left[ \,
    \Theta_{1}(x) \, \frac{(b_2-a_2)}{4(a_2-x)(b_2-x)}
    + 
    \Theta_{2}(x) \, \frac{(b_1-a_1)}{4(x-a_1)(x-b_1)}
    \,\right] + \text{const} 
\end{equation}
which should be compared with the analytic expression reported in 
Eq.\,(A64) of \cite{Kudler-Flam:2019nhr}.
The integral of 
$\tilde{\mathsf{r}}_{A} \big[\mathcal{E}_{\textrm{\tiny F}}\big](x)$ on $A$ gives \eqref{disj-fermions-neg}.

\begin{figure}[t!]
\vspace{-.5cm}

\includegraphics[width=1\textwidth]{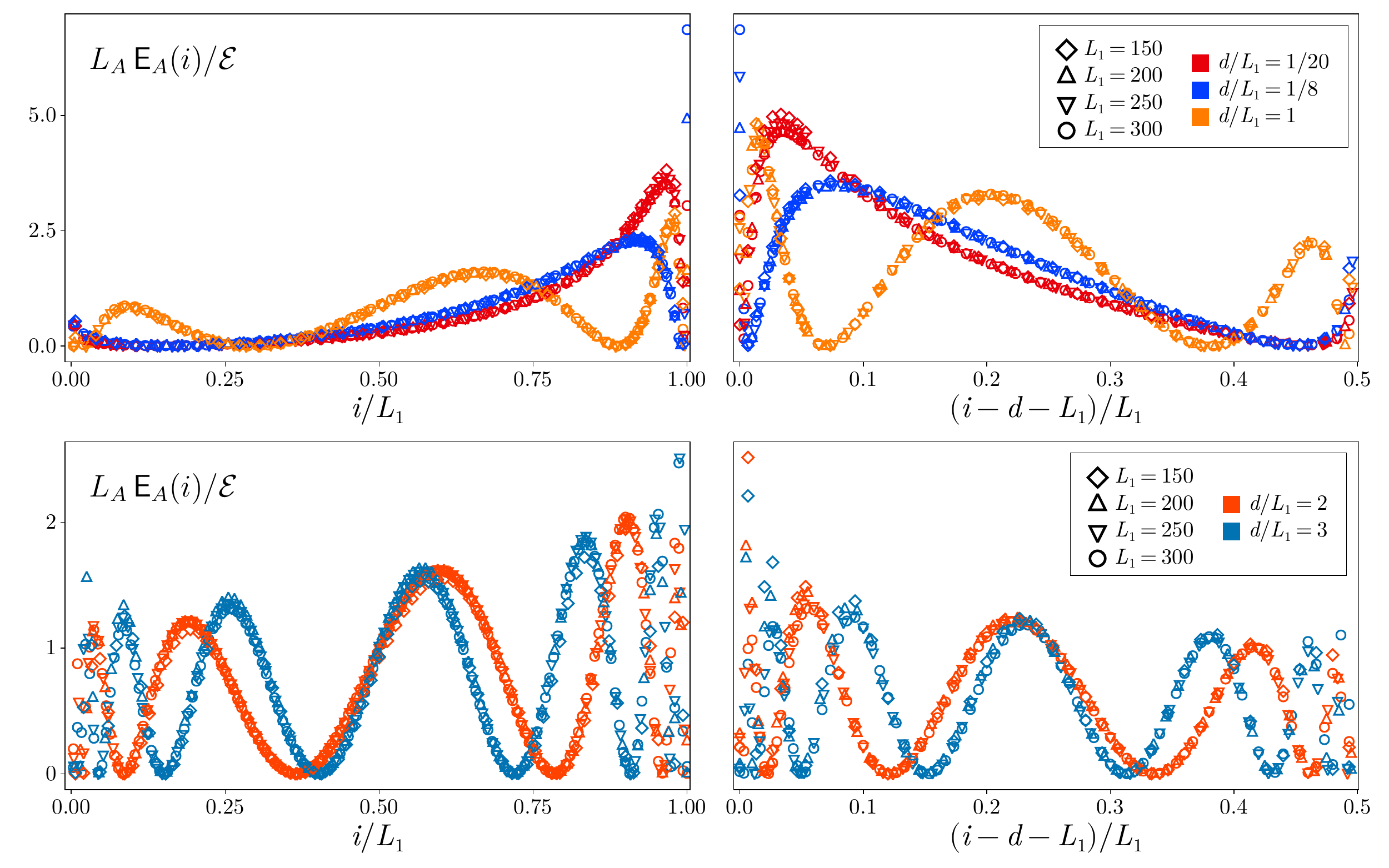}

\caption{\small
Contour function $\mathsf{E}_A(i)$ for the logarithmic negativity 
of two disjoint blocks (from \eqref{contour-function-NEG-projector})
for increasing separation distance, 
in the setup described in the caption of Fig.\,\ref{fig-disjoint}. 
The blocks $A_1$ and $A_2$ correspond to the 
left and right panels respectively.
}
\label{fig-disjoint-far}
\end{figure}

A different analytic expression
that we find it worth comparing with the numerical results from the lattice
is obtained by employing (\ref{Ryu-deriv-neg-E-generic})
and the analytic expression (found in \cite{Calabrese:2012nk}) 
of the logarithmic negativity of two disjoint intervals 
for the massless scalar in the asymptotic regime 
where the separation distance vanishes 
while the lengths of the intervals are kept fixed, 
i.e. as $\eta  \to 1^-$.
This expression reads
\begin{equation}
\label{neg-near}
    \mathcal{E}_{\textrm{\tiny B}}
    \,\equiv\,
    \mathcal{E}_{\textrm{\tiny F}} - \frac{1}{2} 
    \log\!\big(K(\eta)\big) - \log\mathcal{P}_1  
\end{equation}
where $\mathcal{E}_{\textrm{\tiny F}}$ is (\ref{disj-fermions-neg}), 
the constant $\mathcal{P}_1$ is written through the Glaishers constant $\mathscr{A}$ 
as follows
\begin{equation}
    \mathcal{P}_1 
    \equiv 
    \frac{2^{7/6} \,\textrm{e}^{1/2}}{\mathscr{A}^6}
    \simeq  0.832056  
    \;\;\; \qquad \;\;\;
    \mathscr{A} \equiv 
    \exp \! \big(1/12-\zeta'(-1) \big)
    \simeq 1.282427
\end{equation}
and all the $o(1)$ terms as $\eta  \to 1^-$ have been discarded.
This gives
\begin{equation} 
\label{disjoint-bosons}
        \mathsf{b}_{A}(x) 
        \,\equiv\, 
\tilde{\mathsf{r}}_{A}
    \big[\mathcal{E}_{\textrm{\tiny B}}\big](x)
        + 
        \text{const}
        \,=\, 
        \tilde{\mathsf{r}}_{A}
    \big[\mathcal{E}_{\textrm{\tiny F}}\big](x) 
    - \frac{1}{2}\;
    \tilde{\mathsf{r}}_{A}
    \big[\log\!\big(K(\eta)\big)\big](x) 
        + \text{const} \,.
\end{equation}
The derivation of this expression tells us that 
it can at most describe the data points for $\mathsf{E}_A(i)$ close to the endpoints at $b_1$ or $a_2$.

In Fig.\,\ref{fig-disjoint}, 
the analytic expressions 
(\ref{disjoint-fermions}) and (\ref{disjoint-bosons})
have been employed to obtain 
the magenta dashed curve and the black dashed curves, respectively. 
The additive constants have been fitted trying to maximize
the domain where the analytic curves  
overlap with the lattice data points.
We remark that, for two disjoint intervals the analytic expressions coming from the fermionic and bosonic models 
(namely (\ref{disjoint-fermions}) and (\ref{disjoint-bosons}) respectively)
are significantly different,
in contrast with the case of the contour functions 
associated to two adjacent intervals (see Sec.\,\ref{sec:cft}).

From the left panel of Fig.\,\ref{fig-disjoint}, 
it is evident that we cannot find a domain where 
\eqref{disjoint-fermions} captures the behavior 
of the curves provided by the lattice data points
independently of the choice of the additive constant.
Instead, the proposal \eqref{disjoint-bosons},
that provides the dashed black curves, 
reasonably describe the numerical data points 
in the region near to the endpoints at $b_1$ and $a_2$, 
which corresponds to its domain of applicability. 
From the right panel of Fig.\,\ref{fig-disjoint}
and in Fig.\,\ref{fig-disjoint-far}, we observe that, 
when the distance between the two intervals increases, 
its agreement with the lattice data points gets worse,
as expected from the fact that 
\eqref{disjoint-bosons} is based on (\ref{neg-near}), 
which holds only when $\eta \simeq 1^-$.

\begin{figure}[t!]
\vspace{-.5cm}

\includegraphics[width=1\textwidth]{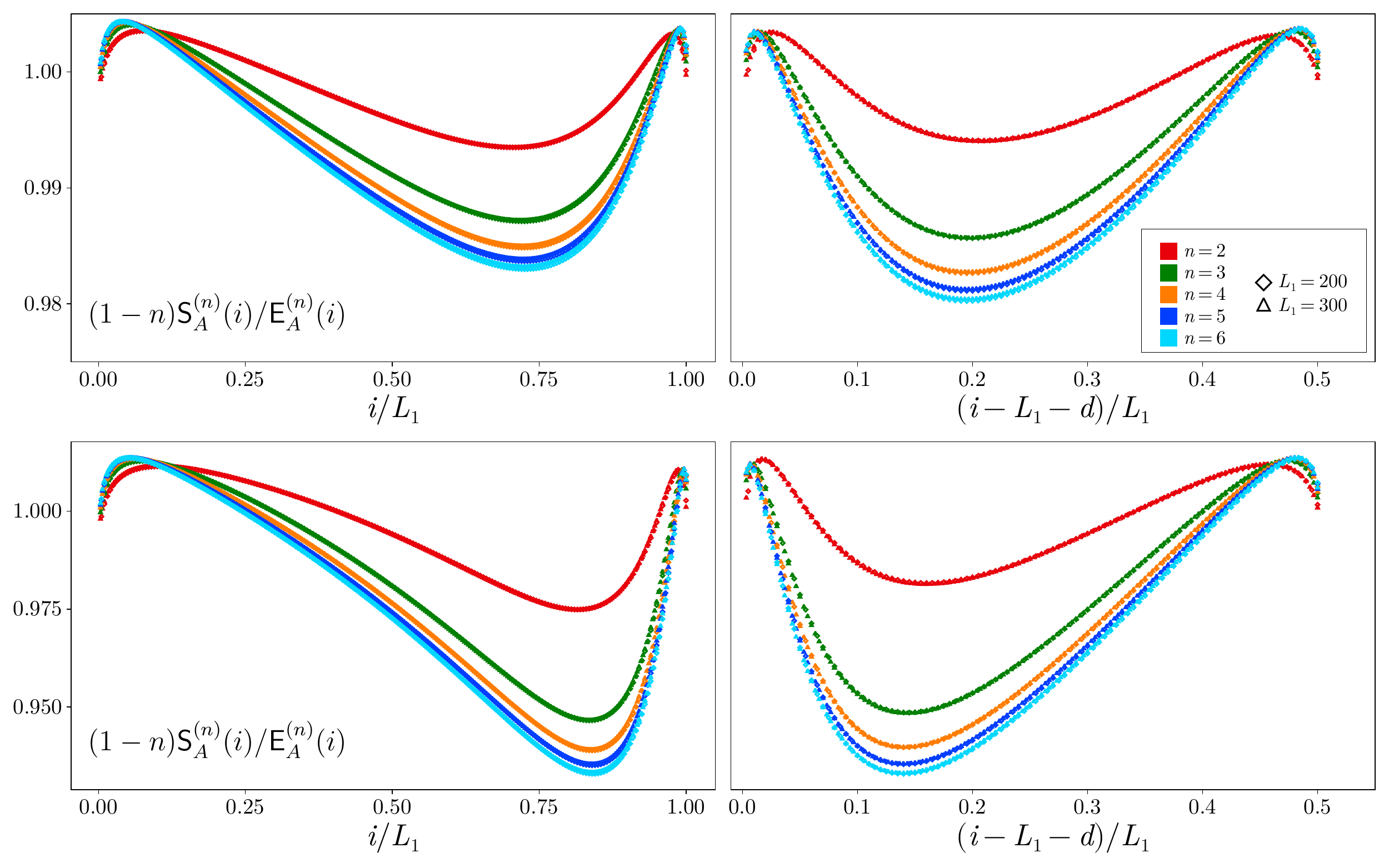}

\caption{\small
Ratio between the contour functions 
$\mathsf{S}_A^{(n)}(i)$ and  $\mathsf{E}_A^{(n)}(i)$,
from \eqref{contour-function-EE-projector} and 
\eqref{contour-function-NEG-projector} respectively, 
of two disjoint blocks with $L_1 = 2 L_2$
and in the massless regime (here $\omega L_1 = 10^{-10}$),
for different values of $n$.
The blocks $A_1$ and $A_2$ correspond to the 
left and right panels respectively, 
which share the same vertical axis, 
while $d/L_1 = 1/2$  and $d/L_1 = 1/8$ 
to the top and bottom panels respectively.
}
\label{fig-renyi-disjoint}
\end{figure}

In Fig.\,\ref{fig-disjoint-far} we report some
numerical results for the contour function of the logarithmic negativity
(obtained from (\ref{contour-function-NEG-projector})) 
for increasing values of the separation distance $d$, while the lengths of the two blocks are kept fixed, for the same setup considered for Fig.\,\ref{fig-disjoint}.
We find it convenient to normalize this contour function by using the logarithmic negativity $\mathcal{E}$ in order to 
display the data points corresponding to very different values 
of $d$.
Indeed, $\mathcal{E}$ decays exponentially as $\eta \to 0^+$
\cite{Marcovitch:2008sxc, Calabrese:2012ew, Calabrese:2012nk,Arias:2026bqh,Wichterich:2008vfx,Calabrese:2013mi,Klco:2020rga,Klco:2021cxq,Parez:2022ind,Parez:2023xpj}.
Remarkably, an oscillatory behaviour is observed 
when $d$ increases while $L_1/L_2$ is kept fixed, 
and the frequency of the oscillations increases 
with $d$.
It would be very interesting to explain these oscillations 
through an analytic approach.

In Fig.\,\ref{fig-renyi-disjoint}, we compare 
the contour functions $\mathsf{E}_A^{(n)}(i)$ 
with the contour functions $\mathsf{S}_A^{(n)}(i)$, 
from \eqref{contour-function-EE-projector} and 
\eqref{contour-function-NEG-projector} respectively, 
by considering their ratio, 
for various values of in the index $n$ and 
the massless regime of the harmonic chain. 
Also in this case $L_1 = 2L_2$ and the results 
for two different values of $L_1$ are reported. 
By considering this ratio for the 
corresponding analytic expressions discussed in Sec.\,\ref{sec:cft}
for these contour functions in the continuum 
(see (\ref{contour-entropies-cft-N-int}) 
and (\ref{contour-neg-CFT-2int-disjoint-tilde})), 
we find the constant function given by $1$ identically;
hence, in particular, the divergencies occurring in the numerator 
simplify with the ones in the denominator. 
In Fig.\,\ref{fig-renyi-disjoint},
the data corresponding to different sizes of $A$ nicely overlap,
indicating that the resulting curves could be obtained 
from a suitable analysis in the continuum, within the 
CFT$_2$ model of the massless scalar field on the line 
and in its ground state. 
Our numerical results confirm that this ratio does not display divergencies. 
Since the range of the numerical values on the vertical axes 
is very small and close to $1$, one concludes that the two curves in the continuum are very close to each other. 
Moreover, the slight discrepancy with respect to the constant function equal to $1$ obtained from the analytic expression of Sec.\,\ref{sec:cft}
is due to the model dependent part that cannot be found 
through the analysis performed in Sec.\,\ref{sec:cft}.
By comparing the top and the bottom panels 
of Fig.\,\ref{fig-renyi-disjoint}, 
where $d/L_1 = 1/2$  and $d/L_1 = 1/8$ respectively, 
one observes that 
$\mathsf{S}_A^{(n)}$ and $\mathsf{E}_A^{(n)}$ 
in the continuum 
tend to collapse as the separation distance between the two intervals becomes very large. 
The fact that the discrepancy between 
$\mathsf{S}_A^{(n)}$ and $\mathsf{E}_A^{(n)}$ increases 
as the two intervals get close to each other is expected;
indeed, the limiting regime of adjacent intervals 
emphasizes the qualitative different nature 
of the reduced density matrix  with respect to 
the operator obtained through its partial transposed.

\section{Thermal state}
\label{sec-thermal-state}

In this section we briefly investigate the contour function 
of the logarithmic negativity $\mathsf{E}(i)$ 
in \eqref{contour-function-NEG-projector}
for the harmonic chain  on the circle made by $N$ sites
and at finite temperature $T$,
when the circle is bipartite into two regions. 

While in Sec.\,\ref{sec-GS-line-adjacent} 
and Sec.\,\ref{sec-GS-line-disjoint}
the mixed state is given by the reduced density matrix 
$\rho_A$ of the union of two blocks, 
here the mixed state is the thermal state of the whole system, 
which is bipartite by two adjacent blocks $A_1$ and $A_2$. 
Notice that two entangling points occur in this case. 
Since $A = A_1 \cup A_2$ is the entire circle
and the two blocks $A_1$ and $A_2$ 
(containing $L_1$ and $L_2$ consecutive sites respectively) 
are adjacent, 
in order to uniformize the notation with the case of 
two adjacent blocks considered 
in Sec.\,\ref{sec:cft} and Sec.\,\ref{sec-GS-line-adjacent},
we set $N=L_A$ and denote by $p_1$ and $p_2$ 
the positions of the two entangling points, 
with $p_1<p_2$.

\begin{figure}[t!]
\vspace{-.5cm}
\includegraphics[width=1\textwidth]{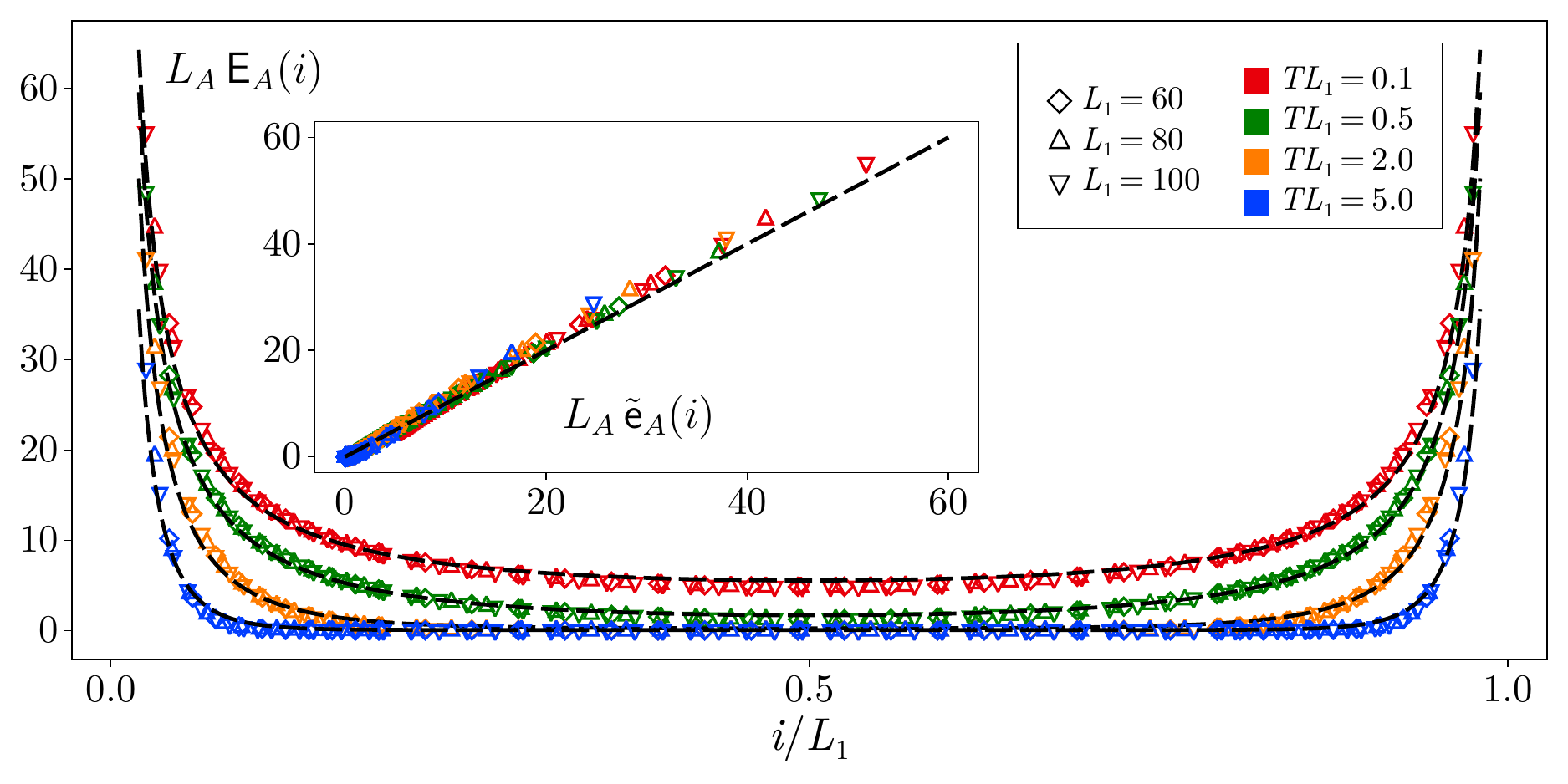}
\caption{\small
Contour function $\mathsf{E}_A(i)$ 
for the logarithmic negativity of the harmonic chain 
at finite temperature $T$ and on a circle 
made by $L_A$ sites bipartite in two blocks $A_1$ and $A_2$,
of lengths $L_1$ and $L_2$ respectively,
from \eqref{contour-function-NEG-projector} and 
(\ref{qq-corr-temp})-(\ref{pp-corr-temp}).
Here $L_A = 10\,L_1$, $\omega L_A = 10^{-10}$ and $i \in A_1$.
The dashed curve corresponds to the analytic expression  
\eqref{CFT-thermal}.
}
\label{fig-single-thermal}
\end{figure}

In Fig.\,\ref{fig-single-thermal}, 
we show our numerical results for $L_A/L_1 = 10$,
when the harmonic chain on the circle is in the massless regime. 
These results are obtained from 
\eqref{contour-function-NEG-projector}, 
when the correlators characterizing the Gaussian state are given by (\ref{qq-corr-temp}) and (\ref{pp-corr-temp}).
Considering different values of $T$ and $L_1$, 
we find that the data sets corresponding to the same value of the dimensionless parameter $T L_1$ nicely collapse. 
For a given bipartition of an harmonic chain of a certain length,
i.e. for fixed $L_1$ and $L_A$, the logarithmic negativity decreases as $T$ increases, as shown by the fact that, 
in Fig.\,\ref{fig-single-thermal}, 
the curve at $T_1 L_1$ is entirely above the one at $T_2 L_1$ when $T_1 < T_2$. 
This is expected from the fact that the classical correlations associated with the temperature destroy the entanglement provided by the quantum correlations \cite{Anders:2008lus,Yu_2009}.
From Fig.\,\ref{fig-single-thermal}, we observe that
$\mathsf{E}_A(i)$ diverges at the entangling points, 
as already observed for the case of two adjacent blocks on the line when the harmonic chain is in the ground state (see Sec.\,\ref{sec-GS-line-adjacent}).
%

In order to provide an ansatz for the analytic curve describing the data sets shown in Fig.\,\ref{fig-single-thermal}, 
we observe that a CFT$_2$ analysis for $\mathcal{E}$ in this setup has been reported in \cite{Calabrese:2014yza}. 
Combining the CFT$_2$ results of \cite{Calabrese:2014yza} 
with the observations made in 
Sec.\,\ref{sec:cft} and Sec.\,\ref{sec-GS-line-adjacent}
about the case of adjacent intervals 
and the behaviour of the contour function for the logarithmic negativity at the entangling points, 
a reasonable ansatz reads
\begin{equation}
\label{CFT-thermal}
    \mathsf{e}_A(x) = \tilde{\mathsf{e}}_A(x) + \text{const} \;\;\;\qquad\;\;\;
    \tilde{\mathsf{e}}_A(x) = 
    \frac{\textrm{e}^{-\alpha \,T|x-p_1|}}{2|x-p_1|}+\frac{\textrm{e}^{-\alpha \,T|p_2-x|}}{2|p_2-x|} \;.
\end{equation}
This expression exhibits the expected divergent behaviour at the entangling points discussed in Sec.\,\ref{sec:cft}, 
while the exponential damping factors are suggested by the CFT$_2$ expression for the logarithmic negativity found in \cite{Calabrese:2014yza}.
Notice that (\ref{CFT-thermal}) does not depend on the finite size of the system;
hence it should be compared with numerical data points corresponding to $L_1 \ll L_A$.
The parameter $\alpha$ is fitted to obtain the best match with the numerical data points
from the lattice and, in particular, 
 $\alpha = 1.8 \pi$ for the black dashed curves
in Fig.\,\ref{fig-single-thermal}.
The inset of Fig.\,\ref{fig-single-thermal}
highlights that the ansatz (\ref{CFT-thermal}) 
describes the numerical data
(see also the inset of Fig.\,\ref{fig-adj-mass} for a similar comparison),
although the still unknown analytic expression 
in the continuum is expected to be more complicated 
(e.g. it must depend also on the finite size of the system).

\section{Conclusions}
\label{sec:conclusions}

In this paper we have studied specific 
contour functions  $\mathsf{E}_A(i)$ and $\mathsf{E}_A^{(n)}(i)$,
for the logarithmic negativity  and 
the logarithm of the moments of the partial transpose respectively,
associated to the multimode bosonic Gaussian states 
of continuous-variable systems.
These contour functions satisfy the  conditions 
(\ref{neg-density-property-intro}) 
and (\ref{neg-density-positivity-property-intro}).
They are constructed through the expressions 
given between (\ref{NEG-prob-distribution-def}) 
and (\ref{contour-function-NEG-projector}) included,
which have been obtained by adapting 
to the partial transpose of the reduced density matrix 
the procedure employed in \cite{Coser:2017dtb} 
to obtain some contour functions for the entanglement entropies 
in the same kind of quantum systems and states. 
By adapting the properties imposed to 
the contour functions for the entanglement entropies
\cite{ChenVidal2014,Coser:2017dtb},
some reasonable properties for the contour functions
for the logarithmic negativity  and 
for the logarithm of the moments of the partial transpose
have been proposed (see the appendix\;\ref{app-CV-neg-contour}). 
The contour functions introduced in Sec.\,\ref{sec:contour-neg} 
satisfy a subset of these properties, 
but this list does not allow to find 
the contour function of the logarithmic negativity in a unique way. 
It is important to explore further the set of properties 
accompanying the contour functions, 
both for the entanglement entropies and for the entanglement negativity, 
in order to constraint further their construction. 

As the contour functions for $\mathcal{E}$ and $\mathcal{E}^{(n)}$
in the continuum, 
some insights about these functions in CFT$_2$ 
are discussed in  Sec.\,\ref{sec:cft}, 
finding expressions 
that describe their divergencies,
both in the ground state
(see (\ref{neg-contour-adjacent-cft}), (\ref{neg-adj-proposal}) and
(\ref{contour-neg-CFT-2int-disjoint}))
and in the thermal state (see (\ref{CFT-thermal})).

A numerical analysis of the contour functions 
$\mathsf{E}_A(i)$ and $\mathsf{E}_A^{(n)}(i)$ constructed 
in Sec.\,\ref{sec:contour-neg} has been performed 
by specifying them to the harmonic chains 
either in their ground state or in the thermal state.
In particular, for the harmonic chain on the line and in its ground state, 
either adjacent blocks or disjoint blocks have been considered 
(see Sec.\,\ref{sec-GS-line-adjacent} 
and Sec.\,\ref{sec-GS-line-disjoint} respectively), 
while for the harmonic chain on the circle and at finite temperature 
the bipartition of the whole chain into two blocks has been studied
(see Sec.\,\ref{sec-thermal-state}).
We have observed that, 
while the contour functions for $\mathcal{E}^{(n)}$ 
diverge at all the boundary points of the two subsystems of the bipartition 
(see Fig.\,\ref{fig-adj-ren} 
and Fig.\,\ref{fig-renyi-adj-mass}), 
the contour function for $\mathcal{E}$ 
is divergent only at the entangling point 
(see Fig.\,\ref{fig-adj}, Fig.\,\ref{fig-adj-mass} 
and Fig.\,\ref{fig-single-thermal}).
In particular, for two disjoint blocks the contour function for the logarithmic negativity does not display any divergence, 
as shown in Fig.\,\ref{fig-disjoint}, 
Fig.\,\ref{fig-disjoint-far}
and Fig.\,\ref{fig-renyi-disjoint}.
The role of the endpoints of the subsystems 
of the bipartition has been explored also by investigating 
the mode participation function (\ref{tilde-pk-i-def}) 
in the regime of large mass
(see Fig.\,\ref{mpf-adj} and Fig.\,\ref{fig-mpf-disj-mass}).

Another result is the analysis of the quantity (\ref{c-neg}),
discussed in Sec.\,\ref{subsec-GS-line-adjacent-logneg},
which is UV finite in a relativistic quantum field theory
and proportional to the central charge in a CFT$_2$.
Our numerical results in the massive regime of the harmonic chain on the line and in its ground state show (see Fig.\,\ref{c_theo})
that this quantity monotonically decreases 
as the mass increases while the lengths of the adjacent blocks are kept fixed.
This suggests that the quantity (\ref{c-neg}) might be useful 
to explore RG flows; hence it is worth exploring it further 
by considering other models, 
in order to check whether its monotonic behaviour is a general property.

The results discussed in this manuscript can be useful for future studies in various ways, which involve lattice or quantum field theory computations. 
For instance, 
by employing the construction proposed in this manuscript or also other proposals
(see e.g. \cite{Nozaki:2013wia, Kudler-Flam:2019nhr, SinghaRoy:2019urc,Roy:2021qxf}),
it would be interesting to explore 
the contour functions of $\mathcal{E}$ 
and $\mathcal{E}^{(n)}$ in higher dimensions 
\cite{Eisler_2016, De_Nobili_2016},
in quantum systems with boundaries or defects
\cite{Calabrese:2004eu, Sorensen:2009zqb,Mintchev:2020uom,Mintchev:2020jhc,Rogerson:2022yim, Estienne:2023ekf, Rottoli:2022plr},
in inhomogeneous systems 
\cite{Vitagliano:2010db, Ramirez:2015yfa, Dubail:2016tsc,Tonni:2017jom,Tonni:2025xvj},
in non-relativistic quantum field theories 
\cite{MohammadiMozaffar:2017nri,Angel-Ramelli:2020wfo,Mintchev:2022xqh, Mintchev:2022yuo}
and in time-dependent scenarios 
\cite{Calabrese:2005in,Coser:2014gsa,DiGiulio:2019lpb, Eisler:2022hej}.
%
A more refined construction of the contour function for the logarithmic negativity could be found by considering specific models, where the moments of the partial transpose are known analytically in the continuum limit 
\cite{Calabrese:2013mi,Alba:2013mg,Coser:2015dvp,Coser:2015eba,Coser:2015mta, DeNobili:2015dla,Rockwood:2022hey}.
This includes massive free fields, 
where it is worth looking for analytic results 
regarding the logarithmic negativity and other quantities 
obtained from the partial transpose of the reduced density matrix 
\cite{Bianchini:2016mra,Castro-Alvaredo:2019irt}.
Finally, we find it worth studying these quantities 
also in the context of the AdS/CFT correspondence
\cite{Ryu:2006bv,Ryu:2006ef,Freedman:2016zud,Agon:2018lwq},
as done e.g. in \cite{Nozaki:2013wia,Nozaki:2013vta,Bhattacharya:2014vja,Tonni18:bariloche-talk, Wen:2018whg,Kudler-Flam:2019oru,Mintchev:2022fcp, Caggioli:2024uza}
for the contour function for the holographic entanglement entropy.

\vskip 32pt 
\centerline{\bf Acknowledgments } 
\vskip 8pt

We are deeply grateful to
Francesco Gentile and Bego$\tilde{\textrm{n}}$a Mula
for collaboration at the initial stage of this project
and for insightful discussions. 
E.T. would like to thank the Isaac Newton Institute for Mathematical Sciences, Cambridge, for support and hospitality during the programme {\it Quantum Field Theory with Boundaries, Impurities, and Defects} where work on this paper was undertaken (this work was supported by EPSRC grant no EP/R014604/1).
E.T. acknowledges also the 
Yukawa Institute for Theoretical Physics, Kyoto,
within the workshop {\it Extreme Universe 2025} (YITP-T-25-01)
and the long-term workshop {\it Progress of Theoretical Bootstrap}, for hospitality and financial support 
during part of this work.
This work was funded by the European Union -- NextGenerationEU, Mission 4, Component 2, Inv.1.3, in the framework of the PNRR Project National Quantum Science and Technology Institute (NQSTI) PE00023; CUP: G93C22001090006.

\appendix

\section{Properties of the contour functions}
\label{app-contour-properties}

In this appendix, 
we further discuss the properties of the contour functions 
\eqref{contour-function-EE-projector} and \eqref{contour-function-NEG-projector}, 
besides the normalization conditions 
(see (\ref{entropy-density-property}) and (\ref{neg-density-property-intro})) 
and the positivity conditions
(see (\ref{entropy-contour-positivity})
and (\ref{neg-density-positivity-property-intro})),
following the analyses reported in \cite{ChenVidal2014,Coser:2017dtb} 
for the contour functions of the entanglement entropies. 
In particular, 
in Sec.\,\ref{C-V-Gaussian} we specify the properties introduced in \cite{ChenVidal2014} to the case of 
the contour functions for the multimode bosonic Gaussian states and show that 
the contour function \eqref{contour-function-EE-projector} 
breaks them in general.
Indeed, this contour function satisfies the properties  
considered in \cite{Coser:2017dtb}, which require further constraints 
with respect to the general ones introduced in \cite{ChenVidal2014}.
In Sec.\,\ref{app-CV-neg-contour} 
we propose some properties for 
the contour functions for the entanglement negativity,
by adapting the ones considered in \cite{ChenVidal2014,Coser:2017dtb}
and show that the contour function \eqref{contour-function-NEG-projector}
satisfies a subset of these constraining conditions.

\subsection{Contour function for the entanglement entropies}
\label{C-V-Gaussian}

The unitary transformations, 
considered in \cite{ChenVidal2014} to discuss 
the general requirements for contour functions of the entanglement entropy,
do not preserve the Gaussian nature of a state in general;
hence we must restrict to Gaussian unitary transformations $U_{\textrm{\tiny G}}$,
namely to  the unitary transformations 
that act within the space of Gaussian states. 
The action of a Gaussian unitary transformation on the density matrix 
is implemented at the level of the covariance matrix by 
a symplectic matrix $M$.
This leads us to specify the properties of \cite{ChenVidal2014}
to the contour function $\mathsf{S}_{A}(i)$ 
of the multimode bosonic Gaussian states 
as follows:

(a) {\it Symmetries.}
The set of spatial symmetries that exchange the sites $i,j \in A$ is associated with symplectic transformations of the type 
\begin{equation}
\label{symplectic-symmetry}
    M = \mathcal{P}
    \Bigg( \,
    \begin{array}{cc}
        \boldsymbol{0} & \mathsf{M}_{i \mapsto j}
        \\
        \rule{0pt}{.45cm}
        \mathsf{M}_{j \mapsto i} & \boldsymbol{0}
    \end{array} 
    \,\Bigg) \mathcal{P}^{\textrm{t}}
    \oplus \mathcal{M}
\end{equation}
where $\mathsf{M}_{i \mapsto j}$ and $\mathsf{M}_{j \mapsto i}$ are the $2 \times 2$ matrices implementing the exchange between the two sites,  $\mathcal{M}$ acts on $A\,\setminus \{i,j\}$ 
and $\mathcal{P}$ is a $4 \times 4$ permutation matrix such that $\mathcal{P}\,(q_i,p_i,q_j,p_j)^{\textrm{t}} = (q_i,q_j,p_i,p_j)^{\textrm{t}}$.
This transformation is a symmetry when the matrix $M$ satisfies the relation 
$\gamma_A = M\,\gamma_A \,M^{\textrm{t}}$. 
If such a symmetry exists, then the contour function must satisfy the 
condition $\mathsf{S}_{A}\left(i\right) = \mathsf{S}_{A}(j)$. 
This property extends the corresponding one considered in \cite{Coser:2017dtb}, where the symplectic matrix $M$ was taken to be also orthogonal.

(b) {\it Local unitary transformations.} 
Consider symplectic transformations that act in a non trivial way 
only on $G \subseteq A$, i.e transformations that take the form $M = M_G \,\oplus\,\boldsymbol{1}_{A\setminus G}$. 
Denoting by $\mathsf{S}^{'}_{A}(i)$ the contour function 
obtained after this transformation, we require that 
$ \mathsf{S}_{A}'(G) = \mathsf{S}_{A}(G)$. 
Again, this condition is stronger than the corresponding one considered in \cite{Coser:2017dtb} because $M$ is not orthogonal in general.

(c) {\it Upper bound}.
Consider an ordering of the degrees of freedom in $A$ such that the first $|\tilde{A}|$ ones are associated with the subregion $\tilde{A}\subseteq A$. 
Let us also assume that there exists a symplectic transformation 
$M$ such that $M\,\gamma_{A}\,M^{\textrm{t}} = \gamma_{\Omega_A}\oplus\gamma_{\widetilde{\Omega}_A}$ 
that  exhibits the following block structure 
        \begin{equation}
        \label{third-blocks}
        V = P\,
        \Bigg( \,
        \begin{array}{cc}
            M_{\Omega_A ;\tilde{A}} \; & M_{\Omega_A ;A \setminus \tilde{A}} 
            \\
            \rule{0pt}{.45cm}
            \boldsymbol{0} & M_{\widetilde{\Omega}_A; A \setminus \tilde{A}}
        \end{array}
        \, \Bigg)\,P^{\textrm{t}}
    \end{equation}
    where the blocks in the r.h.s. are rectangular matrices in general 
    and the permutation $P$ reorders the variables such that each block acts 
    on the spatial subset specified in the second part of its subscript. 
    Under these conditions, $\mathsf{S}_A(i)$ must satisfy the bound (\ref{bound-CV-A}). 
    This extends the bound considered in \cite{Coser:2017dtb} because 
    $V$ is not an orthogonal matrix in general.

The properties (a), (b) and (c) above are more general than the corresponding one considered in \cite{Coser:2017dtb} because the orthogonality condition for the symplectic matrices implementing the various transformations is not required. 
As remarked in \cite{Coser:2017dtb}, 
this orthogonality condition is a necessary condition in order to have that 
a projector $X^{i}$ is transformed into another projector.

    In the subsequent analysis, we employ the transformation 
    of $W$ and $K$,
    occurring in \eqref{williamson-dec-gamma-A} 
    and \eqref{contour-function-EE-projector} respectively, 
     when $\gamma_A$ changes as $\gamma_A \mapsto \gamma_A' = M\,\gamma_A\,M^{\textrm{t}}$. These transformations read 
\begin{equation}
\label{non-local-K}
    W' = WM^{\textrm{t}} 
    \;\;\;\qquad\;\;\; 
    K' = W\big(M^{\textrm{t}}M\big)^{-\frac{1}{2}}
    W^{-1}  KM^{\textrm{t}}  \,.
\end{equation}
From the expression of $K'$, we observe that the mode participation function $p_k(i)$ in \eqref{mpf-proj-ent} transforms locally 
only when $M$ is also orthogonal. 
Indeed, by considering a matrix $M$ that acts in a non trivial way 
only on a spatial subset $G \subseteq A$ (see the property (b)),
we have that \eqref{mpf-proj-ent} is modified everywhere
and not only for $i\in G$.
In the following we show that 
the contour function \eqref{contour-function-EE-projector} 
does not satisfy (a), (b) and (c).

(a) A counterexample is obtained by considering 
the covariance matrix for a subsystem $A$ made by two sites given by 
\begin{equation}
    \gamma_A = 
            \bigg( 
        \begin{array}{cc}
             2 \;& 1 \\
             1 \;& 7
    \end{array} 
    \bigg)
    \oplus 
    \bigg(
        \begin{array}{cc}
             3 \;& 1 \\
             1 \;& 6/7
    \end{array} 
    \bigg)
\end{equation}
that admits a symmetry in the form of \eqref{symplectic-symmetry} with 
\begin{equation}
    \mathsf{M}_{1 \to 2} = \mathsf{M}_{2 \to 1} 
    =
        \bigg(
        \begin{array}{cc}
             0 \;& 1/\sqrt{3.5} \\
             \sqrt{3.5} \;& 0
    \end{array} 
    \,\bigg) \,.
\end{equation}
Constructing the contour function $\mathsf{S}_A(i)$ in this setting, 
we find that $\mathsf{S}_A(1) \neq \mathsf{S}_A(2)$.

(b) Various counterexamples that break this property can be constructed. 
Indeed, a covariance matrix $\gamma_A$ can be obtained 
from its Williamson decomposition \eqref{williamson-dec-gamma-A} 
by choosing a symplectic matrix $W \in \text{Sp}(|A|)$ 
and symplectic eigenvalues $\sigma_\mu \geqslant 1/2$ for 
$1\leqslant\mu\leqslant|A|$.
The freedom in the choice of the symplectic spectrum 
and the fact that it must be invariant under a symplectic transformation 
can be exploited to find that  
a necessary and sufficient condition for the property (b) is given by 
\begin{equation}
\label{app-b-condition}
    \sum_{i \in G} \,p'_k(i) = \sum_{i \in G} \, p_k(i)
\end{equation}
where $p_k(i)$ and $p_k'(i)$ are the mode participation functions 
\eqref{mpf-proj-ent} associated 
with $\gamma_A$ and $\gamma'_A$ respectively.
From the non local nature of the transformation of $p_k(i)$ when $M$ is not orthogonal (see \eqref{non-local-K}), it is reasonable to assume that
the relation (\ref{app-b-condition}) is not satisfied in general. 
This can easily checked numerically through various counterexamples.
Indeed, by constructing 
the matrices $W$ and $M_G$ as random symplectic matrices with integer coefficients,
a violation of (\ref{app-b-condition}) is typically found.

(c) 
A counterexample can be obtained by considering 
a matrix $V$ in \eqref{third-blocks} that is symplectic but not orthogonal. 
Indeed, in this case, $\mathsf{S}_A(\tilde{A})$ depends also on the symplectic eigenvalues associated to $\widetilde{\Omega}_A$. 
By increasing the amplitude of these eigenvalues, it is straightforward to obtain an example where the upper bound is violated.

We remark that these analyses for $\mathsf{S}_A(i)$ 
can be easily adapted to $\mathsf{S}_A^{(n)}(i)$ with $n\geqslant 2$.

\subsection{Contour function for the logarithmic negativity}
\label{app-CV-neg-contour}

As for the contour function for $\mathcal{E}$ given by 
\eqref{contour-function-NEG-projector} and defined on $A = A_1 \cup A_2$,
a formal set of properties similar to the ones proposed in \cite{ChenVidal2014} for the contour functions of the entanglement entropies is not available in the literature.
We remark that, 
while a contour function for the entanglement entropies 
originates from the reduced density matrix of 
only one of the two regions identified by the spatial bipartition, a contour function for $\mathcal{E}$ is constructed from the partial transpose 
$\rho_A^{\textrm{\tiny $\Gamma_2$}}$
that involves both of them. 
This implies that the constraints for contour functions of the logarithmic negativity should be related to unitary transformations acting on all the degrees of freedom of $A$. 
The transformations that leaves the entanglement between $A_1$ and $A_2$ invariant correspond to unitary matrices that can be written as $U = U_1 \oplus U_2$, where $U_i$ is a unitary transformation acting only on $A_i$, for $j \in \{1,2\}$. 
By employing this observation, in the following 
we discuss a set of properties to restrict the possible constructions of the 
contour functions for $\mathcal{E}$ 
within the setting provided by the multimode bosonic Gaussian states.  
Since these properties are inspired from the ones discussed in 
Sec.\,\ref{C-V-Gaussian}
for the contour functions of the entanglement entropies
\cite{ChenVidal2014, Coser:2017dtb}, we denote them in a similar way.
We introduce the following three requirements:

($\mathsf{a}$) \textit{Symmetries.} 
A spatial symmetry that exchanges the sites $i,j \in A_1$  without 
modifying the structure of the bipartite entanglement 
corresponds to a symplectic matrix of the form
\begin{equation}
\label{symplectic-symmetry-neg}
    M = M_1 \oplus M_2 
    \;\;\;\;\qquad\;\;\;\;
    M_1 = \mathcal{P}
    \Bigg( \,
    \begin{array}{cc}
        \boldsymbol{0} & \mathsf{M}_{i \mapsto j}
        \\
        \rule{0pt}{.45cm}
        \mathsf{M}_{j \mapsto i} & \boldsymbol{0}
    \end{array} 
    \,\Bigg) \mathcal{P}^{\textrm{t}}
    \oplus \mathcal{M}_1
\end{equation}
where $M_i$ acts on $A_i$, with $i \in \{1,2\}$,
the matrix $\mathcal{M}_1$ acts on the degrees of freedom of 
$A_1 \, \setminus\{i,j\}$,
and $\mathcal{P}$ implements the permutation 
defined below \eqref{symplectic-symmetry}.
Since $M$ represents a symmetry, the condition $M\,\gamma_A\,M^{\textrm{t}} = \gamma_A$ must be satisfied. 
A contour function for $\mathcal{E}$ 
consistent with this symmetry must satisfy $\mathsf{E}_A(i) = \mathsf{E}_A(j)$. 
It is straightforward to write an analogous relation 
for the case where the two sites that are exchanged by the symmetry 
belong to $A_2$.

($\mathsf{b}$) {\it Local unitary transformations.} 
Considering a symplectic transformation of the form $M=M_G\,\oplus\boldsymbol{1}_{A\setminus G}$, 
where $G \subseteq A_1$,
and denoting by $\mathsf{E}_A'(i)$ the contour function for $\mathcal{E}$
associated with $(\gamma_A^{\textrm{\tiny $\Gamma_2$}} )' \equiv \mathcal{R}_2 \,M\,\gamma_A \, M^{\textrm{t}}\, \mathcal{R}_2$, 
we must have that  $\mathsf{E}_A'(G) = \mathsf{E}_A(G)$ 
for the quantity introduced in (\ref{ent-contour-subregion-neg}). 
An analogous relation is imposed when $G \subseteq A_2$.

($\mathsf{c}$) {\it Upper bounds.}
Consider a symplectic matrix of the form $V = V_{1}\oplus V_{2}$ such that 
    \begin{equation}
    \label{V-gammaA-V-factor}
        V\,\gamma_A\,V^{\textrm{t}} 
        = \gamma_{\Omega_1\Omega_2}\oplus \gamma_{\widetilde{\Omega}_1\widetilde{\Omega}_2}
    \end{equation}
    where $V_{i}$ acts only on $A_i$ and $\mathcal{H}_{A_i} = \mathcal{H}_{\Omega_i} \otimes \mathcal{H}_{\widetilde{\Omega}_i}$. 
    This transformation transforms $\gamma_A^{\textrm{\tiny $\Gamma_2$}}$ as follows
     \begin{equation}
     \label{dec-neg}
        \widetilde{V}\,
        \gamma_A^{\textrm{\tiny $\Gamma_2$}} \,
        \widetilde{V}^{\textrm{t}} 
        = 
        \gamma_{\Omega}^{\textrm{\tiny $\Gamma_2$}}\oplus\gamma_{\widetilde{\Omega}}^{\textrm{\tiny $\Gamma_2$}} \;\;\;\;\qquad\;\;\;\; 
        \widetilde{V} \equiv \mathcal{R}_2\,V\,\mathcal{R}_2 
    \end{equation}
    where $\gamma_{\Omega}^{\textrm{\tiny $\Gamma_2$}}$ and 
    $\gamma_{\widetilde{\Omega}}^{\textrm{\tiny $\Gamma_2$}} $ 
    correspond to the partial transition 
    of the covariance matrices of $\gamma_{\Omega_1\Omega_2}$ 
    w.r.t. $\Omega_2$ 
    and of $\gamma_{\widetilde{\Omega}_1\widetilde{\Omega}_2}$ 
    w.r.t. $\widetilde{\Omega}_2$. 
    From \eqref{dec-neg} and the fact that $\widetilde{V}$ is symplectic, 
    one finds that the total logarithmic negativity can be written as $\mathcal{E}_A = \mathcal{E}_{\Omega}+\mathcal{E}_{\widetilde{\Omega}}$,
    where $\mathcal{E}_\Omega$ and $\mathcal{E}_{\widetilde{\Omega}}$ 
    are obtained from 
    $\gamma_{\Omega}^{\textrm{\tiny $\Gamma_2$}}$ 
    and $ \gamma_{\widetilde{\Omega}}^{\textrm{\tiny $\Gamma_2$}} $ respectively
    in the standard way. 
    Consider $\tilde{A} \subseteq A_1$ 
    and $V_{1}$ (see the text above (\ref{V-gammaA-V-factor})) 
    having the following block structure
    \begin{equation}
        \label{third-blocks}
        V_1 = \widetilde{\mathcal{P}}\,
        \Bigg( \,
        \begin{array}{cc}
            M_{\Omega_1 ;\tilde{A}} \; & 
            M_{\Omega_1; A_1 \setminus \tilde{A}} 
            \\
            \rule{0pt}{.45cm}
            \boldsymbol{0} & M_{\widetilde{\Omega}_1; A_1 \setminus \tilde{A}}
        \end{array}
        \, \Bigg) \, \widetilde{\mathcal{P}}^{\textrm{t}}
    \end{equation}
where $\widetilde{\mathcal{P}}$ corresponds to the reordering of the variables such that each block acts on the spatial subset specified in its second subscript. 
In this case, we impose the following bound
\be
\mathsf{E}_A(\tilde{A}) \leqslant \mathcal{E}_\Omega \,.
\ee
An analogous bound must hold when $\tilde{A} \subseteq A_2$
and $V_{2}$ exhibits the block structure similar to the one showed 
in the r.h.s. of \eqref{third-blocks},
 and therefore $A_1$ is replaced by $A_2$.

Since the construction of the contour functions in 
\eqref{contour-function-EE-projector} and \eqref{contour-function-NEG-projector}, 
for the entanglement entropies and the entanglement negativity respectively, 
are very similar
and \eqref{contour-function-EE-projector} 
do not satisfy (a), (b) and (c) (see Sec.\,\ref{C-V-Gaussian}),
one expects that also \eqref{contour-function-NEG-projector} 
do not satisfy the properties ($\mathsf{a}$), ($\mathsf{b}$) and ($\mathsf{c}$) above, by means of arguments similar to the ones discussed in Sec.\,\ref{C-V-Gaussian}.
Hence, in the following we introduce a weaker version of these properties,
obtained by imposing that the symplectic matrices characterizing the various transformations are also orthogonal, as done in \cite{Coser:2017dtb} 
for the properties discussed in Sec.\,\ref{C-V-Gaussian}.
The construction \eqref{contour-function-NEG-projector} 
satisfies this weaker version of the above properties,
as shown in the following:

($\mathsf{a}$) \textit{Symmetries.} 
When the symplectic $M$ in \eqref{symplectic-symmetry-neg} is also orthogonal, 
the transformation of $\gamma_A^{\textrm{\tiny $\Gamma_2$}}$ is given by
$(\gamma_A^{\textrm{\tiny $\Gamma_2$}} )' = \widetilde{M}\,
\gamma_A^{\textrm{\tiny $\Gamma_2$}}\,\widetilde{M}^{\textrm{t}} $ 
with $\widetilde{M} \equiv M_1 \,\oplus R_2\,M_2\,R_2$ and $\widetilde{M}$ and $\widetilde{M}_2 \equiv R_2\,M_2\,R_2$ are orthogonal and symplectic matrices. 
This implies that the matrix $\widetilde{W}'$, 
occurring in the Williamson decomposition of 
$(\gamma_A^{\textrm{\tiny $\Gamma_2$}} )'$ 
(see \eqref{williamson-dec-gamma-A-transpose}), 
and the corresponding $\widetilde{K}'$, 
which provides the contour  for $(\gamma_A^{\textrm{\tiny $\Gamma_2$}} )'$
(see \eqref{mpf-proj-neg}), 
can be written as follows
\begin{equation}
\label{orth-easy}
    \widetilde{W}' = \widetilde{W}\,\widetilde{M}^{\textrm{t}} 
    \;\;\;\;\qquad\;\;\;\; 
    \widetilde{K}' = \widetilde{K}\,\widetilde{M}^{\textrm{t}}
\end{equation}
in terms of the symplectic matrix $\widetilde{W}$
in the Williamson decomposition \eqref{williamson-dec-gamma-A-transpose} 
of $\gamma_A^{\textrm{\tiny $\Gamma_2$}} $
and of the symplectic and orthogonal matrix $\widetilde{K}$,
providing the contour function for $\mathcal{E}$
through the mode participation function in \eqref{mpf-proj-neg}. 
By using (\ref{orth-easy}) in \eqref{mpf-proj-neg}
and the fact that $X^{(i)} = \widetilde{M}^{\textrm{t}}\,X^{(j)}\,\widetilde{M}$ 
and $X^{(j)} = \widetilde{M}^{\textrm{t}}\,X^{(i)}\,\widetilde{M}$, 
we find 
\begin{equation}
\label{mpf-symm}
    \tilde{p}_k'(i) = \tilde{p}_k(j) 
    \;\;\;\;\qquad\;\;\;\;\; 
    \tilde{p}'_k(j) = \tilde{p}_k(i)
\end{equation}
where $\tilde{p}_k(i)$ and $\tilde{p}'_k(i)$ are
the mode participation function associated 
$\gamma_A^{\textrm{\tiny $\Gamma_2$}}$ and $(\gamma_A^{\textrm{\tiny $\Gamma_2$}} )'$
respectively. 
Since $M$ must be a symmetry,
$\tilde{p}_k(i) = \tilde{p}_k(j)$ must hold in 
\eqref{mpf-symm}. By employing the resulting relations and 
\eqref{contour_constr_neg}, 
it is straightforward to conclude that 
$\mathsf{E}_A(i)=\mathsf{E}_A(j)$.

($\mathsf{b}$) {\it Local unitary transformations.}
In this property we impose that the symplectic matrix 
$M_G$ is also orthogonal.
Since the symplectic matrix
$M = M_G \, \oplus \boldsymbol{1}_{A\,\setminus G}$ 
acts non trivially only on $G \subseteq A_1$, 
the matrices $M$ and $\mathcal{R}_2$ commute.
By employing in \eqref{mpf-proj-neg}
the relation \eqref{orth-easy} and that 
$M^{\textrm{t}}\big( \sum_{i \in G}\,X^{(i)}\big) M 
= \big(\sum_{i \in G}\,X^{(i)}\big)$, 
which is obtained from the fact that $M_G$ is orthogonal, 
one finds that 
$\sum_{i \in G}\, \tilde{p}'_k(i) = \sum_{i \in G}\, \tilde{p}_k(i)$. Then, since 
$(\gamma_A^{\textrm{\tiny $\Gamma_2$}} )'$ and $\gamma_A^{\textrm{\tiny $\Gamma_2$}} $ have the same symplectic spectrum, 
from \eqref{contour_constr_neg} we conclude that 
$\mathsf{E}_A'(G) = \mathsf{E}_A(G)$.

($\mathsf{c}$) {\it Upper bounds.}
Following the proof of the corresponding property reported in \cite{Coser:2017dtb} for the contour function of the entanglement entropies, 
let us consider a symplectic and orthogonal matrix $V = V_1\,\oplus \, V_2$. 
We also need the following Williamson decompositions 
\bea
    & &
    \gamma_{\Omega}^{\textrm{\tiny $\Gamma_2$}} = 
    W_\Omega \big(D_\Omega \oplus D_\Omega \big) W_{\Omega} 
    \;\;\;\qquad \;\;\;
    K_{\Omega} \equiv 
    \big(W_{\Omega}W_\Omega^{\textrm{t}} \big)^{-\frac{1}{2}}\, W_{\Omega}
\\
\rule{0pt}{.6cm}
& &
    \gamma_{\widetilde{\Omega}}^{\textrm{\tiny $\Gamma_2$}} 
    = 
    W_{\widetilde{\Omega}} 
    \big(D_{\widetilde{\Omega}} \oplus D_{\widetilde{\Omega}} \big) W_{\widetilde{\Omega}} 
    \;\;\;\qquad \;\;\;
    K_{\widetilde{\Omega}} \equiv \big(W_{\widetilde{\Omega}}W_{\widetilde{\Omega}}^{\textrm{t}} \big)^{-\frac{1}{2}}\, W_{\widetilde{\Omega}} \,.
\eea
Since also the matrix $\widetilde{V} = V_1 \, \oplus R_2 \, V_2 \,R_2$ 
occurring in \eqref{dec-neg} is symplectic and orthogonal, 
for the matrices $\widetilde{D}$, $\widetilde{W}$
and $\widetilde{K}$,
occurring in \eqref{williamson-dec-gamma-A-transpose} and \eqref{mpf-proj-neg}
respectively, we get 
\begin{equation}
    \widetilde{D} = D_\Omega \oplus D_{\widetilde{\Omega}} 
    \;\;\;\qquad\;\;\;
    \widetilde{W} = \left(W_\Omega \,\oplus\, W_{\widetilde{\Omega}} \right)\, \widetilde{V} 
    \;\;\;\qquad\;\;\;
    \widetilde{K} = \left(K_\Omega \,\oplus\, K_{\widetilde{\Omega}} \right)\, \widetilde{V} \,.
\end{equation}
By using these relations and the ones reported in 
\eqref{contour-function-NEG-projector} 
for the contour function of the logarithmic negativity we find 
\begin{eqnarray}
        \mathsf{E}_A(\tilde{A}) 
        &=& 
        \frac{1}{2} \Tr\! 
        \left[ \left(\,\sum\nolimits_{i \in \tilde{A}} 
        X^{(i)}\right) \widetilde{K}^{\textrm{t}} \,
        \widetilde{F}_1\big( \widetilde{D} \oplus \widetilde{D}\big)  \,\widetilde{K} \,\right]
        \nonumber
        \\
        \rule{0pt}{.65cm}
        &=&  
        \frac{1}{2} \Tr\! 
        \left[\,
        \widetilde{V} 
        \left(\,\sum\nolimits_{i \in \tilde{A}} 
        X^{(i)}\right)\widetilde{V}^{\textrm{t}} \left( K_{\Omega} \oplus K_{\widetilde{\Omega}} \right)^{\textrm{t}} 
        \widetilde{F}_1\big( \widetilde{D} \oplus \widetilde{D}\big) 
        \big( K_{\Omega} \oplus K_{\widetilde{\Omega}} \big)  \right]
        \nonumber
        \\
        \label{ineq-c-property-neg-appA2-step0}
        \rule{0pt}{.65cm}
        &=&  
        \frac{1}{2}\Tr\! 
        \left[ 
        \left(M_{\Omega_1; \tilde{A}} 
        M^{\textrm{t}}_{\Omega_1; \tilde{A}}\,\oplus \boldsymbol{0}_{\Omega_2}\right) K_{\Omega}^{\textrm{t}} \,
        \widetilde{F}_1\big( D_{\Omega} \oplus D_{\Omega}\big) 
        \, K_{\Omega} \,\right]
         \\
         \label{ineq-c-property-neg-appA2}
        \rule{0pt}{.65cm}
        &\leqslant& 
        \frac{1}{2}\Tr\! \left[K_{\Omega}^{\textrm{t}}\;
        \widetilde{F}_1\big( D_{\Omega} \oplus D_{\Omega}\big)  
        \, K_{\Omega} \,\right] 
        = 
        \mathcal{E}(\Omega)
\end{eqnarray}
where it has been used that, 
since $ \displaystyle V_1$ is orthogonal, the following relation holds 
\begin{equation}
    M_{\Omega_1; \tilde{A}} \,M^{\textrm{t}}_{\Omega_1; \tilde{A}} 
    = \boldsymbol{1}_{\Omega_1}-M_{\Omega_1; A_1 \setminus \tilde{A}}\,M_{\Omega_1; A_1 \setminus \tilde{A}}^{\textrm{t}}
\end{equation}
and the inequality in (\ref{ineq-c-property-neg-appA2}) 
is due to the fact that 
the negative contribution in 
(\ref{ineq-c-property-neg-appA2-step0}),
namely
\begin{equation}
    -
        \frac{1}{2}\Tr\! 
        \left[ \Big(M_{\Omega_1; A_1 \setminus \tilde{A}}^{\textrm{t}}\,\oplus \boldsymbol{0}_{\Omega_2}\Big)  
        K_{\Omega}^{\textrm{t}} 
        \,\widetilde{F}_1 \big( D_{\Omega} \oplus D_{\Omega}\big)  K_{\Omega} 
        \Big(M_{\Omega_1; A_1 \setminus \tilde{A}}\,\oplus\boldsymbol{0}_{\Omega_1} \Big)\right]
        \leqslant 0
\end{equation}
has been discarded.

As for the contour functions $\mathsf{E}_A^{(n)}(i)$ with $n \geqslant 2$ 
in \eqref{contour-function-NEG-projector},
the discussions about the properties 
($\mathsf{a}$) and ($\mathsf{b}$) can be adapted straightforwardly,
while the property ($\mathsf{c}$) cannot be formulated 
because these functions of the position $i$ 
do not have a well defined sign in general.

 \section{A relation between contour functions for pure states}
\label{app-relation}

In this appendix we prove the validity of the relation \eqref{neg_ent_loc},
which holds for any pure Gaussian state associated to the bipartite 
spatial domain $A= A_1 \cup A_2$, 
for the special case of the contour functions given by 
\eqref{contour-function-EE-projector} and \eqref{contour-function-NEG-projector}.

In the following analysis, 
we employ a different convention \cite{Botero_2003,Weedbrook2012}
for the order of the operators 
with respect to the one introduced at the beginning of Sec.\ref{sec:contour-entropies}. 
In particular, let us consider the vector 
$\hat{\boldsymbol{r}}\equiv 
\big(\hat{q}_1, \hat{p}_1, \dots ,\hat{q}_{N} , \hat{p}_{N} \big)^{\textrm{t}}$
with $N = |A_1| + |A_2|$,
where the first $2|A_1|$ elements and the remaining $2|A_2|$ elements
correspond to the position and momenta operators 
associated with the sites in $A_1$ and $A_2$, respectively.
Hereafter, we assume $|A_1| \leqslant |A_2|$ without loss of generality. 
In this convention, also the Williamson decomposition takes a slightly different form.
We denote by $\gamma_A$, $\gamma_{A_1}$ and 
$\gamma_{A_2}$ the covariance matrices 
corresponding to the domains $A$, $A_1$ and $A_2$ respectively.

Since $\gamma_A$ is associated to a pure state, its Williamson decomposition reads 
\begin{equation}
\label{gamma-A-Will-dec-app}
    \gamma_A = \frac{1}{2}\,W^{\textrm{t}} \, W
\end{equation}
where $W\in \textrm{Sp}(N)$, namely
\begin{equation}
        W \left(\oplus_{i=1}^{N} J_2\right) W^{\textrm{t}} 
        = 
        \oplus_{i=1}^{N} J_2
        \;\;\;\qquad\;\;\;  
        J_2 \equiv 
    \Bigg( 
    \begin{array}{cc}
            0 \; & 1 \\
            -1 \; & 0
    \end{array}
    \, \Bigg) \,.
\end{equation}

In \cite{Botero_2003} it has been shown that the purity of the state $\gamma_A$
implies that $\gamma_{A_1}$ and $\gamma_{A_2}$ share $|A_1|$ symplectic eigenvalues $\{\sigma_\mu\,; 1\leqslant \mu \leqslant |A_1|\}$,
which provide the symplectic spectrum of $\gamma_{A_1}$,
and that the remaining symplectic eigenvalues of $\gamma_{A_2}$ 
are all equal to $1/2$. 
Thus, the Williamson decompositions of $\gamma_{A_1}$ and $\gamma_{A_2}$
are given respectively by 
\begin{equation}
\label{gamma-A1-A2-dec-app}
    \gamma_{A_1} = 
    W_{A_1}^{\textrm{t}} \left(\oplus_{\mu=1}^{|A_1|} \sigma_{\mu} \boldsymbol{1}_{2} \right) W_{A_1}
    \;\;\qquad \;\;
    \gamma_{A_2} = 
    W_{A_2}^{\textrm{t}} 
    \left[
    \left( \oplus_{\mu=1}^{|A_1|} \sigma_{\mu} \boldsymbol{1}_{2} \right) \oplus \frac{1}{2}\, \boldsymbol{1}_{2(|A_2|-|A_1|)}
    \right] 
    W_{A_2}
\end{equation}
where $W_{A_1} \in $ Sp($|A_1|$) and $W_{A_2} \in $ Sp($|A_2|$). 
Another result of \cite{Botero_2003}, 
that will be useful in our discussion, 
is that $\gamma_A$ can be factorized as follows
\begin{equation}
\label{modewise_ent}
         \gamma_A 
         = 
         \big( W_{A_1} \oplus W_{A_2} \big)^{\textrm{t}} \left[\mathcal{P}^{\textrm{t}}\left( \oplus_{\mu=1}^{|A_1|} \gamma_\mu \right) \mathcal{P} \oplus 
         \frac{1}{2}\,\boldsymbol{1}_{2(|A_2|-|A_1|)}
         \right] 
         \big( W_{A_1} \oplus W_{A_2} \big)
\end{equation}
in terms of the $4 \times 4$ matrices $\gamma_\mu $ 
and $2 \times 2$ matrix $\tau_z$, 
defined respectively by 
\begin{equation}
    \gamma_\mu \equiv 
    \Bigg( 
    \begin{array}{cc}
             \sigma_\mu \boldsymbol{1}_2 \; & y_\mu \tau_z
             \\
             \rule{0pt}{.45cm}
             y_\mu \tau_z \; & \sigma_\mu \boldsymbol{1}_2 
    \end{array}
    \Bigg)
         \;\;\;\qquad \;\;\;
         \tau_z \equiv 
    \Bigg( 
    \begin{array}{cc}
            1 \; & 0 \\
            0 \; & -1
    \end{array}
    \Bigg)
    \;\;\;\qquad \;\;\;
    y_\mu \equiv \sqrt{\sigma_\mu^2-1/4}
\end{equation}
where  $\mathcal{P}$ is a permutation matrix. 
Notice that the factorization of $\gamma_A$ in (\ref{modewise_ent}) 
is different from its Williamson decomposition (\ref{gamma-A-Will-dec-app}).
From (\ref{gamma-A1-A2-dec-app}) and \eqref{modewise_ent}, 
one observes \cite{Botero_2003} that 
the modes defined by the Williamson decompositions 
of $\gamma_{A_1}$ and $\gamma_{A_2}$ in (\ref{gamma-A1-A2-dec-app})
and associated to the shared symplectic eigenvalue $\sigma_\mu$ 
are coupled in  $\gamma_A$ and that such pairing is enforced by 
the permutation matrix $\mathcal{P}$ and the matrices $\gamma_\mu$. 
The decomposition \eqref{modewise_ent} also tells us that 
the remaining $|A_2|-|A_1|$ modes, 
corresponding to the trivial symplectic eigenvalues equal to $1/2$, 
are not coupled among themselves 
and are completely contained in $A_2$. 

Our strategy to get \eqref{neg_ent_loc} is based on the following two steps
(see also (\ref{contour_constr}) and (\ref{contour_constr_neg})):
(I) relate $F_n(\sigma_\mu)$ to $\widetilde{F}_n(\widetilde{\sigma}_\mu)$;
(II) relate the mode participation functions $p^{(1)}_\mu(i)$ and $p^{(2)}_\mu(i)$,
providing the contour functions for the entanglement entropies in $A_1$ and $A_2$ respectively, to the mode participation function $\widetilde{p}_\mu(i)$.

As for the step (I), 
the relation between the spectra of $\rho_{A_1}$ and $\rho_{A_2}$ 
and the spectrum of $\rho_A^{\textrm{\tiny $\Gamma_2$}}$ in this setup
has been explored in previous works 
\cite{Eisler:2015tgq,Ruggiero:2016yjt, DeNobiliThesis}. 
In particular,  
it has been shown \cite{DeNobiliThesis} that, 
for each symplectic eigenvalue $\sigma_\mu$
shared between $\gamma_{A_1}$ and $\gamma_{A_2}$, 
there are two symplectic eigenvalues 
$\tilde{\sigma}_\mu^{<}$  and $\tilde{\sigma}_\mu^{>}$
of $\gamma_{A}^{\textrm{\tiny $\Gamma_2$}}$ 
such that
\begin{equation}
\label{spectrum-pure}
    \tilde{\sigma}_\mu^{<} \equiv \sigma_\mu - \sqrt{\sigma_\mu^2-1/4} 
    \;\;\;\qquad \;\;\;\;
    \tilde{\sigma}_\mu^{>} \equiv \sigma_\mu + \sqrt{\sigma_\mu^2-1/4}
\end{equation}
while the remaining symplectic spectrum of 
$\gamma_{A}^{\textrm{\tiny $\Gamma_2$}}$ is degenerate,
with all the symplectic eigenvalues equal to $1/2$. 
From (\ref{F1-entropy-def})-(\ref{Fn-entropy-def}) and 
(\ref{F1-neg-def})-(\ref{Fn-neg-def}) combined with (\ref{contour_constr}) and (\ref{contour_constr_neg}), it is straightforward to realize that the symplectic eigenvalues equal to $1/2$ do not play any role in \eqref{neg_ent_loc}; 
hence they are ignored in the following discussion. 
Notice that if $\sigma_\mu \neq 1/2$ then $\tilde{\sigma}_\mu^{<} < \tilde{\sigma}_\mu^{>}$. The relation \eqref{spectrum-pure} 
leads us to the relation between $F_n(\sigma_\mu)$ and $\widetilde{F}_n(\tilde{\sigma}_\mu)$ that we are looking for, namely
\begin{equation}
\label{mode_ent_neg}
    \frac{1}{2}
    \left( \widetilde{F}_n(\tilde{\sigma}_\mu^{>}) 
    + 
    \widetilde{F}_n (\tilde{\sigma}_\mu^{<} )
    \right) 
    = 
    \left\{
    \begin{array}{ll}
        \displaystyle
         \; \frac{1-n_{\textrm{\tiny o}}}{2}\,  F_{n_{\textrm{\tiny o}}}(\sigma_\mu)
         \hspace{1.5cm} & 
         \text{odd } n=n_{\textrm{\tiny o}}
         \\
         \rule{0pt}{.7cm}
         \displaystyle
         \left(1-\frac{n_{\textrm{\tiny e}}}{2}\right) 
         F_{n_{\textrm{\tiny e}}/2}(\sigma_\mu)
         & 
         \text{even }n=n_{\textrm{\tiny e}} \,.
    \end{array}
    \right.
\end{equation}
As for the step (II), 
consider the Williamson decomposition of $\gamma_A^{\textrm{\tiny $\Gamma_2$}}$, that reads
\begin{equation}
\label{W-dec-T2-app}
\gamma_A^{\textrm{\tiny $\Gamma_2$}}  
= 
\widetilde{W}^{\textrm{t}} \!
\left( \oplus_{\mu=1}^{N} \tilde{\sigma}_\mu \boldsymbol{1}_2\right) 
\widetilde{W} \,.
\end{equation}
Plugging \eqref{modewise_ent} into (\ref{gamma-T2-from-gamma}), 
we find the following factorization 
\begin{equation}
\label{gamma-T2-factorisation-app}
        \gamma_A^{\textrm{\tiny $\Gamma_2$}}  
        = 
        \left( W_{A_1} \oplus W'_{A_2} \right)^{\textrm{t}} 
        \left[\,
        \mathcal{P}^{\textrm{t}}\left( \oplus_{\mu=1}^{|A_1|} \gamma'_\mu \right) \mathcal{P}
        \oplus 
        \frac{1}{2}\,\boldsymbol{1}_{2(|A_2|-|A_1|)} 
        \, \right] 
        \left( W_{A_1} \oplus W'_{A_2} \right)
\end{equation}
where
\begin{equation}
\label{part_transp_2}
W'_{A_2} \equiv R_2 \, W_{A_2} \,R_2
\;\;\;\qquad\;\;\;
\gamma'_{\mu} \equiv 
    \Bigg( 
    \begin{array}{cc}
             \sigma_\mu \boldsymbol{1}_2 \;& y_\mu \boldsymbol{1}_2
             \\
             \rule{0pt}{.45cm}
             y_\mu \boldsymbol{1}_2 \; & \sigma_\mu \boldsymbol{1}_2 
    \end{array}
    \Bigg)
\end{equation}
and it can easily checked that $W'_{A_2} \in \textrm{Sp}(|A_2|)$. 
As for the $4 \times 4$ matrices $\gamma_\mu'$, 
let us introduce the permutation matrix $\mathcal{P}'$ defined by the condition  $\mathcal{P}'(q_1,p_1,q_2,p_2)^\textrm{t} = (q_1,q_2,p_1,p_2)^\textrm{t}$,
that allows to write $\gamma_\mu'$ as 
\begin{equation}
\label{block-gamma-mu}
    \gamma_\mu' = \mathcal{P}'
    \big( \mathcal{V}_\mu\,\oplus\mathcal{V}_\mu \big)
    \mathcal{P}' 
    \;\;\;\qquad\;\;\; 
\mathcal{V}_\mu \equiv
    \bigg(\, 
    \begin{array}{cc}
             \sigma_\mu  \;& y_\mu 
             \\
             \rule{0pt}{.3cm}
             y_\mu \; & \sigma_\mu 
    \end{array}
    \bigg)
\end{equation}
where the $2 \times 2$ matrix $\mathcal{V}_\mu$ 
can be diagonalized through the  orthogonal matrix $O_\mu$ as follows 
\begin{equation}\label{T-diag}
    \mathcal{V}_\mu = O_\mu^{\textrm{t}} 
    \bigg( 
    \begin{array}{cc}
             \tilde{\sigma}_\mu^{<}  \;& 0 
             \\
             \rule{0pt}{.3cm}
             0 \; & \tilde{\sigma}_\mu^{>}
    \end{array}
    \bigg)
     \,O_\mu 
    \;\;\;\;\; \qquad  \;\;\;\;\;
    O_\mu = \frac{1}{\sqrt{2}}
    \bigg( 
    \begin{array}{cc}
             -1  \;& 1 
             \\
             \rule{0pt}{.3cm}
             1 \; & 1
    \end{array}
    \bigg)
\end{equation}
and its eigenvalues are given in (\ref{spectrum-pure}).
Hence, for (\ref{block-gamma-mu}) we get
\begin{equation}\label{W-dec-mu}
    \gamma_\mu'=
    k_\mu^{\textrm{t}} \left( \tilde{\sigma}_\mu^{<}\,\boldsymbol{1}_2 \oplus \tilde{\sigma}_\mu^{>}\,\boldsymbol{1}_2 \right) k_\mu 
\;\;\;\qquad\;\;\; 
    k_\mu \equiv  \mathcal{P}'\,(O_\mu\oplus O_\mu)\,\mathcal{P}'
\end{equation}
where $k_\mu$ is an orthonormal symplectic matrix, 
and therefore \eqref{W-dec-mu} provides 
the Williamson decomposition of $\gamma_\mu'$. 
Combining (\ref{W-dec-T2-app}) and (\ref{gamma-T2-factorisation-app}),
we find that the symplectic matrix $\widetilde{W}$ 
can be written in the following factorized form
\begin{equation}\label{fact-W}
        \widetilde{W} =  \left[{\mathcal{P}^{\textrm{t}}}\left( \oplus_{\mu=1}^{|A_1|} k_\mu\right) \mathcal{P} \oplus \boldsymbol{1}_{2(\ell_2-|A_1|)}\right]\left( W_{A_1} \oplus W'_{A_2} \right) .
\end{equation}
Substituting this expression in \eqref{K-from-W-tilde}, 
we get 
\begin{equation}\label{fact}
        \widetilde{K} =  \left[\mathcal{P}^{\textrm{t}} \left( \oplus_{\mu=1}^{|A_1|} k_\mu\right) \mathcal{P} \oplus\boldsymbol{1}_{2(|A_2|-|A_1|)}\right]\left( K_{A_1} \oplus K'_{A_2} \right) 
        \;\;\;\qquad\;\;
        K'_{A_2} \equiv  R_2 \,K_{A_2} \,R_2
\end{equation}
where $K_{A_1}$ and $K_{A_2}$ are the orthogonal 
and symplectic matrices defined 
by plugging $W_{A_1}$ and $W_{A_2}$ respectively
in the r.h.s. of \eqref{K-from-W}.

We find it convenient to order the relevant symplectic eigenvalues of $\gamma_A^{\textrm{\tiny $\Gamma_2$}}$ in such a way that the modes labeled by  $2\mu-1$ and $2\mu$ are associated with the symplectic eigenvalues $\tilde{\sigma}_\mu^{<}$ and $\tilde{\sigma}_\mu^{>}$ respectively,  
for $1\leqslant \mu \leqslant  |A_1|$.

At this point, all the ingredients to write an explicit relation between the mode participation functions $p_\mu^{(1)}$ and $p_\mu^{(2)}$, 
that appear in the construction of $\mathsf{S}_{A_1}(i)$ 
and $\mathsf{S}_{A_2}$(i) (see \eqref{mpf-proj-ent}), 
and the mode participation function $\tilde{p}_\mu(i)$ associated with $\mathsf{E}_A(i)$ (see \eqref{mpf-proj-neg}) are available. 
Thus, from \eqref{mpf-proj-ent}, \eqref{mpf-proj-neg} and \eqref{fact}, 
we find 
\begin{eqnarray}
\label{rel-MPF-tilde-app-II}
    \Tilde{p}_{2\mu-1}(i) 
    &=& 
    \Tilde{p}_{2\mu}(i) =
    \frac{1}{2}\, p^{(1)}_{\mu}(i)
    \;\;\quad\;\;\;
    i \in A_1 
    \\
\rule{0pt}{.7cm}
    \Tilde{p}_{2\mu-1}(i) 
    &=& 
    \Tilde{p}_{2\mu}(i) =
    \frac{1}{2}\, p^{(2)}_{\mu}(i)
    \;\;\quad\;\;\;
    i \in A_2
\end{eqnarray}
which can be used to prove \eqref{neg_ent_loc}. 
Indeed, in the step (I) we have obtained \eqref{mode_ent_neg}, 
which provides a relation between the functions $F_n(\sigma_\mu)$ and $\widetilde{F}_n(\tilde{\sigma}_\mu)$ occurring in the constructions 
\eqref{contour-function-EE-projector} 
and \eqref{contour-function-NEG-projector}, 
while the step (II) has provided 
\eqref{rel-MPF-tilde-app-II}, which  is the relation between the mode participation functions appearing in the same expressions. 
For instance, when 
$i \in A_1$ and $n=n_{\textrm{\tiny e}}$ is even, 
these relations give \eqref{neg_ent_loc} specialized to this case 
as follows 
\begin{equation}
\label{neg-cont-ps-app-ne-A1}
        \mathsf{E}_A^{(n_{\textrm{\tiny e}})}(i) 
        =
        \sum_{\mu = 1}^{2|A_1|} 
        \Tilde{p}_{\mu}(i) \,
        \widetilde{F}_{n_{\textrm{\tiny e}}}(\tilde{\sigma}_\mu) 
        = 
        \sum_{\mu = 1}^{|A_1|} 
        \frac{p^{(1)}_{\mu}(i)}{2} 
        \left(
        \widetilde{F}_{n_{\textrm{\tiny e}}} ( \tilde{\sigma}_\mu^{>})
        +
        \widetilde{F}_{n_{\textrm{\tiny e}}} ( \tilde{\sigma}_\mu^{<})
        \right)
        =
        \left(
        1-\frac{n_{\textrm{\tiny e}}}{2}
        \right) 
        \mathsf{S}^{(n_{\textrm{\tiny e}}/2)}_{A_1}(i)
        \,.
\end{equation}
The steps performed in (\ref{neg-cont-ps-app-ne-A1}) 
can be easily adapted to obtain the relation 
\eqref{neg_ent_loc} in the remaining cases,
namely for $\displaystyle i \in A_2$ or $n=n_{\textrm{\tiny o}}$ odd.

\section{Algorithm for the Williamson decomposition}
\label{app-W-matrix}

In this appendix we describe the algorithm employed 
to compute numerically the Williamson decomposition, 
which has been constructed by 
adapting the procedure discussed in \cite{Houde:2024mkj}.
This algorithm provides the symplectic matrices $W$ and $\widetilde{W}$
for $\gamma_A$ and $\gamma_A^{\textrm{\tiny $\Gamma_2$}}$ respectively,
which determine the symplectic and orthogonal matrices $K$ and $\widetilde{K}$, 
through (\ref{K-from-W}) and (\ref{K-from-W-tilde}) respectively.
Finally, $K$ and $\widetilde{K}$ give the mode participation functions 
$p_k(i)$ and $\tilde{p}_k(i)$
(see (\ref{K-matrix-def})-(\ref{mode-part-W}) 
and (\ref{tilde-K-def})-(\ref{tilde-pk-i-def})), 
that characterize the contour functions 
\eqref{contour-function-EE-projector} 
and \eqref{contour-function-NEG-projector}
explored in this manuscript. 

In the following, this procedure is described for a generic real, symmetric and positive definite $(2L_A)\times (2L_A)$ matrix $\gamma$;
hence it can be adapted to $\gamma_A$ and $\gamma_A^{\textrm{\tiny $\Gamma_2$}}$
in a straightforward way. 
Given $\gamma$,
one first constructs the following real and antisymmetric matrix 
    \begin{equation}
        G
        \equiv 
        \big( \sqrt{\gamma}\, \big)^{-1}  
        J \,
        \big( \sqrt{\gamma}\, \big)^{-1}
    \end{equation}
    in terms of the 
    $(2L_A)\times (2L_A)$ antisymmetric matrix (\ref{J-mat-def}).
Then, we determine the normal form of $G$, namely
\cite{Horn_Johnson_1985}
        \begin{equation}
        \label{schur-dec-M}
        G \,=\, 
        O \, \mathcal{G} \,  O^{\textrm{t}}
             \;\;\;\;\qquad\;\;\;
             \mathcal{G} 
             \,\equiv\, 
             \bigoplus_{k=1}^{L_A} 
        \begin{pmatrix}
        0 & g_k \\
        - g_k & 0
             \end{pmatrix}
        \end{equation}
where $g_k > 0$ and $O$ is orthogonal,
which can be found by employing the real Schur decomposition. 
The orthogonal matrix $O$ is not unique \cite{Horn_Johnson_1985}.
However, only the matrix $W$ occurring in the Williamson decomposition depends on $O$ (see \eqref{D-W-from-schur} below) and it is known that $W$ is not unique as well
\cite{de2006symplectic}. 
Then, let us introduce the 
permutation matrix $\mathcal{P}_1$ such that
\begin{equation}
\mathcal{P}_1^{\textrm{t}}  \;
\mathcal{G}\,
\mathcal{P}_1
\,=\, 
    \Bigg( 
    \begin{array}{cc}
                    0 & \bigoplus_{k=1}^{L_A}g_{k} 
                    \\
                    -\bigoplus_{k=1}^{L_A}g_{k} & 0
    \end{array}
    \Bigg)
\end{equation}
or alternatively
\begin{equation}
    \mathcal{P}_1^{\textrm{t}}  \,
    \big(
    \hat{q}_1, \hat{p}_1, \dots  , \hat{q}_{L_A} , \hat{p}_{L_A} \big)^{\textrm{t}}
    = 
    \big(
    \hat{q}_1, \dots \hat{q}_{L_A} , \hat{p}_1, \dots \hat{p}_{L_A} \big)^{\textrm{t}} \,.
\end{equation}
Finally, the matrices $D$ and $W$ providing the Williamson decomposition of $\gamma_A$ in (\ref{williamson-dec-gamma-A})
are given respectively by 
    \begin{equation}
    \label{D-W-from-schur}
    {
    D = \left( \oplus_{k=1}^{L_A} g_k \right) ^{-1}
    }
    \;\;\;\qquad\;\;\;\;
    W = 
    \big(  D \oplus D \big)^{-1/2}\, \mathcal{P}_1^{\textrm{t}}\,O^{\textrm{t}}\,\sqrt{\gamma}
    \end{equation}
which satisfy the relation 
$\gamma = W^{\textrm{t}} (  D \oplus D ) W$
and the condition $W \in \textrm{Sp}(L_A)$.

We remark that, 
since all the covariance matrices considered in our analysis are block diagonal, 
a faster algorithm is obtained by adapting the above procedure to this case.


\bibliographystyle{nb}

\bibliography{refsNegContour}

\end{document}